\documentclass[final,3p,times,preprint]{elsarticle}

\usepackage{hyperref}
\hypersetup{
     colorlinks=true,       
     linkcolor=blue,          
     citecolor=blue,        
     urlcolor=blue,           
}

\usepackage{amssymb}
\usepackage{amsmath}
\usepackage{bm}
\usepackage[subrefformat=parens]{subcaption}
\usepackage{xcolor}
\usepackage{slashed}
\usepackage{hyperref}
\usepackage{ulem}

\biboptions{square}

\journal{Annals of Physics}

\cortext[cor1]{YITP-24-63, J-PARC-TH-0305}

\begin{document}

\begin{frontmatter}



\title{
Electromagnetic response of dense quark matter around color-superconducting phase transition and  QCD critical point}

\author[label1,label2]{Toru Nishimura}
\author[label2,label3]{Masakiyo Kitazawa}
\author[label2]{Teiji Kunihiro}

\affiliation[label1]{
organization={Department of Physics, 
Osaka University},
addressline={}, 
city={Toyonaka},
postcode={560-0043}, 
state={Osaka},
country={Japan}
}

\affiliation[label2]{
organization={Yukawa Institute for Theoretical Physics, 
Kyoto University},
addressline={}, 
city={Kyoto},
postcode={606-8502}, 
state={Kyoto},
country={Japan}
}

\affiliation[label3]{
organization={J-PARC Branch, KEK Theory Center, 
Institute of Particle and Nuclear Studies, KEK},
addressline={}, 
city={Tokai},
postcode={319-1106}, 
state={Ibaraki},
country={Japan}
}

\begin{abstract}
We explore how the electric conductivity and associated relaxation time are modified near the QCD critical point and the phase transition to a color superconducting phase using the two-flavor Nambu-Jona-Lasinio model with finite current quark masses.
We give a comprehensive account of the nature of the soft modes associated with these phase transitions and how they affect the photon self-energy when 
the system approaches these phase transitions in a combined way with an emphasis on the common and different aspects in the two transitions.
The formalism developed for describing the paraconductivity in metallic superconductors is used for the analysis of the photon self-energy.
We show that the transport coefficients calculated from the self-energy show anomalous enhancements in both cases with different critical exponents for the individual transitions. 
We briefly discuss the possibility of detecting the enhancements in the relativistic heavy-ion collisions  
 in the present and future facilities.
\end{abstract}

\begin{keyword}
color superconductivity \sep 
QCD critical point \sep 
soft mode \sep 
electric conductivity
\end{keyword}

\end{frontmatter}

\section{Introduction}
\label{sec:Introduction}

Revealing the phase structure of hot and dense matter governed 
by QCD is one of the central subjects of 
the present theoretical and experimental nuclear physics. 
It is expected that such a matter under extreme conditions admits rich phase transitions~\cite{Fukushima:2010bq} including the first-order phase transition in the high-density 
and low-temperature region, 
the QCD critical point(s) (QCD-CP) as the endpoint(s) of the first-order transition line~\cite{Asakawa:1989bq,Kitazawa:2002jop}, 
and the onset of color superconducting (CSC) phases~\cite{Alford:2007xm}. 
The hot and dense matter can be created in the intermediate stage of the 
relativistic heavy-ion collisions (HIC)~\cite{Yagi:2005yb}. In the
HIC experiments, the baryon number density of the produced matter can be varied by the beam-energy scan, i.e. changing the collision energy~\cite{Bzdak:2019pkr,STAR:2021iop}.
Such experiments are now ongoing at the Relativistic Heavy Ion Collider (RHIC) at BNL, the Super Proton Synchrotron (SPS) at CERN, and the Heavy Ion Synchrotron SIS18 at GSI, and 
more refined experimental studies will be pursued at future facilities~\cite{Galatyuk:2019lcf}.

To explore the phase transitions of QCD in these experiments, it is of basic importance to clarify 
adequate experimental observables for their identification.
Promising candidates that have been discussed widely are observables that are sensitive to 
fluctuations of the order parameter enhanced around the critical point of the second-order phase transitions~\cite{Stephanov:1998dy}.
In the case of the QCD-CP, 
active studies have been being made to extract fluctuations of various observables including their higher-order cumulants by the event-by-event analyses~\cite{Asakawa:2015ybt,Bzdak:2019pkr,Bluhm:2020mpc}.

Because of the penetrating property in the strongly interacting medium, 
the electromagnetic probes would also be useful to detect the fluctuations enhanced at the phase transitions.  
In fact, the present authors have recently shown~\cite{Nishimura:2022mku,Nishimura:2023oqn} that
the dilepton-production rate (DPR) is a promising probe for the search for not only the QCD-CP but also the two-flavor superconductor (2SC), which is one of the CSC phases.
The critical fluctuations near a second-order phase transition form the soft modes, i.e. the collective modes that become massless at the transition point. In Refs.~\cite{Nishimura:2022mku,Nishimura:2023oqn}, 
it was shown that the DPR is significantly affected by the soft modes near the phase transition to the 2SC (2SC-PT) and the QCD-CP.
More specifically, a prominent enhancement of the DPR occurs near these phase transitions, especially in the low energy/momentum region, triggered by the scatterings of photons via the soft modes. 
If such enhancements 
were observed in the HIC,
it could be a signature of these 
phase transitions~\cite{Nishimura:2023not}.

The purpose of the present paper is twofold.
First, using the two-flavor Nambu--Jona-Lasinio (NJL) model as an effective model of QCD,
we give a systematic and extended analysis of the soft modes of the 2SC-PT and QCD-CP.
Although most of the calculational procedures are 
based on the formulation given in Refs.~\cite{Kitazawa:2005vr,Fujii:2003bz,Fujii:2004jt,Nishimura:2022mku,Nishimura:2023oqn}, 
an emphasis will be put on the common and different properties of 
the soft modes associated with these phase transitions.
Moreover, some extensions of the formalism are made, such as an inclusion of the finite quark mass in the analysis of the diquark mode.
The second purpose is the analysis of the electric conductivity and associated relaxation time. 
We shall explore how these transport coefficients are affected by the soft modes near the 2SC-PT and QCD-CP.
Although the electric conductivity of hot and dense hadronic matter has been investigated 
in various studies~\cite{Arnold:2000dr,Arnold:2003zc,Teaney:2006nc,Cassing:2013iz,Greif:2014oia,Aarts:2020dda,Kaczmarek:2022ffn}, its quantitative behavior near the 2SC-PT and QCD-CP has not been studied so far to the best of our knowledge. 
These coefficients are important inputs for the hydrodynamic models for the spacetime evolution of heavy-ion reactions~\cite{Hirono:2012rt,Nakamura:2022wqr,Mayer:2024kkv}. 
Moreover, since they are directly connected to the low energy-momentum structure of the DPR, their analysis will provide insights into the quantitative understanding of the DPR near the 2SC-PT and QCD-CP.

To investigate the effects of the soft modes on the transport coefficients, 
we employ a similar formalism used for the analysis of the DPR in Refs.~\cite{Nishimura:2022mku,Nishimura:2023oqn}. 
We calculate the modification of the photon self-energy due to
the Aslamazov–Larkin (AL)~\cite{AL:1968}, Maki–Thompson (MT)~\cite{Maki:1968,Thompson:1968} and density of states (DOS) terms~\cite{Larkin:book},
all of which involve the propagators of the soft modes.
These terms have been used in condensed matter physics to investigate anomalous excesses of conductivity near the critical temperature $T_c$ of metallic superconductors.
Although they incorporate effects of the soft modes only at the mean-field level in the sense that nonlinear effects are not included, they are known to describe the excess of conductivity near $T_c$ well in metallic superconductors~\cite{Larkin:book,tinkham2004introduction}. Therefore, their analysis should be a good starting point to investigate the transport coefficients near the 2SC-PT and QCD-CP quantitatively. 
Whereas the formalism to obtain the transport coefficients is basically the same as our previous study, we perform some extensions of the analysis and elucidate detailed calculational procedures skipped in Refs.~\cite{Nishimura:2022mku,Nishimura:2023oqn}.
Through the simultaneous formalisms for the 2SC-PT and QCD-CP, we also clarify the similarities and differences in the analysis and physical properties.

We show that the transport coefficients are divergent at the 2SC-PT and QCD-CP. Their critical exponents are calculated analytically. It will be shown that the electric conductivities
have different critical exponents for the QCD-CP and 2SC-PT. 
The behavior of the coefficients on the QCD phase diagram will also be calculated numerically.

The organization of the present paper is as follows. 
Section~\ref{sec:Model} is devoted to the explanation of the model and its phase diagram. 
In Sec.~\ref{sec:Soft-modes}, we discuss the propagators of the soft modes associated with the 2SC-PT and the QCD-CP. Effective descriptions for their low energy-momentum behavior are then introduced in Sec.~\ref{sec:Effective-theory}.
In Sec.~\ref{sec:Self-energy}, we calculate the photon self-energy 
including the soft modes. 
We then show the analytical and numerical results of the transport coefficients in Sec.~\ref{sec:Transport-coeff}. 
Finally, we summarize and conclude in Sec.~\ref{sec:Summary}.

\section{Model and phase diagram}
\label{sec:Model}

To investigate the phase transitions in dense quark matter and their effects on the transport coefficients,
we employ the 2-flavor NJL model~\cite{Hatsuda:1994pi, Buballa:2003qv} 
\begin{align}
\mathcal{L} &= \bar{\psi} i (\slashed{\partial} - m) \psi 
+G_S [(\bar\psi \psi)^2 + (\bar\psi i \gamma_5 \vec{\tau} \psi)^2]
+G_D (\bar\psi i \gamma_5 \tau_2 \lambda_A \psi^C)
                       (\bar\psi^C i \gamma_5 \tau_2 \lambda_A \psi),
\label{eq:Lagrangian}
\end{align}
where $\psi(x)$ is the quark field and 
$\psi^C (x) = i \gamma_2 \gamma_0 \bar\psi^T (x)$.
$\vec{\tau} = (\tau_1, \tau_2, \tau_3)$ are the Pauli matrices for the flavor $SU(2)_f$
and $\lambda_A~(A = 2,5,7)$ are the antisymmetric components 
of the Gell-mann matrices for the color $SU(3)_c$.
The scalar coupling constant $G_S = 5.50~\rm{GeV^{-2}}$ and 
the three-momentum cutoff $\Lambda = 631~{\rm MeV}$ are determined
so as to reproduce the pion mass $m_{\pi} = 138~\rm{MeV}$ and
the pion decay constant $f_{\pi} = 93~\rm{MeV}$
at the current quark mass $m = 5.5~{\rm MeV}$~\cite{Hatsuda:1994pi}.
Since the value of the diquark coupling 
$G_D$ is not constrained by the vacuum property, we treat it as a free parameter and vary within $G_D/G_S = 0.6$--$0.7$, which is an intermediate range in various estimates~\cite{Buballa:2003qv}.

\begin{figure}[t]
\centering
\includegraphics[keepaspectratio, scale=0.47]{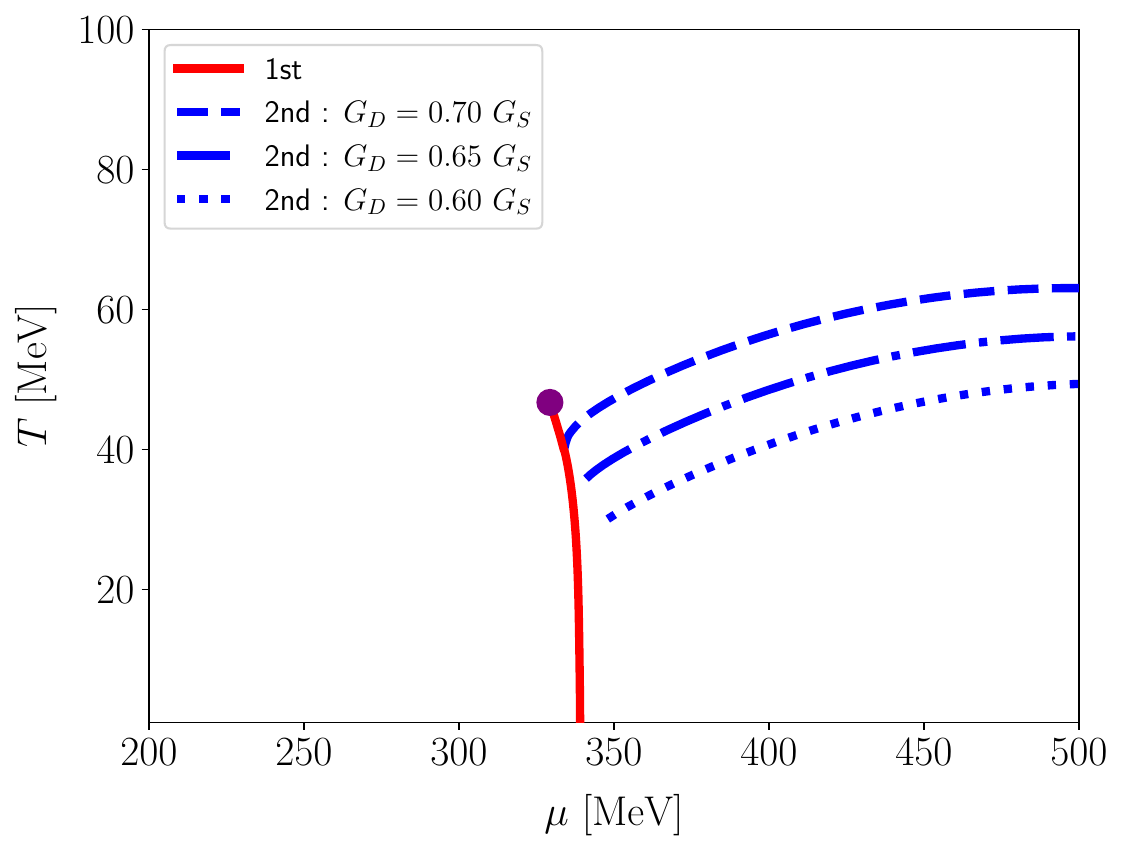}
\caption{
Phase diagram calculated by the mean-field approximation 
in the 2-flavor NJL model (\ref{eq:Lagrangian}). 
The solid line shows the first-order phase transition calculated with $G_D=0.70G_S$.
The dashed, dash-dotted, and dotted lines are the second-order 2SC-PT 
for $G_D/G_S = 0.70$, $0.65$, and $0.60$, respectively.
The QCD-CP is indicated by the big dot 
located at $(T_{\rm CP}, \mu_{\rm CP}) \simeq (46.712, 329.34)$~MeV.
}
\label{fig:Phase-D}
\end{figure}

In the mean-field approximation (MFA) assuming the nonzero chiral condensate $\langle \bar\psi \psi \rangle$
and diquark condensate $\langle \bar\psi^C i \gamma_5 \tau_2 \lambda_A \psi \rangle$, the thermodynamic potential per unit volume 
$\omega_{\rm MFA} = \Omega_{\rm MFA}/V$ at temperature $T$ and quark chemical potential $\mu$ is calculated to be~\cite{Kitazawa:2005vr}
\begin{align}
\omega_{\rm MFA} =&\ \frac{(M-m)^2}{4G_S} + \frac{|\Delta|^2}{4G_D}
-4 \int \frac{d^3p}{(2\pi)^3} 
\bigg\{ E_{\bm{p}} + T{\rm log} 
\big( 1 + e^{-\xi_+ / T} \big) \big( 1 + e^{-\xi_- / T} \big) \nonumber \\
&\hspace{0.8cm} +\ \epsilon_+ + {\rm sgn}(\xi_-) \epsilon_-
+ 2T {\rm log} \big( 1 + e^{-\epsilon_+ / T} \big) 
\big( 1 + e^{- {\rm sgn}(\xi_-) \epsilon_- / T} \big)
\bigg\},
\label{eq:thermodynamic-potential} \\
&\hspace{0.3cm}
E_{\bm{p}} = \sqrt{\bm{p}^2 + M^2}, 
\qquad
\xi_\pm = E_{\bm{p}} \pm \mu ,
\qquad
\epsilon_\pm = \sqrt{\xi_\pm^2 + |\Delta|^2} ,
\end{align}
where $M = m-2G_S\langle \bar\psi \psi \rangle$ and
$\Delta = -2G_D\langle \bar\psi^C i \gamma_5 \tau_2 \lambda_A \psi \rangle$
are the constituent quark mass and the gap of the 2SC, respectively.
The expectation values $\langle \bar\psi \psi \rangle$ and $\langle \bar\psi^C i \gamma_5 \tau_2 \lambda_A \psi \rangle$ for a given set of $T$ and $\mu$ 
are given by minimizing $\omega_{\rm MFA}$.
The stationary condition gives the gap equations $\partial \omega_{\rm MFA}/\partial M = 0$ and
$\partial \omega_{\rm MFA}/\partial \Delta = 0$, 
which are explicitly given by 
\begin{align}
M-m =&~8G_S M \int \frac{d^3p}{(2\pi)^3} \frac{1}{E_{\bm{p}}}
\bigg\{
1 - n(\xi_+) - n(\xi_-) + 
\frac{\xi_+}{\epsilon_+} {\rm tanh} \frac{\epsilon_+}{2T}
+ \frac{\xi_-}{\epsilon_-} {\rm tanh} \frac{\epsilon_-}{2T}
\bigg\},
\label{eq:gap-chiral} \\
\Delta =&~8G_D \Delta \int \frac{d^3p}{(2\pi)^3} 
\bigg\{
\frac{1}{\epsilon_+} {\rm tanh} \frac{\epsilon_+}{2T}
+ \frac{1}{\epsilon_-} {\rm tanh} \frac{\epsilon_-}{2T}
\bigg\},
\label{eq:gap-diquark}
\end{align}
where $n (x) = 1 / (1 + e^{x/T})$ is the Fermi distribution function.
In Fig.~\ref{fig:Phase-D}, we show the phase diagram 
in the $T$--$\mu$ plane obtained by the MFA.
The solid line shows the first-order transition line 
and the circle marker at 
$(T_{\rm CP},\, \mu_{\rm CP}) \simeq (46.712,\, 329.34)~{\rm MeV}$ 
indicates the QCD-CP. 
The dashed, dash-dotted, and dotted lines show the second-order 2SC-PT 
for $G_D/G_S = 0.70$, $0.65$, and $0.60$, respectively.
The critical temperature of the 2SC-PT increases as $G_D$ becomes larger.

The phase transition to the 2SC in the MFA is of second order as shown in Fig.~\ref{fig:Phase-D}.
The order of this phase transition, however, may become a first order
owing to the fluctuations of gluon fields~\cite{Matsuura:2003md,Giannakis:2004xt,Fejos:2019oxz} that are not included in our analysis.
The transition to 2SC can also be a crossover, because of the absence of global symmetry distinguishing the 2SC phase from the normal one~\cite{Alford:2007xm}.
However, since a definite conclusion on the order of the 2SC-PT has not been obtained so far to the best of the author's knowledge, 
we use the result of the MFA in the following.

\section{Soft modes}
\label{sec:Soft-modes}

In this section, we examine the collective excitations of 
the diquark $\bar\psi^C i \gamma_5 \tau_2 \lambda_A \psi$ 
and chiral $\bar\psi\psi$ fields, which couple to the soft modes of the 2SC-PT and QCD-CP, respectively~\cite{Kitazawa:2005vr,Fujii:2003bz,Fujii:2004jt}.
Most of this section is 
a combined recapitulation of Refs.~\cite{Kitazawa:2005vr,Fujii:2003bz,Fujii:2004jt,Nishimura:2022mku,Nishimura:2023oqn}, with
an emphasis on the common and different properties of the soft modes associated with these phase transitions.
Some extensions of the formalism are also made, such as an inclusion of the finite quark mass in the analysis of the diquark field.

\subsection{Response functions}
\label{response}

According to the linear response theory~\cite{fetter2012quantum},
dynamical properties of the fluctuations of a field operator ${\cal O}(\bm{x},t)$ are encoded in the response function, which is equivalent to the retarded Green's function 
\begin{align}
    D^R(\bm{k},\omega) =&\ \int d^3xdt e^{i\omega t - i\bm{k}\cdot\bm{x}} D^R(\bm{x},t),
    \label{eq:Xi^R}\\
    D^R(\bm{x},t) =&\ -i \langle [ {\cal O}(\bm{x},t),{\cal O}(\bm{0},0)] \rangle \theta(t),
    \label{eq:Xi^R(x)}
\end{align}
for a bosonic operator ${\cal O}(\bm{x},t)$ with $[A,\,B]=AB-BA$ being the commutator.
The poles of $D^R(\bm{k},\omega)$ at $\omega=\omega(\bm{k})$, 
which exist in the lower-half complex-energy plane, represent the collective modes that couple to ${\cal O}(\bm{x},t)$. 
When the imaginary part of $\omega(\bm{k})$ is small, the dynamical structure factor defined by
\begin{align}
S(\bm{k},\omega)=-\frac1\pi \frac1{e^{\omega/T}-1} {\rm Im} D^R(\bm{k},\omega) ,
\end{align}
has a peak around $\omega={\rm Re}\, \omega(\bm{k})$,
which conspicuously shows the existence of the collective mode.
The response functions that couple to the soft modes of the 2SC-PT and QCD-CP are given by substituting 
\begin{align}
\hat{\delta}_A(\bm{x},t) = \bar\psi^C(\bm{x},t) i \gamma_5 \tau_2 \lambda_A \psi(\bm{x},t) 
\qquad \mbox{and} \qquad \hat\sigma(\bm{x},t) = \bar\psi(\bm{x},t)\psi(\bm{x},t),
\label{eq:delta-sigma}
\end{align}
into the operator ${\cal O}(\bm{x},t)$ 
in Eq.~\eqref{eq:Xi^R(x)}, respectively,
which are denoted as $D^R_D(\bm{k},\omega)$ and $D^R_S(\bm{k},\omega)$ in what follows.

\begin{figure}[t]
\centering
\includegraphics[keepaspectratio, scale=0.5]{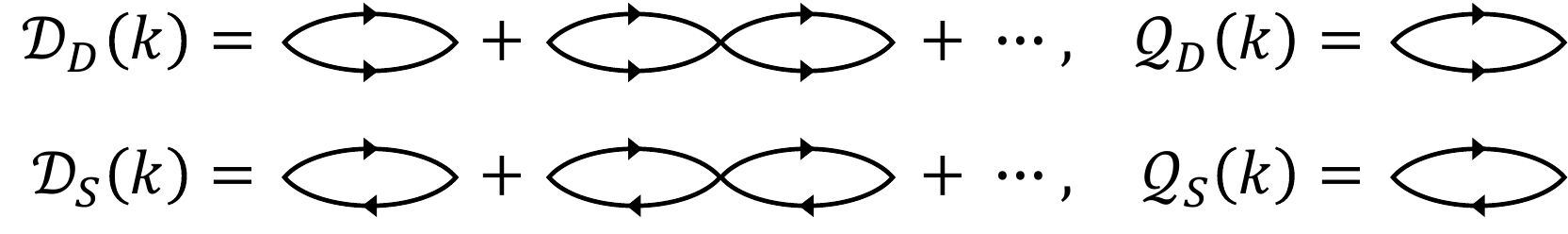}
\caption
{
Diagrammatic representation of Eq.~(\ref{eq:D}).
The single lines denote the quark propagator.
}
\label{fig:D-Q}
\end{figure}

To calculate $D^R_D(\bm{k},\omega)$ and $D^R_S(\bm{k},\omega)$, it is convenient to use the imaginary-time (Matsubara) Green's functions corresponding to the respective response functions
\begin{align}
    {\cal D}_D(k) =&\ {\cal D}_D(\bm{k},i\nu_n)
    = \int d^3xdt e^{i\nu_n\tau-i\bm{k}\cdot\bm{x}}
    \langle T_\tau \hat\delta_A(\bm{x},\tau) \hat\delta_A^\dagger(\bm{0},0)\rangle ,
    \label{eq:XiDelta}
    \\
    {\cal D}_S(k) =&\ {\cal D}_S(\bm{k},i\nu_n)
    = \int d^3xdt e^{i\nu_n \tau-i\bm{k}\cdot\bm{x}}
    \langle T_\tau \hat\sigma(\bm{x},\tau)\hat\sigma(\bm{0},0) \rangle,
    \label{eq:XiS}
\end{align}
where $T_\tau$ is the time-ordering operator and $k=(\bm{k},i\nu_n)$ is the 
four-momentum in imaginary time with $\nu_n=2\pi n/T$ 
being the Matsubara frequency for bosons.
The response functions are obtained from Eqs.~\eqref{eq:XiDelta} 
and~\eqref{eq:XiS} by the analytic continuation
\begin{align}
    D^R_\gamma(\bm{k},\omega) =&\ {\cal D}_\gamma(k)|_{i\nu_n\to\omega+i\eta}, \quad \gamma=D, S,
\end{align}
with $\eta$ being a positive infinitesimal number.

\subsection{Random-phase approximation}
\label{sec:RPA}

We calculate Eqs.~\eqref{eq:XiDelta} and~\eqref{eq:XiS}
in the random-phase approximation (RPA), where the Green's functions are given by
\begin{align}
    {\cal D}_\gamma(k) = \frac{{\cal Q}_\gamma(k)}{ 1 + G_\gamma{\cal Q}_\gamma(k)}
    = {\cal Q}_\gamma(k) - {\cal Q}_\gamma(k) G_\gamma {\cal Q}_\gamma(k) + \cdots,
    \label{eq:D}
\end{align}
with the unperturbed correlation functions
\begin{align}
    {\cal Q}_D(k) 
    =&\ \int d^3xdt e^{i\nu_n \tau-i\bm{k}\cdot\bm{x}}
    \langle T_\tau \hat\delta_A(\bm{x},\tau) \hat\delta_A^\dagger(\bm{0},0) \rangle_{\rm free}
    \notag \\
    =&\
    \int_p {\rm Tr}_{c,f,D} [ i\gamma_5 \tau_2 \lambda_A {\cal G}_0(p) i\gamma_5 \tau_2 \lambda_A {\cal G}_0(k-p)^T]
    \notag \\
    =&\ -2N_f(N_c-1) \int_p {\rm Tr}_D[ {\cal G}_0(p) {\cal G}_0(k-p)],
    \label{eq:Q_Delta}
    \\
    {\cal Q}_S(k) 
    =&\ \int d^3xdt e^{i\nu_n \tau-i\bm{k}\cdot\bm{x}}
    \langle T_\tau \hat\sigma(\bm{x},\tau)\hat\sigma(\bm{0},0) \rangle_{\rm free}
    \notag \\
    =&\
    \int_p {\rm Tr}_{c,f,D} [ {\cal G}_0(p) {\cal G}_0(k+p)]
    \notag \\
    =&\ -2N_f N_c \int_p {\rm Tr}_D [ {\cal G}_0(p) {\cal G}_0(k+p)].
    \label{eq:Q_S}
\end{align}
Here, $\langle\cdot\rangle_{\rm free}$ denotes the expectation value 
in the non-interacting system and
\begin{align}
\mathcal{G}_0 (p) = \mathcal{G}_0 (\bm{p}, i\omega_m) 
= \frac{1}{(i\omega_m + \mu) \gamma_0 - \bm{p} \cdot \bm{\gamma}-M} ,
\nonumber
\end{align}
is the free-quark propagator with $\omega_m=(2m+1)\pi/T$ 
being the Matsubara frequency for fermions.
${\rm Tr}$ denotes the trace over the color ($c$), flavor ($f$) and Dirac ($D$) indices, 
and $\int_p = T\sum_m \int d^3p/(2\pi)^3$. 
Equations~\eqref{eq:Q_Delta} and~\eqref{eq:Q_S}
are diagrammatically represented by the one-loop graphs as shown
in Fig.~\ref{fig:D-Q}, while ${\cal D}_D (k)$ and ${\cal D}_S (k)$ are given 
by the sum of their products as in the figure.
We note that the direction of a quark propagator is opposite in ${\cal Q}_D (k)$ and ${\cal Q}_S (k)$.
The RPA for ${\cal D}_D(k)$ is also referred to as the $T$-matrix approximation in the literature~\cite{Larkin:book}.

\begin{figure}[t]
\centering
\includegraphics[keepaspectratio, scale=0.55]{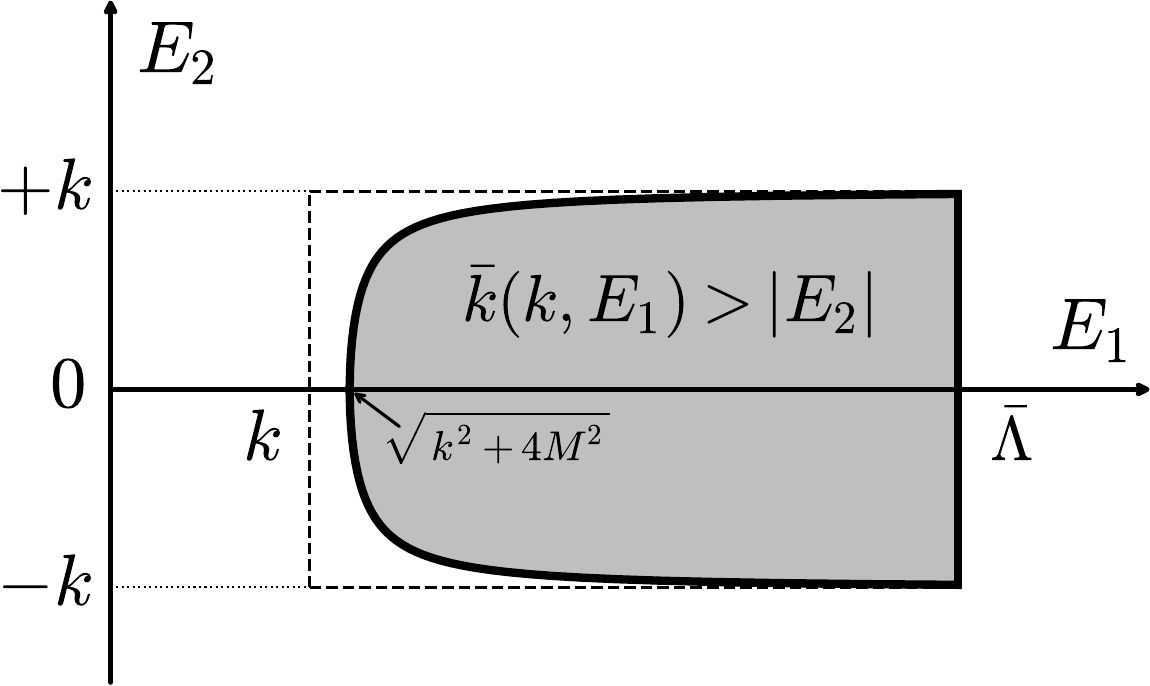}
\caption
{
Integral region of Eq.~\eqref{eq:integral} on the $E_1$--$E_2$ plane.
The rectangular area shown by the dashed lines represents the massless case.
}
\label{fig:Integral-region}
\end{figure}

By taking the Matsubara summation of Eqs.~\eqref{eq:Q_Delta} and~\eqref{eq:Q_S}, we obtain
\begin{align}
\mathcal{Q}_D (\bm{k}, i\nu_n) 
=&\ -\frac{N_f (N_c - 1)}{2} \int \frac{d^3p}{(2\pi)^3} 
\frac{1}{e_1e_2}
\sum_{s, t = \pm}
\bigg\{
\frac{E_1^2 - \bm{k}^2}{E_1 - t(i\nu_n+2\mu)}
{\rm tanh} \frac{E_1 + sE_2 -2t\mu}{4T} 
\nonumber \\ 
&-\frac{E_2^2 - \bm{k}^2}{E_2 - t(i\nu_n+2\mu)}
{\rm tanh} \frac{E_2 + sE_1 -2t\mu}{4T}
\bigg\},
\label{eq:Q-explicit_2SC}
\\
\mathcal{Q}_S (\bm{k}, i\nu_n) 
=&\ \frac{N_f N_c}{2} \int \frac{d^3p}{(2\pi)^3} 
\frac{1}{e_1e_2}
\sum_{s, t = \pm}
\bigg\{
\frac{E_1^2 - \bm{k}^2 - 4M^2}{E_1 - it\nu_n}
{\rm tanh} \frac{E_1 + s(E_2 - 2t\mu)}{4T} 
\nonumber \\ 
&-\frac{E_2^2 - \bm{k}^2 - 4M^2}{E_2 - it\nu_n}
{\rm tanh} \frac{E_2 + s(E_1 - 2t\mu)}{4T}
\bigg\},
\label{eq:Q-explicit_CSC}
\end{align}
where $E_1 = e_1 + e_2$ and $E_2 = e_1 - e_2$
with $e_1 = \sqrt{\bm{p}^2 + M^2}$ and $e_2 = \sqrt{(\bm{k} - \bm{p})^2 + M^2}$. 
The momentum integral in Eqs.~(\ref{eq:Q-explicit_2SC}) and~(\ref{eq:Q-explicit_CSC}) are nicely transformed as follows
\begin{align}
\int \frac{d^3p}{(2\pi)^3} 
=&\ \frac{1}{(2\pi)^2} \int \bm{p}^2 d|\bm{p}| \int^1_{-1} d({\rm cos} \theta)
= \frac{1}{2(2\pi)^2 |\bm{k}|} \int^{\bar\Lambda}_{\bar{k} (|\bm{k}|, 0)} dE_1
\int^{\bar{k} (|\bm{k}|, E_1)}_{-\bar{k} (|\bm{k}|, E_1)} dE_2 e_1e_2,
\label{eq:integral}
\\
& \hspace{0.35cm}
\bar{k}(|\bm{k}|, \omega) =\ |\bm{k}| \sqrt{1 - 4M^2 / (\omega^2-\bm{k}^2)} , 
\quad \bar\Lambda = 2\sqrt{\Lambda^2+M^2},
\notag
\end{align}
where we have introduced the UV cutoff $\bar\Lambda$ on the $E_1$ integral
in the far right-hand side of Eq.~\eqref{eq:integral}. 
This cutoff is chosen so that the range of the three-momentum integral is consistent with that used for Eq.~(\ref{eq:thermodynamic-potential}) at $\bm{k}=\bm{0}$.
The integral region of Eq.~\eqref{eq:integral} is shown in Fig.~\ref{fig:Integral-region}.

\begin{figure*}[t]
\centering
\begin{tabular}{cc}
    \centering
    \includegraphics[height=0.3\textheight]{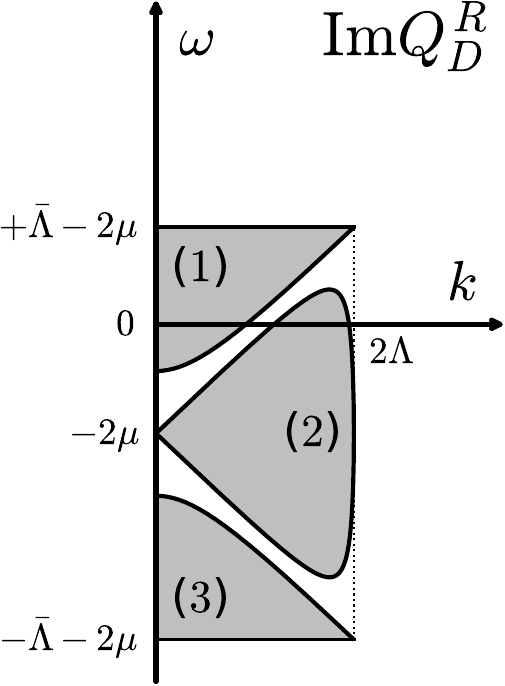}
    \hspace{1.5cm}
    \includegraphics[height=0.3\textheight]{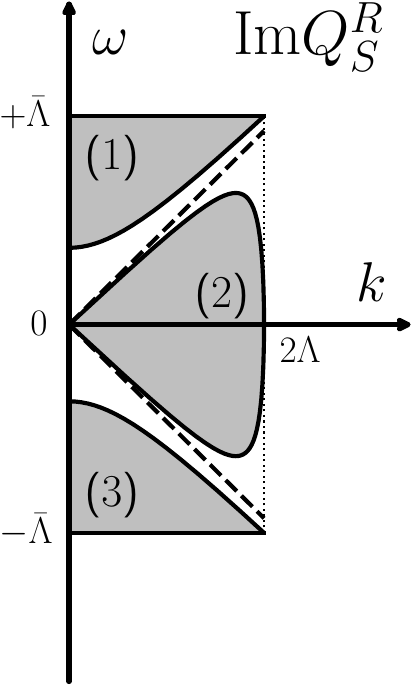}
\end{tabular}
\caption
{
Supports of ${\rm Im} Q^R_D (\bm{k}, \omega)$ (left)
and ${\rm Im} Q^R_S (\bm{k}, \omega)$ (right) on the energy--momentum plane.
The regions (1) and (3) represent the support of the first line of  Eqs.~\eqref{eq:ImQ_2SC} and~\eqref{eq:ImQ_QCDCP}, while the region (2) corresponds to the second line. The dashed line in the right panel shows the light cone.
}
\label{fig:Support-region}
\end{figure*}

The retarded Green's functions $Q_D^R(\bm{k},\omega)$ and $Q_S^R(\bm{k},\omega)$ corresponding to Eqs.~(\ref{eq:Q-explicit_2SC}) and~(\ref{eq:Q-explicit_CSC}), respectively, are obtained by the analytic continuation $i\nu_n \rightarrow \omega+i\eta$.
Their imaginary parts are calculated to be
\begin{align}
{\rm Im} Q^R_D (\bm{k}, \omega) 
=& -\frac{N_f (N_c - 1) T}{4 \pi} \frac{(\omega + 2\mu)^2 - \bm{k}^2}{|\bm{k}|}
\nonumber \\
&\times
\bigg\{
\theta \bigl( \bar\Lambda - |\omega+2\mu| \bigr) 
\theta \bigl( |\omega+2\mu| - \sqrt{\bm{k}^2 + 4M^2} \bigr) 
F_D \bigl( \omega, \bar{k}(|\bm{k}|, \omega+2\mu) \bigr)
\nonumber \\ 
&\qquad +
\theta \bigl( \bar{k}(|\bm{k}|, \bar\Lambda) - |\omega+2\mu| \bigr) 
\Big[ 
F_D \bigl( \omega, \bar{k}(|\bm{k}|, \omega+2\mu) \bigr) - 
F_D \bigl( \omega, \bar\Lambda \bigr)
\Big]
\bigg\},
\label{eq:ImQ_2SC}
\\
F_D (\omega, x) =& \ 2 \sum_{s=\pm} s~{\rm log}~{\rm cosh} ([\omega+sx]/4T) ,
\end{align}
and
\begin{align}
{\rm Im} Q^R_S (\bm{k}, \omega) 
=& -\frac{N_f N_c T}{4 \pi} \frac{\omega^2 - \bm{k}^2 - 4M^2}{|\bm{k}|}
\nonumber \\
&\times
\bigg\{
\theta \bigl( \bar\Lambda - |\omega| \bigr) 
\theta \bigl( |\omega| - \sqrt{\bm{k}^2+4M^2} \bigr) 
F_S \bigl( \omega, \bar{k}(|\bm{k}|, \omega) \bigr)
\nonumber \\ 
&\qquad +
\theta \bigl( \bar{k}(|\bm{k}|, \bar{\Lambda}) - |\omega| \bigr) 
\Big[ 
F_S \bigl( \omega, \bar{k}(|\bm{k}|, \omega) \bigr) - 
F_S \bigl( \omega, \bar{\Lambda} \bigr)
\Big]
\bigg\},
\label{eq:ImQ_QCDCP} 
\\
F_S (\omega, x) =& \sum_{s, t=\pm} s~{\rm log}~{\rm cosh} ([\omega + sx - 2t\mu]/4T) .
\end{align}
Although Eqs.~\eqref{eq:ImQ_2SC} and~\eqref{eq:ImQ_QCDCP} have 
seemingly similar structures with each other, 
their analytic structures are significantly different due to the different locations of the terms $2\mu$,
which lead to a difference in the supports of Eqs.~\eqref{eq:ImQ_2SC} and~\eqref{eq:ImQ_QCDCP}.
The first (second) term in the curly bracket in Eq.~(\ref{eq:ImQ_2SC}) takes a nonzero value at 
$|\omega + 2\mu| > \sqrt{\bm{k}^2 + 4M^2}$
($|\omega + 2\mu| < \bar{k}(|\bm{k}|, \bar\Lambda)$),
while that in Eq.~(\ref{eq:ImQ_QCDCP}) is nonzero at 
$|\omega| > \sqrt{\bm{k}^2 + 4M^2}$
($|\omega| < \bar{k}(|\bm{k}|, \bar\Lambda)$).
As is shown in Fig.~\ref{fig:Support-region}, these supports are deviated by $2\mu$ in the $\omega$--$|\bm{k}|$ plane.
We also note that the supports of the terms in Eq.~\eqref{eq:ImQ_QCDCP} are in the time- and space-like regions.

From explicit calculations one can easily show that $Q_D^R(\bm{k},\omega)$ and $Q_S^R(\bm{k},\omega)$ are not analytic 
at the boundary of the supports. 
Accordingly, $Q_S^R(\bm{k},\omega)$ is not analytic at the origin $(|\bm{k}|,\omega)=(0,0)$ 
in the $\omega$--$|\bm{k}|$ plane, while $Q_D^R(\bm{k},\omega)$ is continuous there.
The non-analyticity of $Q_S^R(\bm{k},\omega)$ at the origin is readily understood by the fact that the limiting value of ${\rm Im} Q^R_S(\bm{k}, \omega)$ 
at the origin depends on the direction to approach,
\begin{align}
\lim_{|\bm{k}| \rightarrow 0} {\rm Im} Q^R_S(\bm{k}, a|\bm{k}|)
=a\frac{N_f N_c M^2}{2\pi} \sum_{t=\pm}
\bigg\{
{\rm tanh} \frac{\lambda_a - 2t\mu}{4T} -
{\rm tanh} \frac{\bar\Lambda - 2t\mu}{4T}
\bigg\}
\theta \bigg( \frac{2\Lambda}{\bar\Lambda} -|a| \bigg),
\label{eq:ImQ-limitation_QCDCP}
\end{align}
with $\lambda_a = \sqrt{4M^2 / (1-a^2)}$. 

To calculate the real parts of $Q^R_D(\bm{k},\omega)$ and $Q^R_S(\bm{k},\omega)$, 
it is convenient to use the Kramers-Kronig relation 
\begin{align}
{\rm Re} Q^R_D(\bm{k}, \omega) =&\ \frac{P}{\pi}
\int^{\bar\Lambda-2\mu}_{-\bar\Lambda-2\mu} 
d\omega' \frac{{\rm Im} Q^R_D (\bm{k}, \omega')}{\omega' - \omega},
\label{eq:ReQ_2SC}
\\
{\rm Re} Q^R_S(\bm{k}, \omega) =&\ \frac{P}{\pi}
\int^{\bar\Lambda}_{-\bar\Lambda} 
d\omega' \frac{{\rm Im} Q^R_S (\bm{k}, \omega')}{\omega' - \omega},
\label{eq:ReQ_QCDCP}
\end{align}
with Eqs.~\eqref{eq:ImQ_2SC} and~\eqref{eq:ImQ_QCDCP}, where $P$ denotes the principal value.
The integral regions in Eqs.~\eqref{eq:ReQ_2SC} and~\eqref{eq:ReQ_QCDCP} 
are chosen so as to be consistent with the cutoff for the $E_1$ integral in Eq.~\eqref{eq:integral}. 
In fact, one can easily verify that Eqs.~\eqref{eq:ReQ_2SC} and~\eqref{eq:ReQ_QCDCP} 
agree with the retarded functions obtained by the analytic continuation 
of Eqs.~\eqref{eq:Q-explicit_2SC} and~\eqref{eq:Q-explicit_CSC} with Eq.~\eqref{eq:integral}.

\subsection{Thouless criterion}
\label{sec:Thouless}

Since $D^R_D (\bm{k}, \omega)$ and $D^R_S (\bm{k}, \omega)$ possess 
all the information of the collective modes coupled to
the operators in Eq.~(\ref{eq:delta-sigma}), 
they encode the properties of the respective soft modes.
As mentioned already the soft modes become massless at the second-order phase transition. This property is ensured by the Thouless criterion~\cite{Thouless} given as a peculiar property of 
$D^R_D (\bm{k}, \omega)$ and $D^R_S (\bm{k}, \omega)$ 
that these functions have a pole 
at the origin in the complex $\omega$ plane at $\bm{k} = \bm{0}$ at the critical point.

\begin{figure}[t]
\centering
\includegraphics[keepaspectratio, scale=0.6]{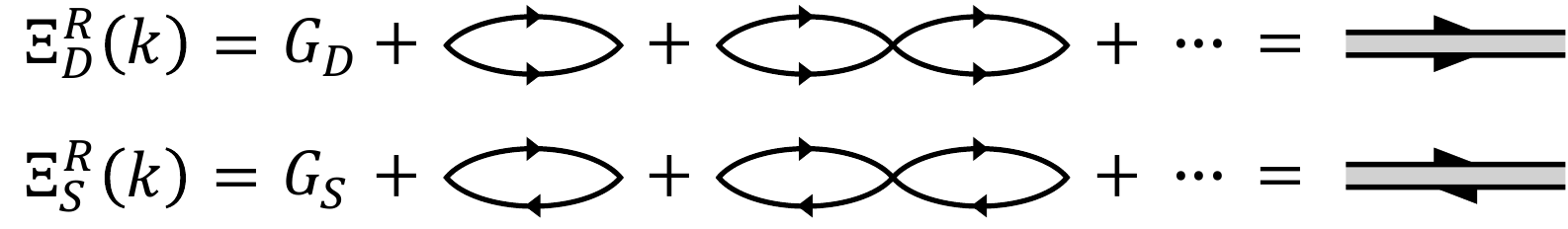}
\caption
{
Diagrammatic representation of $\Xi^R_\gamma (\bm{k}, \omega)$.
}
\label{fig:Xi}
\end{figure}

To show the Thouless criterion,
it is convenient to introduce 
the retarded $T$-matrix $\Xi^R_\gamma(\bm{k},\omega)$ 
defined by 
\begin{align}
\Xi^R_\gamma(\bm{k},\omega) 
=& \frac1{G_\gamma^{-1}+Q^R_\gamma(\bm{k},\omega)}
=  G_\gamma - G_\gamma D^R_\gamma(\bm{k},\omega) G_\gamma
\notag \\
=&\, G_\gamma - G_\gamma Q^R_\gamma(\bm{k},\omega) G_\gamma + G_\gamma Q^R_\gamma(\bm{k},\omega) G_\gamma Q^R_\gamma(\bm{k},\omega) G_\gamma -\cdots,
\label{eq:genf-T-matrix}
\end{align}
which also means
$D^R_\gamma(\bm{k},\omega)=G_\gamma^{-1} Q^R_\gamma (\bm{k},\omega)\Xi^R_\gamma (\bm{k},\omega)$.
The diagrammatic representations of $\Xi^R_D(\bm{k},\omega)$ and $\Xi^R_S(\bm{k},\omega)$ are given in Fig.~\ref{fig:Xi}.
Then, it is readily confirmed that these functions satisfy
\begin{align}
{\Xi^R_D}^{-1}(\bm{0}, 0) =
2 \frac{\partial^2 \omega_{\rm MFA}}{\partial \Delta^2},
\qquad
{\Xi^R_S}^{-1}(\bm{0}, 0) = 2 \frac{\partial^2 \omega_{\rm MFA}}{\partial M^2}.
\label{eq:Thouless}
\end{align}
Since the thermodynamic potential satisfies
$\partial^2 \omega_{\rm MFA}/\partial\Delta^2=0$ at the 2SC-PT 
and $\partial^2 \omega_{\rm MFA}/\partial M^2=0$ at the QCD-CP,
we obtain 
\begin{align}
{\Xi^R_\gamma}^{-1}(\bm{0}, 0) =0 
\quad \mbox{and} \quad
{D^R_\gamma}^{-1}(\bm{0}, 0) =0 ,
\label{eq:Thouless_gen}
\end{align}
at the respective critical points.
Equation~(\ref{eq:Thouless_gen}) 
proves the existence of a pole in $D_\gamma^R(\bm{k},\omega)$ at the origin $(|\bm{k}|,\omega)=(0,0)$, 
i.e. the Thouless criterion surely holds.

The pole at $\omega=0$ at the critical point
moves continuously in the lower-half complex-energy plane as a function of $T$ and $\mu$.
This means the existence of the soft mode, i.e.
the collective mode near the origin that eventually becomes massless at the critical point.

In the above argument, we derived the Thouless criterion based on the MFA and the RPA. However, the validity of the Thouless criterion generally holds even beyond the MFA, since the second-order phase transition is characterized by the flattening of the effective potential~\cite{Larkin:book}. This means that the appearance of the soft modes is guaranteed even beyond the MFA as long as the phase transition is of second order. 

When the effects beyond the MFA are incorporated, however, it is possible that the second-order phase transition becomes first-order or crossover. In this case, the soft mode does not become massless because of the absence of the second-order transition. Even in this case, however, the softening of the order-parameter field is expected if the first-order transition is weak or the crossover transition takes place sharply. In later sections, we calculate the transport coefficients assuming the second-order phase transition. While this always applies to the QCD-CP, the 2SC-PT may not be of second-order~\cite{Alford:2007xm}. Even in this case, however, the qualitative feature of the following discussion would not change drastically as long as the 2SC-PT is close to the second-order transition and soft collective modes manifest themselves near the phase transition.

As discussed in Sec.~\ref{sec:RPA}, 
$D^R_D(\bm{k},\omega)$ is analytic at the origin, while $D^R_S (\bm{k}, \omega)$ is not. 
This difference leads to
a qualitative difference in the analytic properties of the soft modes of 2SC-PT and QCD-CP.
In terms of the dynamical structure factor,
$S_D(\bm{k},\omega)$ is a smooth function around the origin, while
$S_S (\bm{k}, \omega)$ is not.
In $D^R_S (\bm{k}, \omega)$, it is known that a pole in the space-like region behaves as the soft mode that moves 
toward the origin in the complex $\omega$ plane~\cite{Fujii:2003bz,Fujii:2004jt}, 
while any poles in the time-like region do not show such a softening. 
The latter property can be confirmed easily from the fact that the spectral supports of $S_S (\bm{k}, \omega)$ 
in the time-like region exist only
at $|\omega| > \sqrt{\bm{k}^2 + 4M^2}$.
Since $M$ stays nonzero at the QCD-CP, 
the pole in the time-like region never becomes massless.

Before closing this subsection, let us 
briefly comment on the effect of the current quark mass $m$ on the properties of the soft modes. 
First, it is readily found by an explicit calculation that
the properties of the soft mode of the 2SC-PT are hardly affected by $m$.
This result is understood intuitively from
the fact that 
the soft mode of the 2SC-PT is the diquark mode that
is governed by the excitations near the Fermi surface
that are less affected by the quark mass when $\mu\gg M$.
On the other hand, the properties of the soft mode of the QCD-CP are crucially affected by $m$ since
the QCD-CP becomes a tri-critical point in the $m \to 0$ limit.
One can also argue that the soft mode of the QCD-CP is less affected by the diquark gap $\Delta$, 
while that of the 2SC-PT is 
crucially 
 modified by it.

\section{Effective theory of soft modes}
\label{sec:Effective-theory}

In this section, we derive effective formulas for $\Xi_D^R(\bm{k},\omega)$ and $\Xi_S^R(\bm{k},\omega)$ that describe low-energy behavior of the soft modes near the 2SC-PT or QCD-CP.
These formulas will be utilized in the next section for the analysis of the photon self-energy. 

\subsection{Soft mode of 2SC-PT}
\label{sec:Soft-mode_CSC}

Let us start from the soft mode of the 2SC-PT, which is encoded in the $T$-matrix $\Xi_D^R(\bm{k},\omega)$. Since the excitation energy of the soft mode approaches zero at the critical point, 
it is reasonable to approximate $\Xi_D^R(\bm{k},\omega)$ 
by expanding its inverse ${\Xi^R_D}^{-1} (\bm{k}, \omega)$ with respect to $\omega$ and pick up the first two terms as 
\begin{align}
{\Xi^R_D}^{-1} (\bm{k}, \omega) 
\simeq A_D (\bm{k}) + C_D \omega,
\label{eq:Xi-LEE_2SC}
\end{align}
with
\begin{align}
    A_D (\bm{k}) = G_D^{-1} + Q^R_D (\bm{k}, 0) , \qquad 
    C_D = \frac{\partial Q^R_D (\bm{0}, \omega)}{ \partial \omega} \bigg|_{\omega=0},
\label{eq:Def-A_D-C_D}
\end{align}
which are found to be real and complex numbers, respectively.
Owing to the Thouless criterion~\eqref{eq:Thouless_gen}, 
\[
A_D(\bm{0})=0 ,
\]
at $T=T_c$.
In Eq.~\eqref{eq:Xi-LEE_2SC}, we have neglected the $\bm{k}$ dependence of $C_D$ 
as well as higher order terms of $\omega$. 
We call Eq.~(\ref{eq:Xi-LEE_2SC}) the low-energy (LE) approximation. 
In Ref.~\cite{Nishimura:2022mku}, this formula has been employed for the calculation of the photon self-energy and the DPR near the 2SC-PT.

To simplify the approximation further, one may replace $A_D(\bm{k})$ with its Taylor expansion at $\bm{k}=\bm0$. 
By picking up the first two terms of the expansion, one has
\begin{align}
{\Xi^R_D}^{-1} (\bm{k}, \omega)
\simeq a_D + b_D \bm{k}^2 + c_D \omega
\simeq \tilde{a}_D \epsilon + b_D \bm{k}^2 + c_D \omega,
\label{eq:Xi-TDGL_2SC} 
\end{align}
where the coefficients $a_D$, $b_D$, and $c_D$ are given by
\begin{align}
a_D
=~\frac{1}{G_D} + Q^R_D (\bm{0}, 0) 
, \quad
b_D
=~\frac{\partial Q^R_D (\bm{k}, 0)}{\partial \bm{k}^2} \bigg|_{|\bm{k}|=0}
, \quad
c_D = C_D.
\label{eq:abc_TDGL}
\end{align}
with 
\begin{align}
    \epsilon = \frac{T-T_c}{T_c} ,
    \label{eq:reduced-T}
\end{align}
being the reduced temperature.
In the far-right-hand side of Eq.~\eqref{eq:Xi-TDGL_2SC}, 
we have utilized
the fact that $a_D=0$ at $T=T_c$ and 
made a further approximation 
\begin{align}
a_D \simeq \tilde{a}_D\epsilon
\hspace{0.3cm} {\rm with} \hspace{0.3cm}
\tilde{a}_D = T \frac{\partial Q^R_D (\bm{0}, 0)}{\partial T}\bigg|_{T=T_c}.
\label{eq:a_D}
\end{align}
Equation~\eqref{eq:Xi-TDGL_2SC} is known as the linearized time-dependent Ginzburg-Landau (TDGL) approximation,
as ${\Xi^R_D}^{-1} (\bm{k},\omega)\Delta(\bm{k},\omega) = 0$ with Eq.~\eqref{eq:Xi-TDGL_2SC} corresponds to the linearized TDGL equation~\cite{Larkin:book}. 
In this approximation, the pole of the soft mode determined by $\Xi^R_D(\bm{k},\omega)=0$ is located at $\omega=-(a_D+b_D\bm{k}^2)/c_D$. The stability of the TDGL equation requires ${\rm Im}c_D<0$, which means that the pole lies on the lower-half complex-energy plane.

When the real part of $c_D$ is small, $|{\rm Re}c_D|\ll|{\rm Im}c_D|$, Eq.~\eqref{eq:Xi-TDGL_2SC} is rewritten as 
\begin{align}
\Xi^R_D (\bm{k}, \omega) 
= \frac{1}{a_D + b_D \bm{k}^2 + c_D \omega}
=\frac{i}{|{\rm Im} c_D|}
\frac{1}{\omega +i \tau_{\rm GL}^{-1}
\big(
1 + \xi_D^2 \bm{q}^2
\big)}.
\label{eq:TDGLtauxi}
\end{align}
Here, $\tau_{\rm GL}=|c_D|/a_D$ and $\xi_D=\sqrt{b_D/a_D}$ are interpreted as the relaxation time and the coherence length of the soft mode, respectively.

As was discussed in Sec.~\ref{sec:RPA}, $\Xi_D^R(\bm{k},\omega)$ is analytic in the shaded regions in the left panel of Fig.~\ref{fig:Support-region}. Since Eq.~\eqref{eq:Xi-TDGL_2SC} is obtained from the expansion at the origin, it applies only to the region~(2) in the panel that includes the origin. The same argument also applies to Eq.~\eqref{eq:Xi-LEE_2SC}. Therefore, when we use Eq.~\eqref{eq:Xi-LEE_2SC} or~\eqref{eq:Xi-TDGL_2SC} in the following analyses we assume that $\Xi_D^R(\bm{k},\omega)$ vanishes outside the region~(2), although this treatment hardly modifies the final result since $S_D(\bm{k},\omega)$ is well suppressed there.

\begin{figure}[t]
\centering
\includegraphics[keepaspectratio, scale=0.8]{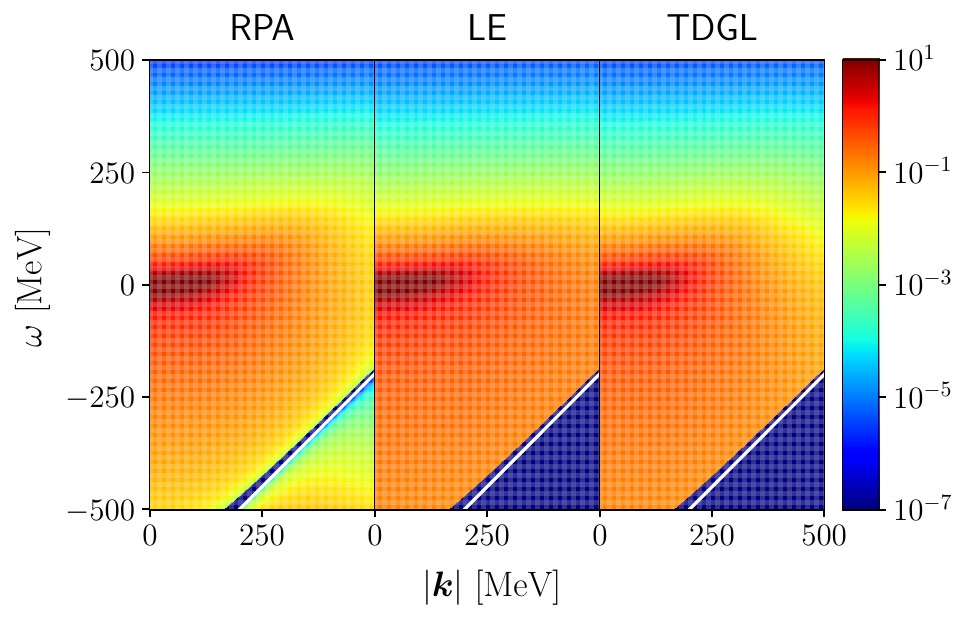}
\caption{
Contour map of the dynamical structure factor of the diquark soft mode
$S_D(\bm{k},\omega)$ at $\mu = 350~{\rm MeV}$ and $T = T_c$ with $G_D = 0.7G_S$.
The left, middle, and right panels are the results of the RPA, 
the LE approximation and the TDGL approximation, respectively.
}
\label{fig:DSF_2SC}
\end{figure}

As discussed in \ref{sec:TDGL}, the coefficients in Eq.~\eqref{eq:Xi-TDGL_2SC} can be calculated analytically in the large $\Lambda$ limit, $\Lambda\gg T+|\mu|$, at $M=0$, which yields
\begin{align}
\tilde{a}_D =& \ \frac{2N_f(N_c-1)}{\pi^2} \mu^2 \bigg( 1 + \frac{\pi^2}3 \frac{T^2}{\mu^2} \bigg) ,
\label{eq:a-tilde_D}
\\
b_D =& \ \frac{N_f(N_c-1)}{4\pi^2} \Big(
\frac{7\zeta(3)}{12\pi^2} \frac{\mu^2}{T^2} + 
\log \frac{\Lambda^2-\mu^2}{4T^2} + 2\gamma_{\rm E} - 2\log\frac{\pi}{4} 
- 1 \Big),
\label{eq:b_D}
\\
{\rm Re} \ c_D =& \ - \frac{N_f(N_c-1)}{\pi^2} \mu
\bigg(
\log \frac{\Lambda^2 - \mu^2}{4 T^2} + 2\gamma_{\rm E} - 2\log \frac{\pi}{4} + 1
\bigg) ,
\label{eq:c-r_D}
\\
{\rm Im} \ c_D =& \ - \frac{N_f(N_c-1)}{4\pi} \frac{\mu^2}T.
\label{eq:c-i_D}
\end{align}
Assuming further $T/\mu\ll1$ and neglecting the logarithmic terms, we obtain 
\begin{align}
\tilde{a}_D = \frac{2N_f(N_c-1)}{\pi^2} \mu^2,
\quad
b_D = \frac{7N_f(N_c-1)\zeta(3)}{48\pi^4} \frac{\mu^2}{T^2},
\quad
c_D = -i\frac{N_f(N_c-1)}{4\pi} \frac{\mu^2}T,
\label{eq:abc_simplified-TDGL}
\end{align}
where in the last equation we neglected the real part of $c_D$ which is parametrically suppressed compared to the imaginary part.
On account of Eq.~\eqref{eq:abc_simplified-TDGL}, $\tau_D$ and $\xi_D$ in Eq.~\eqref{eq:TDGLtauxi}
are reduced to
\begin{align}
\tau_{\rm GL} = \frac\pi{8T} \frac1\epsilon,
\qquad
\xi_D = \sqrt{\frac{7\xi(3)}{96T^2}} \, \frac1{\epsilon^{1/2}} .
    \label{eq:tau-xi_2SC}
\end{align}
This result shows that $\tau_{\rm GL}$ and $\xi_D$ are divergent for $\epsilon \rightarrow 0$. Moreover, the coefficients do not depend on $\mu$, which implies that the property of the soft mode is insensitive to $\mu$.

Finally, let us inspect the validity of the LE and TDGL approximations for $\Xi^R_D(\bm{k},\omega)$ numerically.
The left, middle, and right panels of Fig.~\ref{fig:DSF_2SC} show 
the contour maps of the dynamical structure factor $S_D(\bm{k},\omega)$ at $T=T_c$ with $\mu=350$~MeV
calculated in the RPA, the LE approximation~\eqref{eq:Xi-LEE_2SC}
and the TDGL approximation~\eqref{eq:Xi-TDGL_2SC}. 
One finds that the LE and TDGL approximations reproduce the result 
of the RPA well over a wide range of $\omega$ and $\bm{k}$.

\subsection{Soft mode of QCD-CP}
\label{sec:Soft-mode_QCDCP}

Next, we perform a similar analysis for the soft mode of the QCD-CP.
To this end, we adopt the LE approximation to $\Xi^R_S (\bm{k}, \omega)$ as
\begin{align}
{\Xi^R_S}^{-1} (\bm{k}, \omega) 
\simeq A_S(\bm{k}) + C_S(\bm{k}) \omega,
\label{eq:Xi-LEE_QCDCP}
\end{align}
with 
\begin{align}
A_S(\bm{k}) = G_S^{-1} + Q^R_S(\bm{k}, 0) \quad
\textrm{and} \quad
C_S(\bm{k}) = 
\frac{\partial Q^R_S (\bm{k}, \omega) }{ \partial \omega }\bigg|_{\omega=0},
\label{eq:Def-A_S-C_S}
\end{align}
which are found to be real and pure imaginary, respectively.
Here, in contrast to the case of Eq.~\eqref{eq:Xi-LEE_2SC},
we keep the $\bm{k}$ dependence of $C_S(\bm{k})$,
since an explicit manipulation shows that 
$C_S(\bm{k})$ diverges as $1/|\bm{k}|$ for
$\bm{k} \rightarrow \bm{0}$.
This difference originates from the fact that $\Xi_S^R(\bm{k},\omega)$ is not analytic at the origin as discussed in Sec.~\ref{sec:RPA}, in contrast to $\Xi_D^R(\bm{k},\omega)$ that is analytic there. 
We also note that $\Xi_S^R(\bm{k},\omega)$ is discontinuous at $|\omega|=\bar k(|\bm{k}|,\bar\Lambda)\simeq|\bm{k}|$. Therefore, the LE approximation~\eqref{eq:Xi-LEE_QCDCP} is applicable only for $|\omega|<\bar k(|\bm{k}|,\bar\Lambda)$, i.e. the region~(2) in the right panel of Fig.~\ref{fig:Support-region}, which is within the space-like region. 
Since the soft mode of the QCD-CP has a spectral support only in the space-like region as discussed already, the approximation~\eqref{eq:Xi-LEE_QCDCP} is adequate for describing it.
In Ref.~\cite{Nishimura:2023oqn}, Eq.~\eqref{eq:Xi-LEE_QCDCP} is adopted 
for the analysis of the DPR near the QCD-CP.

To describe the low energy-momentum behavior,
one would further expand 
the coefficients as 
\begin{align}
{\Xi^R_S}^{-1} (\bm{k}, \omega) 
\simeq \ a_S + b_S|\bm{k}|^2 + c_S \frac\omega{|\bm{k}|},
\label{eq:Xi-TDGL_QCDCP}
\end{align}
with
\begin{align}
a_S =\frac{1}{G_S} + Q^R_S (\bm{0}, 0) ,
\qquad
b_S =\frac{\partial Q^R_S (\bm{k}, 0)}{\partial \bm{k}^2} \bigg|_{|\bm{k}|=0},
\qquad
c_S =\frac{\partial(|\bm{k}| Q^R_S (\bm{k}, \omega)  \big|_{|\bm{k}|\to0})}{\partial \omega} 
 \bigg|_{\omega=0} .
\end{align}
We refer to Eq.~\eqref{eq:Xi-TDGL_QCDCP} as the TDGL approximation for $\Xi^R_S(\bm{k},\omega)$ in analogy with Eq.~\eqref{eq:Xi-TDGL_2SC}. 
When we use Eqs.~\eqref{eq:Xi-LEE_QCDCP} or~\eqref{eq:Xi-TDGL_QCDCP}, it is assumed that ${\Xi^R_S}^{-1} (\bm{k}, \omega) =0$ outside the region~(2) in the right panel of Fig.~\ref{fig:Support-region}. We also note that the last term in Eq.~\eqref{eq:Xi-TDGL_QCDCP} is finite in the limit $|\bm{k}|\to0$ because $|\omega|<|\bm{k}|$ is always satisfied in region~(2).

The parameter $a_S$ in Eq.~\eqref{eq:Xi-TDGL_QCDCP} vanishes at the QCD-CP. However, its behavior around there is more 
subtle than the case of the 2SC-PT.
As discussed in \ref{sec:pole}, when $T$ and $\mu$ approach the QCD-CP linearly with a fixed ratio $T-T_{\rm CP}:\mu-\mu_{\rm CP}$, it behaves as 
\begin{align}
a_S \sim
\begin{cases}
~\epsilon_{\rm CP} & \text{parallel to the first-order line,} \\
~\epsilon_{\rm CP}^{2/3} & \text{otherwise,} 
\end{cases}
\label{eq:a_S}
\end{align}
in the MFA with 
\begin{align}
\epsilon_{\rm CP} = \sqrt{
~ \bigg( \frac{T - T_{\rm CP}}{T_{\rm CP}} \bigg)^2 
+ \bigg( \frac{\mu - \mu_{\rm CP}}{\mu_{\rm CP}} \bigg)^2~
}.
\end{align}
We have numerically checked Eq.~\eqref{eq:a_S} in our model.
We will use Eq.~\eqref{eq:a_S} in Sec.~\ref{sec:Transport-coeff} to discuss the behavior of the transport coefficients near the QCD-CP.

\begin{figure}[t]
\centering
\includegraphics[keepaspectratio, scale=0.8]{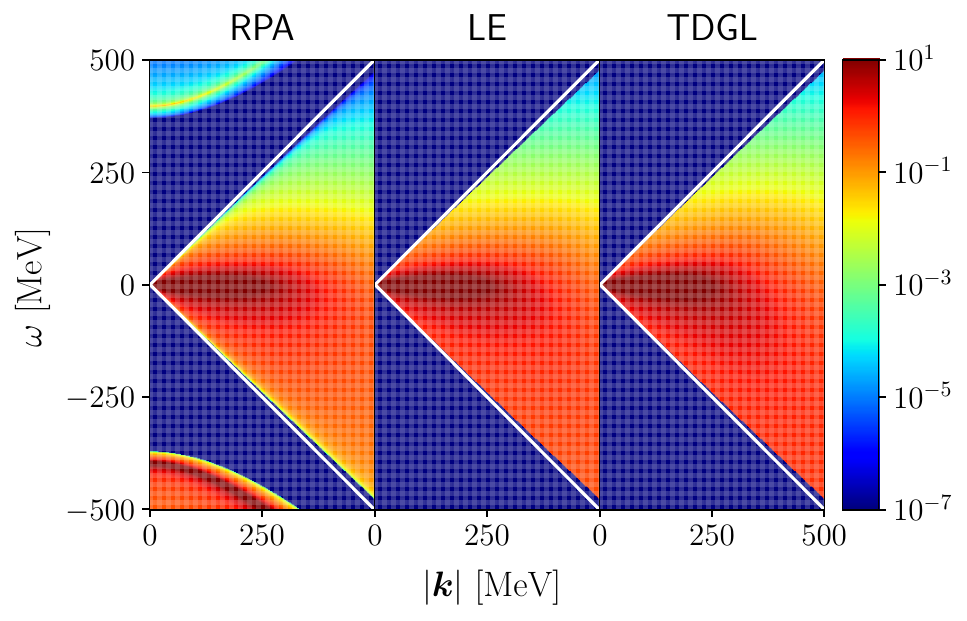}
\caption{
Contour map of the dynamical structure factor 
$S_S(\bm{k},\omega)$ at the QCD-CP ($T_c, \mu_c$).
The left, middle, and right panels are the results of the RPA, 
the LE approximation and the TDGL approximation,
respectively.
}
\label{fig:DSF_QCDCP}
\end{figure}

Figure~\ref{fig:DSF_QCDCP} shows the dynamical structure factor 
$S_S(\bm{k},\omega)$ calculated by the RPA, 
the LE approximation (\ref{eq:Xi-LEE_QCDCP})
and the TDGL approximation (\ref{eq:Xi-TDGL_QCDCP}). 
One finds that both Eqs.~(\ref{eq:Xi-LEE_QCDCP}) and~(\ref{eq:Xi-TDGL_QCDCP}) well reproduce
the RPA result near the origin, while 
the former is slightly better than the latter.
One also finds that the RPA result has a spectral peak in the time-like region corresponding to the mesonic excitation~\cite{Hatsuda:1994pi}. This structure does not exist in the LE and TDGL results since these approximations do not apply to the time-like region. However, since the mesonic excitation stays massive at the CP, it does not contribute to the critical phenomena.
In the next sections, we use Eqs.~(\ref{eq:Xi-LEE_QCDCP}) and~(\ref{eq:Xi-TDGL_QCDCP}) for the analysis of the transport coefficients.

\section{Photon self-energy $\tilde\Pi^{\mu\nu}(k)$}
\label{sec:Self-energy}

In this section, we calculate the photon self-energy $\tilde\Pi^{\mu\nu} (k)$ incorporating the effects of the 2SC-PT and QCD-CP soft modes. 
In these analyses, we use the effective propagators of the soft modes obtained by the LE and TDGL approximations introduced in the previous section.
When the LE approximations~\eqref{eq:Xi-LEE_2SC} or~\eqref{eq:Xi-LEE_QCDCP} is adopted,
the formalism is similar to those in Ref.~\cite{Nishimura:2022mku,Nishimura:2023oqn}, while we shall give detailed accounts 
of the calculational procedures omitted there. 
The difference between the 2SC-PT and QCD-CP and the analyses with the TDGL approximation~\eqref{eq:Xi-TDGL_2SC} or~\eqref{eq:Xi-TDGL_QCDCP} will also be discussed.

At the leading order of electromagnetic interaction, the photon self-energy 
in the imaginary-time formalism is given by the electromagnetic current-current correlation function as~\cite{book_LeBellac}
\begin{align}
    \tilde\Pi^{\mu\nu}(\bm{k},i\nu_l)
    = \tilde\Pi^{\mu\nu}(k) = \int d^4x e^{i\nu_l \tau - i \bm{k}\cdot \bm{x}}\langle T_\tau [ j^\mu(\bm{x},\tau),j^\nu(\bm{0},0)]\rangle ,
    \label{eq:jj}
\end{align}
with the electromagnetic current $j^\mu(\bm{x},\tau)$.
The retarded self-energy $\Pi^{R\mu\nu}(\bm{k},\omega)$ is then obtained through the analytic continuation $\Pi^{R\mu\nu}(\bm{k},\omega) = \tilde\Pi^{\mu\nu}(\bm{k},i\nu_l)|_{i\nu_l\to\omega+i\eta}$.
As we will see in the next section, the transport coefficients are obtained from the low energy-momentum limit of its imaginary part of the spatial components, 
${\rm Im} \Pi^{R ij} (\bm{k},\omega)$.

We assume that $\tilde\Pi^{\mu\nu}(k)$ consists of three parts
\begin{align}
\tilde\Pi^{\mu\nu} (k) = \tilde\Pi^{\mu\nu}_{\rm free} (k) +
\tilde\Pi^{\mu\nu}_D (k) + \tilde\Pi^{\mu\nu}_S (k),
\label{eq:Pi-tot}
\end{align}
where $\tilde\Pi^{\mu\nu}_D (k)$ and $\tilde\Pi^{\mu\nu}_S (k)$ 
are the contributions from the soft modes of the 2SC-PT and QCD-CP, respectively, that will be constructed below and
\begin{align}
\tilde\Pi^{\mu\nu}_{\rm free} (k) = 
N_c C_{\rm em} \int_p {\rm Tr}_D
[\gamma^\mu \mathcal{G}_0 (p+k) \gamma^\nu \mathcal{G}_0 (p)],\quad\quad
(C_{\rm em} \equiv e_u^2 + e_d^2)
\label{eq:Pi-free}
\end{align}
is the self-energy of the free-quark system 
where $e_u = 2|e|/3$  ($e_d = -|e|/3$) is the electric charge of the up (down) quark with $e$ being the electron charge.

From the charge conservation, the total photon self-energy $\tilde\Pi^{\mu\nu} (k)$ in Eq.~\eqref{eq:Pi-tot} should satisfy the Ward-Takahashi (WT) identity
\begin{align}
    k_\mu \tilde{\Pi}^{\mu\nu} (k) = 0.    
    \label{eq:Pi-WT}
\end{align}
We assume that $\tilde\Pi^{\mu\nu}_D (k)$ and $\tilde\Pi^{\mu\nu}_S (k)$ in Eq.~\eqref{eq:Pi-tot} separately satisfy
\begin{align}
    k_\mu \tilde\Pi^{\mu\nu}_D (k) = 
    k_\mu \tilde\Pi^{\mu\nu}_S (k) = 0,
    \label{eq:Pi-WT-DS}
\end{align}
while $k_\mu \tilde\Pi^{\mu\nu}_{\rm free} (k)=0$ is readily shown from Eq.~\eqref{eq:Pi-free}.

\subsection{Contribution of the diquark soft modes}
\label{sec:Self-energy_2SC}

\subsubsection{Aslamazov-Larkin, Maki-Thompson, and Density of states terms}
\label{sec:ALMTDOS_2SC}

\begin{figure}[t]
\centering
\includegraphics[keepaspectratio, scale=0.45]{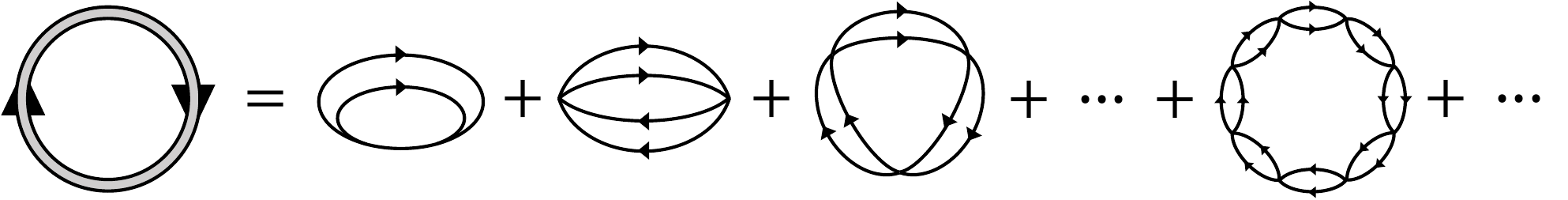}
\caption
{
Contribution of the diquark soft mode to the thermodynamic potential.
}
\label{fig:Omega_2SC}
\end{figure}

In this subsection, we calculate 
$\tilde\Pi^{\mu\nu}_D (k)$.
To construct it 
in accordance with the WT identity~\eqref{eq:Pi-WT-DS},
we start with the set of diagrams representing 
the lowest-order contribution of the soft modes to the thermodynamic potential
\begin{align}
\Omega_D = 3 \int_p \ln [G_D \tilde\Xi_D^{-1} (p)], 
\label{eq:Omega_D}
\end{align}
i.e. the one-loop diagram of $\tilde\Xi_D(k)$ shown in Fig.~\ref{fig:Omega_2SC}, where $\tilde\Xi_D(p)$ is the imaginary-time $T$-matrix corresponding to Eq.~\eqref{eq:genf-T-matrix} and the overall coefficient $3$ comes from 
three anti-symmetric channels of the diquark modes.

\begin{figure*}[t]
    \centering
    \begin{tabular}{cccc}
    \includegraphics[keepaspectratio, scale=0.1]{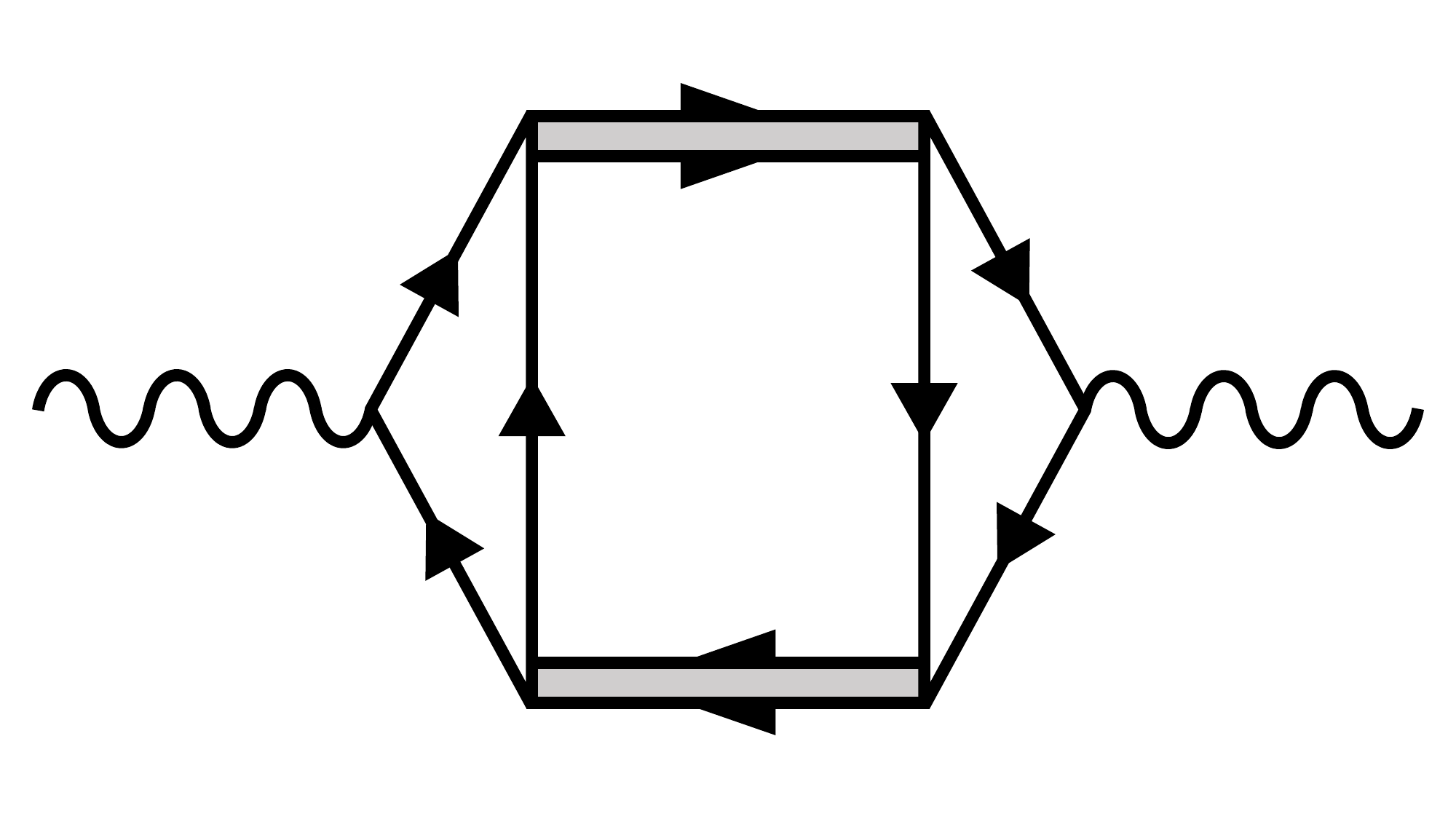} &
    \includegraphics[keepaspectratio, scale=0.1]{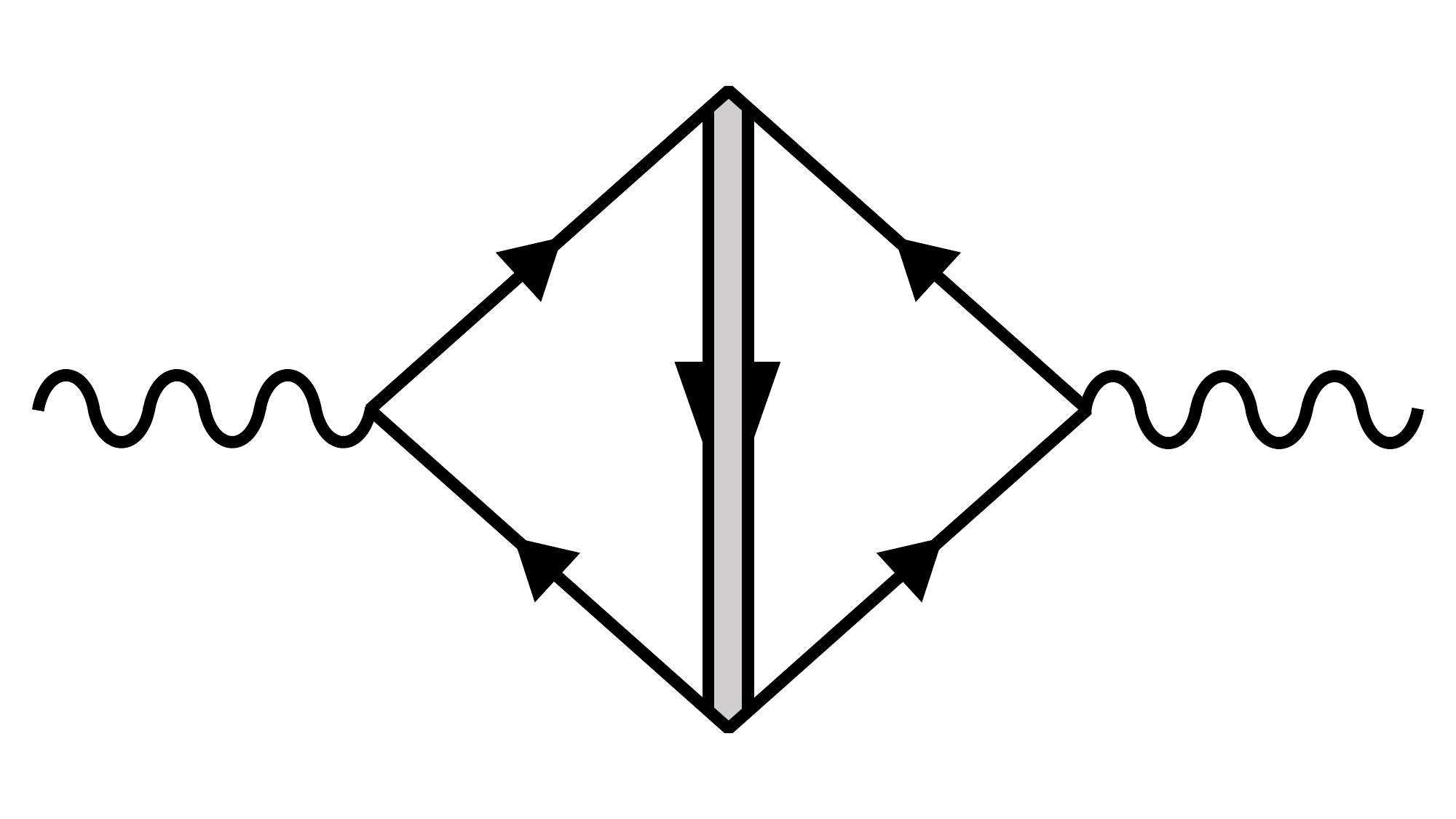} &
    \includegraphics[keepaspectratio, scale=0.1]{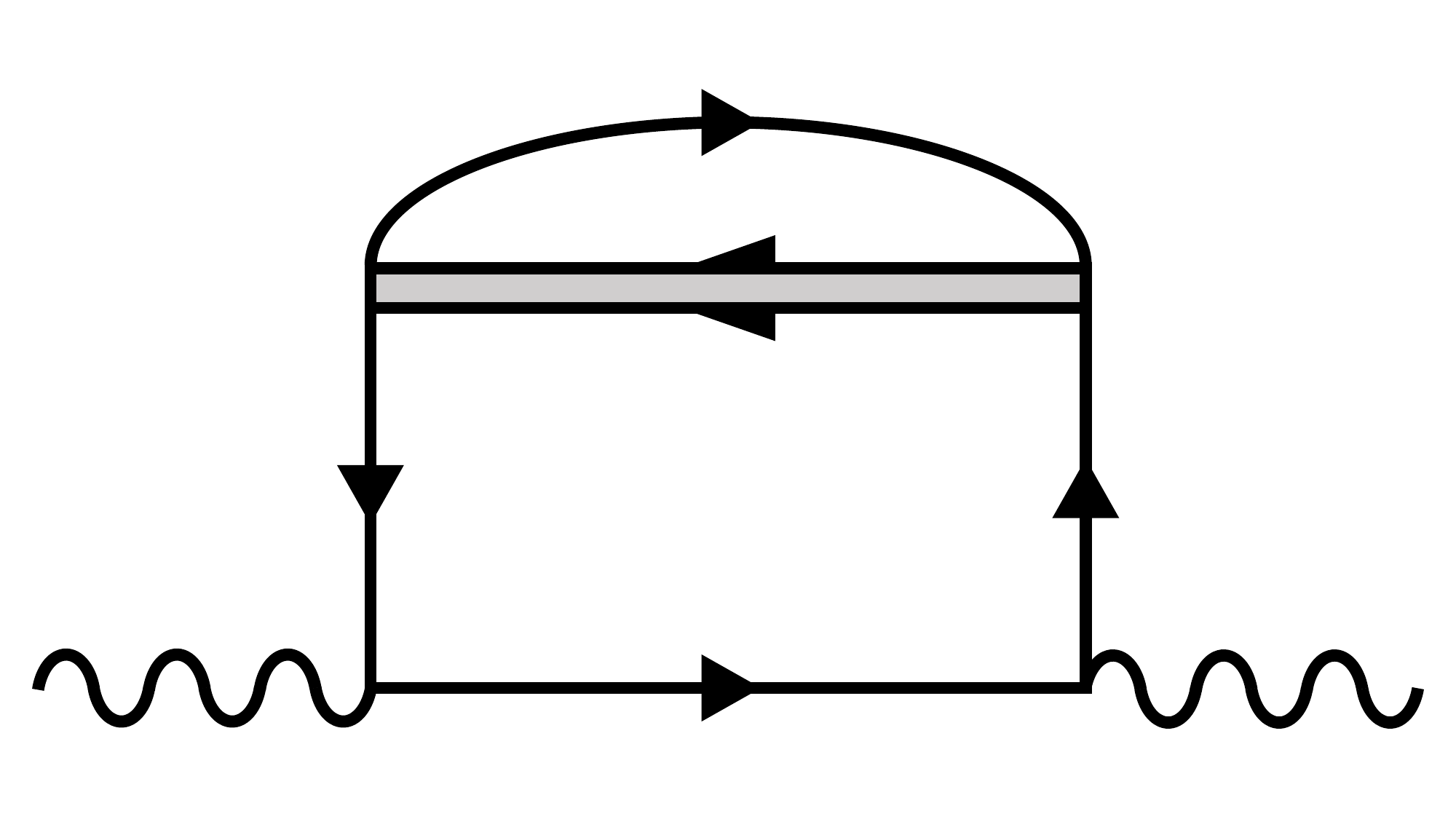} &
    \includegraphics[keepaspectratio, scale=0.1]{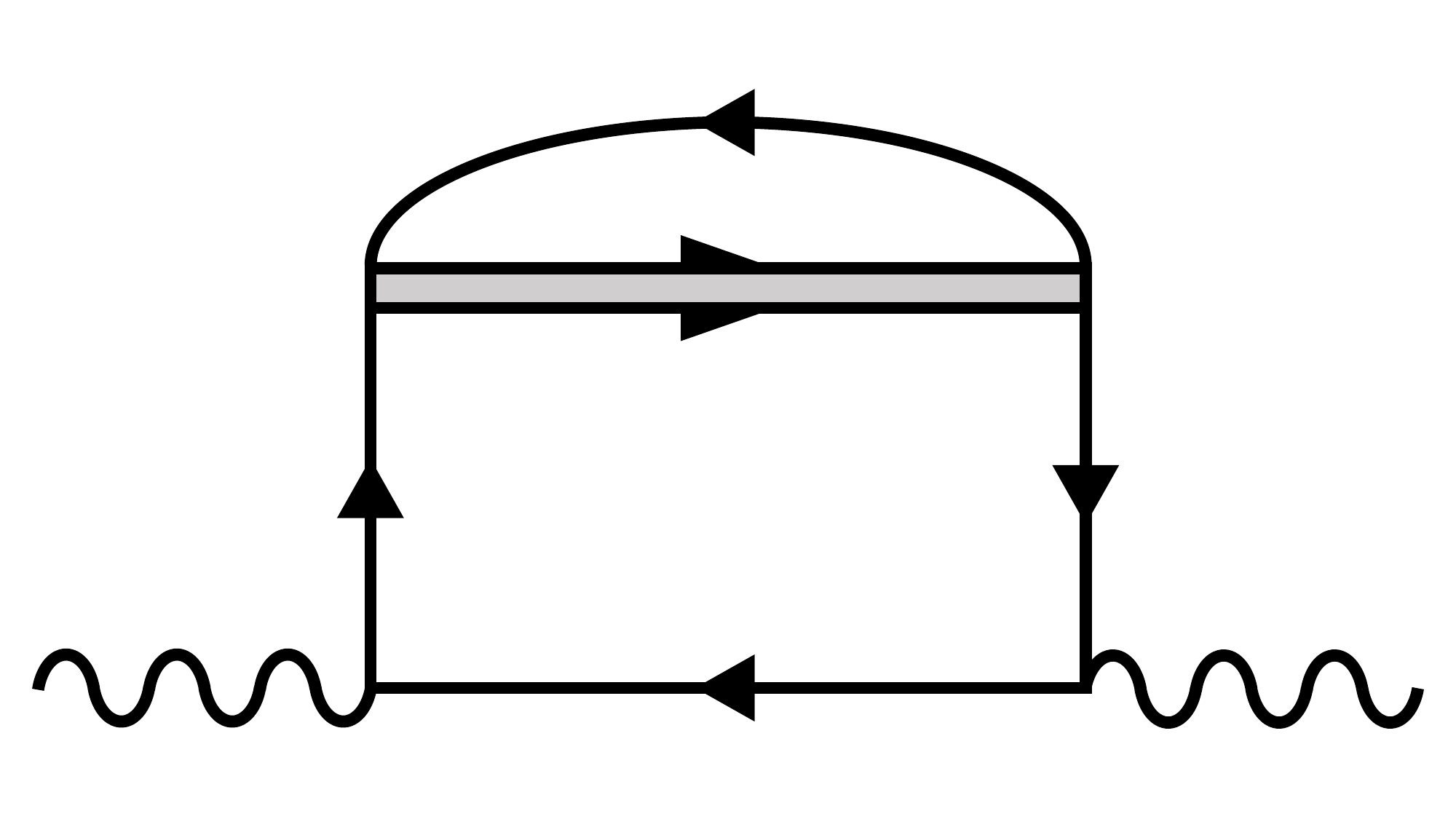}
    \\
    (a) & (b) & (c) & (d)
    \end{tabular}
    \caption{
Diagrammatic representations of the Aslamazov-Larkin (a),
Maki-Thompson (b) and density of states (c, d) terms with the 2SC soft modes
with the wavy lines being the photon ones.
}
\label{fig:Self-energy_2SC}
\end{figure*}

The photon self-energy satisfying Eq.~\eqref{eq:Pi-WT-DS} is then constructed by attaching electromagnetic vertices 
at any two points in the quark lines in $\Omega_D$.
This procedure leads to the four types of diagrams
shown in Fig.~\ref{fig:Self-energy_2SC};
they are called (a) Aslamazov-Larkin (AL)~\cite{AL:1968}, 
(b) Maki-Thompson (MT)~\cite{Maki:1968,Thompson:1968}
and (c, d) density of states (DOS) terms, respectively,
in the theory of metallic superconductivity~\cite{Larkin:book}. 
The contribution of the diquark soft modes to the photon self-energy $\tilde\Pi^{\mu\nu}_D (k)$ is expressed as
\begin{align}
\tilde\Pi^{\mu\nu}_D (k) 
= \tilde\Pi^{\mu\nu}_{{\rm AL},D} (k)
+ \tilde\Pi^{\mu\nu}_{{\rm MT},D} (k) 
+ \tilde\Pi^{\mu\nu}_{{\rm DOS},D} (k) ,
\label{eq:Pi-fluc_2SC}
\end{align}
where 
\begin{align}
\tilde\Pi_{{\rm AL},D}^{\mu\nu} (k) &= 3 \int_q
\tilde\Gamma^\mu_D (q, q+k) \tilde\Xi_D (q+k) 
\tilde\Gamma^\nu_D (q+k, q) \tilde\Xi_D (q),
\label{eq:AL_2SC} \\
\tilde\Pi_{{\rm MT},D}^{\mu\nu} (k) &= 3 \int_q
\tilde\Xi_D (q) \ \mathcal{R}_{{\rm MT}, D}^{\mu\nu} (q, k),
\label{eq:MT_2SC} \\
\tilde\Pi_{{\rm DOS},D}^{\mu\nu} (k) &= 3 \int_q
\tilde\Xi_D (q) \ \mathcal{R}_{{\rm DOS},D}^{\mu\nu} (q, k),
\label{eq:DOS_2SC} 
\end{align}
denote the contributions of the AL, MT, and DOS terms, respectively,
and
$q = (\bm{q}, i\nu_n)$ is the four-momentum of the soft mode.
The vertex functions $\tilde\Gamma^\mu_D (q, k)$, 
$\mathcal{R}_{{\rm MT},D}^{\mu\nu} (q, k)$ and
$\mathcal{R}_{{\rm DOS},D}^{\mu\nu} (q, k)$ 
in Eqs.~(\ref{eq:AL_2SC})--(\ref{eq:DOS_2SC}) are given by
\begin{align}
\tilde\Gamma^\mu_D (q, q+k) 
&= 4 (N_c - 1) e_\Delta \int_p
{\rm Tr}_D [\mathcal{G}_0 (p) \gamma^\mu \mathcal{G}_0 (p+k) \mathcal{G}_0 (q-p)],
\label{eq:Gamma_2SC} \\
\mathcal{R}_{{\rm MT},D}^{\mu\nu} (q, k) 
&= 8 (N_c - 1)~e_u e_d \int_p
{\rm Tr}_D [\mathcal{G}_0 (p) \gamma^\mu \mathcal{G}_0 (p+k) \mathcal{G}_0 (q-p-k) 
\gamma^\nu\mathcal{G}_0 (q-p)],
\label{eq:RMT_2SC} \\
\mathcal{R}_{{\rm DOS},D}^{\mu\nu} (q, k) 
&= 4 (N_c - 1) (e_u^2 + e_d^2) \sum_{s=\pm} \int_p
{\rm Tr}_D [\mathcal{G}_0 (p) \gamma^\mu \mathcal{G}_0 (p+sk) \gamma^\nu    
\mathcal{G}_0 (p) \mathcal{G}_0 (q-p)],
\label{eq:RDOS_2SC}
\end{align}
with $e_\Delta = e_u + e_d$ being the electric charge of the diquarks.
It is nicely shown that Eqs.~\eqref{eq:Gamma_2SC}--\eqref{eq:RDOS_2SC}
satisfy the WT identities 
\begin{align}
k_\mu \tilde{\Gamma}^\mu_D (q, q+k) &
= e_\Delta [\tilde{\Xi}_D^{-1}(q+k) - \tilde{\Xi}_D^{-1}(q)], 
\label{eq:Gamma-WT_2SC} \\
k_\mu \mathcal{R}^{\mu\nu}_D (q, k) &
= e_\Delta [\tilde{\Gamma}^\nu_D (q-k, q) - \tilde{\Gamma}^\nu_D (q, q+k)], 
\label{eq:R-WT_2SC}
\end{align}
with $\mathcal{R}^{\mu\nu}_D (q, k) =
\mathcal{R}_{{\rm MT},D}^{\mu\nu} (q, k)+
\mathcal{R}_{{\rm DOS},D}^{\mu\nu} (q, k)$.
Using Eqs.~\eqref{eq:Gamma-WT_2SC} and~\eqref{eq:R-WT_2SC}, it is easy to show that Eq.~\eqref{eq:Pi-fluc_2SC} does satisfy the WT identity in Eq.~\eqref{eq:Pi-WT-DS}.

In the above procedure, $\tilde\Pi_D^{\mu\nu}(q)$ is constructed from the one-loop diagram in Fig.~\ref{fig:Omega_2SC}. This means that our procedure corresponds to the linear approximation in the sense that the interaction 
between the soft modes is not taken into account. 
The validity of the linear approximation would be
lost when $T$ is quite close to
$T_c$ where the amplitude of the diquark fluctuations is large. 
Our analysis does not apply to such a temperature region, 
where the non-linear effects give rise to the critical exponents of transport coefficients
specific to the dynamical universality class~\cite{halperin1974first}. 
Nevertheless, a formalism analogous
to ours for metallic superconductivity
is applied successfully
to describe the precursory phenomena like the anomalous enhancement 
of the electric conductivity 
in metals in a wide range of $T$ above $T_c$~\cite{Larkin:book,tinkham2004introduction}. 
It is thus expected that our following analysis 
is also capable 
of describing quantitative behaviors of the transport coefficients in a certain 
range of $T$ that is not too close to $T_c$.
To extend the formalism to the close vicinity of $T_c$, one may start from 
the thermodynamic potential with multiple loops of $\tilde\Xi_D(q)$~\cite{Takayama:1971}. 
The properties of the soft mode will also be modified there, which will be amended 
via the self-consistent treatment~\cite{Kagamihara:2019nsz}.

Another issue in the present analysis
is that the equilibrium state is determined in the MFA 
with use of the thermodynamic potential Eq.~\eqref{eq:thermodynamic-potential}, where
the contribution of the fluctuations~\eqref{eq:Omega_D} are not 
taken into account.
In this sense, our formalism is not self-consistent.
In the present study, however, we neglect these effects to investigate the magnitude 
of fluctuations qualitatively in a simple formalism.

\subsubsection{Vertices}
\label{sec:vertex_2SC}

The calculation of Eq.~\eqref{eq:Pi-fluc_2SC} involves three-loop diagrams and is not easy to calculate even when $\tilde\Xi_D(q)$ is a simple propagator. 
We thus have recourse to an approximate method, which should have a good validity near the 2SC-PT.

For the construction of such an approximation, we employ the approximate formula~\eqref{eq:Xi-LEE_2SC} or~\eqref{eq:Xi-TDGL_2SC} for the $T$-matrix $\Xi^R_D(\bm{k},\omega)$,
which reproduce the low energy-momentum behavior of the soft modes as was shown in the previous section.
When the TDGL approximation is adopted, 
since Eq.~\eqref{eq:Xi-TDGL_2SC} 
is governed only by three parameters, $\tilde a_D$, $b_D$, $c_D$, 
the analysis may acquire a generality not restricted to the NJL model.

Once Eq.~\eqref{eq:Xi-LEE_2SC} or~\eqref{eq:Xi-TDGL_2SC} is employed for $\Xi^R_D (\bm{k}, \omega)$, 
the vertices $\tilde\Gamma^\mu_D (q, q+k)$ and $\mathcal{R}^{\mu\nu}_D (q, k)$ 
should
satisfy Eqs.~\eqref{eq:Gamma-WT_2SC} and \eqref{eq:R-WT_2SC} in accordance with the 
approximate form of $\Xi^R_D (\bm{k}, \omega)$. 
To construct such vertices that are valid in the low energy-momentum region, 
we 
utilize the WT identities~\eqref{eq:Gamma-WT_2SC} and~\eqref{eq:R-WT_2SC}, 
instead of a direct calculation of Eqs.~\eqref{eq:Gamma_2SC}--\eqref{eq:RDOS_2SC}.

We first take $\tilde\Gamma^\mu_D (q,q+k)$.
The analytic continuation of the LE approximation~\eqref{eq:Xi-LEE_2SC} is given by $\tilde\Xi_D(q)=A_D(\bm{q})+{\cal C}_D(i\nu_n)$ with 
\begin{align}
    {\cal C}_D(i\nu_n) = 
    \begin{cases}
        C_D i\nu_n & \nu_n>0 ,\\
        0 & \nu_n=0 , \\
        (C_D i\nu_n)^* & \nu_n<0 .
    \end{cases}
    \label{eq:calC}
\end{align}
Substituting Eq.~\eqref{eq:calC} into the right-hand side of Eq.~\eqref{eq:Gamma-WT_2SC}
leads to 
\begin{align}
\tilde\Xi^{-1}_D (\bm{q}+\bm{k},i\nu_n+i\nu_m) - \tilde\Xi^{-1}_D (\bm{q},i\nu_m) 
&= 
[{\cal C}_D (i\nu_n+i\nu_m) - {\cal C}_D (i\nu_n) ]
+ A_D (\bm{q} + \bm{k}) - A_D (\bm{q})
\nonumber \\
&= 
[{\cal C}_D (i\nu_n+i\nu_m) - {\cal C}_D (i\nu_n) ]
+ A_D^{(1)} (\bm{q} + \bm{k}, \bm{q})  (2\bm{q}+\bm{k}) \cdot \bm{k},
\label{eq:Xi-Xi_2SC}
\end{align}
where 
\begin{align}
A_D^{(1)} (\bm{q}_1, \bm{q}_2) 
= \frac{A_D (\bm{q}_1) - A_D (\bm{q}_2)}{ |\bm{q}_1|^2 - |\bm{q}_2|^2 },
\end{align}
is a real function since $A_D(\bm{q})$ is real.
In Eq.~(\ref{eq:Xi-Xi_2SC}), one sees that only the first (second) term has the $\nu_n$ ($\bm{k}$) dependence.
We thus identify the first and second terms in Eq.~\eqref{eq:Xi-Xi_2SC} as $k_0 \tilde\Gamma_D^0 (q, q+k)$ 
and $k_i \tilde\Gamma^i_D (q, q+k)$ in Eq.~(\ref{eq:Gamma-WT_2SC}), respectively, which gives
\begin{align}
\tilde\Gamma^0_D (q, q+k) =&~e_\Delta
\frac{{\cal C}_D (i\nu_n+i\nu_m) - {\cal C}_D (i\nu_m)}{i\nu_n} ,
\label{eq:Gamma-0-omega_2SC} \\
\tilde\Gamma^i_D (q, q+k) =&\ \Gamma^i_D(\bm{q},\bm{q}+\bm{k})=-e_\Delta
A_D^{(1)} (\bm{q} + \bm{k}, \bm{q}) (2q+k)^i,
\label{eq:Gamma-i-omega_2SC} 
\end{align}
where $\tilde\Gamma^0_D (q, q+k)$ and $\tilde\Gamma^i_D (q, q+k)$ 
are complex and real numbers, respectively. 
In Eq.~\eqref{eq:Gamma-i-omega_2SC}, we have introduced the function $\Gamma^i_D(\bm{q},\bm{q}+\bm{k})$ 
to show explicitly that this vertex function does not depend on $\nu_n,\nu_m$, 
and hence it has the same form for the imaginary- and real-time formalisms.
Similarly, when one employs 
the TDGL approximation~\eqref{eq:Xi-TDGL_2SC} for $\tilde\Xi_D(k)$, we have 
\begin{align}
\Gamma^i_D(\bm{q},\bm{q}+\bm{k})=-e_\Delta b_D  (2q+k)^i.
\label{eq:Gamma-i-TDGL_2SC} 
\end{align} 
We note that Eq.~\eqref{eq:Gamma-i-TDGL_2SC} agrees with Eq.~\eqref{eq:Gamma-i-omega_2SC} 
in the small $\bm{k}$ limit.
Also, Eq.~\eqref{eq:Gamma-0-omega_2SC} is obtained by substituting $\bm{k}=\bm0$ into Eq.~\eqref{eq:Gamma-WT_2SC}.
In the analysis of the transport coefficients, we do not use Eq.~\eqref{eq:Gamma-0-omega_2SC} since only the spatial components of $\tilde\Pi^{\mu\nu}(k)$ are necessary for it. 

In general, the form of the vertex functions is not determined uniquely by the WT identity. 
However, it is shown that Eqs.~\eqref{eq:Gamma-0-omega_2SC}--\eqref{eq:Gamma-i-TDGL_2SC} agrees with the explicit calculation of Eq.~\eqref{eq:Gamma_2SC} in the $k^\mu\to0$ limit.
In fact, Eq.~\eqref{eq:Gamma_2SC} in this limit is calculated to be
\begin{align}
\lim_{k_\mu\to0} \tilde\Gamma^\mu_D (q, q+k) =&~4 (N_c - 1) e_\Delta \int_p
{\rm Tr}_{D} [\mathcal{G}_0 (p) \gamma^\mu \mathcal{G}_0 (p) \mathcal{G}_0 (q-p)] 
= e_\Delta \frac{\partial \tilde\Xi^{-1}_D (q)}{\partial q_\mu} .
\label{eq:Gamma-exact_2SC}
\end{align}
Plugging Eq.~(\ref{eq:Xi-LEE_2SC}) into Eq.~(\ref{eq:Gamma-exact_2SC}), one obtains
\begin{align}
\lim_{k_\mu\to 0} \tilde\Gamma^0_D (q, q+k) &=
e_\Delta \frac{\partial {\cal Q}_D (\bm{q}, q_0)}{\partial q_0} = e_\Delta {\cal C}_D(q_0),
\label{eq:Q-derivative-q0_2SC}
\\
\lim_{k_\mu\to 0} \tilde\Gamma^i_D (q, q+k) &=
e_\Delta \frac{\partial Q^R_D (\bm{q}, \omega)}{\partial q_i}
= -e_\Delta \frac{\partial A_D(\bm{q})}{\partial |\bm{q}|^2} 2q^i.
\label{eq:Q-derivative-qi_2SC}
\end{align}
These results are consistent with 
Eqs.~\eqref{eq:Gamma-0-omega_2SC}--\eqref{eq:Gamma-i-TDGL_2SC} for $k^\mu\to0$.

\begin{figure}[t]
\centering
\includegraphics[keepaspectratio, scale=0.40]{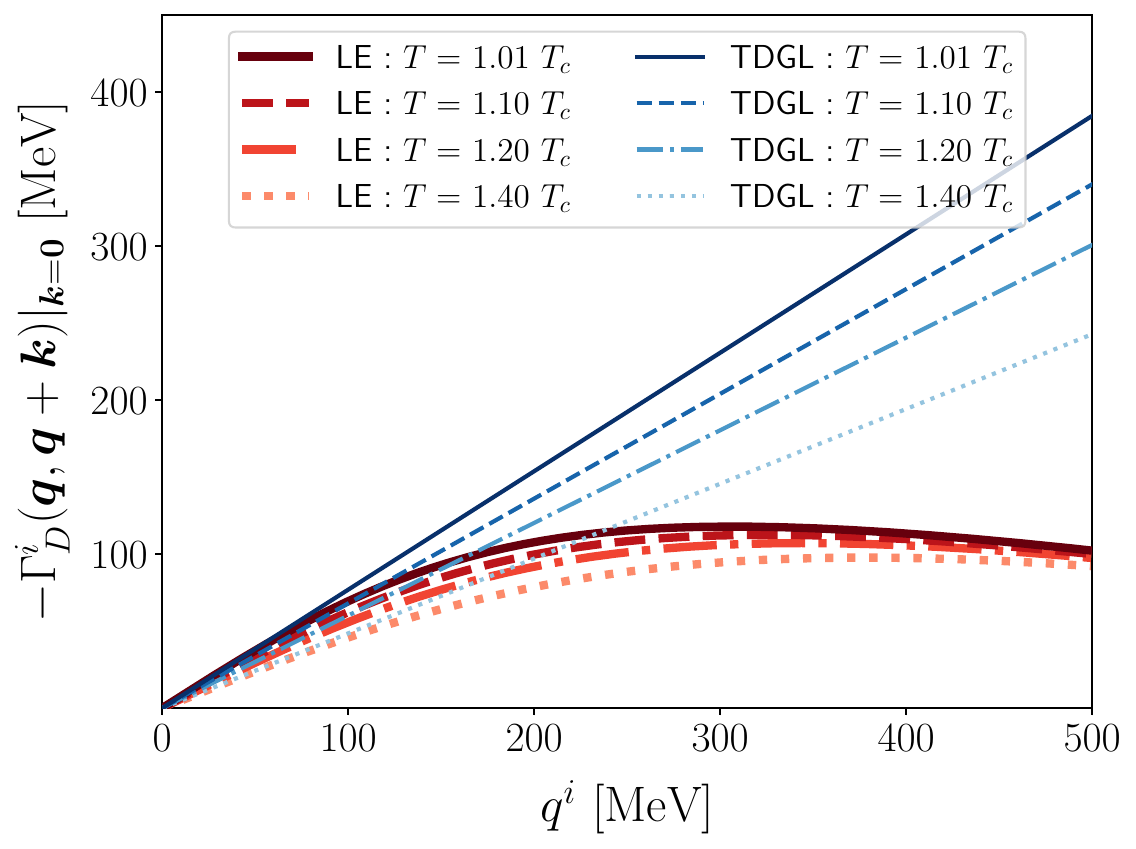}
\caption{
Momentum dependence of the vertex in the AL term 
$\tilde\Gamma^i_D (q, q+k)$ at $\mu = 350~{\rm MeV}$ 
and $G_D = 0.7G_S$ with $q^i=|\bm{q}|$ and $|\bm{k}|=0$
for several temperatures.
The thick (red) and thin (blue) lines correspond to the LE and TDGL approximations, respectively.
}
\label{fig:Gamma_2SC}
\end{figure}

To compare the behaviors of $\Gamma^i_D (\bm{q}, \bm{q}+\bm{k})$ in the LE and TDGL approximations, Eqs.~\eqref{eq:Gamma-i-omega_2SC} and~\eqref{eq:Gamma-i-TDGL_2SC},
in Fig.~\ref{fig:Gamma_2SC} we show $q^i$ dependence of $\Gamma^i_D (\bm{q}, \bm{q})$ for several temperatures near $T_c$ at $\mu=350$~MeV.
The figure shows that the difference between the LE and TDGL approximations is
large at large $|\bm{q}|$, where the former is suppressed while the latter is increasing linearly.
This result indicates that the TDGL approximation tends to overestimate the contribution of the integral~\eqref{eq:AL_2SC} at large $|\bm{q}|$, which is dominated by the non-critical contributions.
This tendency will be confirmed by the numerical analysis in the next section.

Next, we consider the vertices of the MT and DOS terms
$\mathcal{R}^{\mu\nu}_D (q, k)$.
Repeating the same 
manipulation as was done for the AL term 
with the WT-identity (\ref{eq:R-WT_2SC}) 
one obtains 
\begin{align}
\mathcal{R}^{ij}_D (q, k) = R^{ij}_D(\bm{q},\bm{k}) = 4 e_\Delta
\frac{\tilde\Gamma^i_D (q, q+k) - \tilde\Gamma^i_D (q-k, q)}
{(\bm{q} + \bm{k})^2 - (\bm{q} - \bm{k})^2} q^j
= 4 e_\Delta
\frac{\Gamma^i_D (\bm{q}, \bm{q}+\bm{k}) - \Gamma^i_D (\bm{q}-\bm{k}, \bm{q})}
{(\bm{q} + \bm{k})^2 - (\bm{q} - \bm{k})^2} q^j.
\label{eq:R-ii_2SC}
\end{align}
From the fact that $\Gamma_D^i(\bm{q},\bm{q}+\bm{k})$ is a real number that does not depend on the Matubara frequencies for both the LE and TDGL approximations, $R^{ij}_D(\bm{q},\bm{k})$ in Eq.~\eqref{eq:R-ii_2SC} 
is also a real function of momentum only. 
As in the previous calculation for $\tilde\Gamma_D^\mu(q,q+k)$, it is shown by the explicit calculation that the $k^\mu\to0$ limit of Eqs.~\eqref{eq:RMT_2SC} and~\eqref{eq:RDOS_2SC} agrees with that of Eq.~\eqref{eq:R-ii_2SC}.

\subsubsection{Photon self-energy}

Now that the effective forms of the vertices are obtained, 
the next step is the calculations of Eqs.~(\ref{eq:AL_2SC})--(\ref{eq:DOS_2SC}). 
However, using the fact that Eq.~\eqref{eq:R-ii_2SC} is a real function 
of momentum only, it is shown that 
\begin{align}
    {\rm Im}\big[\Pi^{Rij}_{{\rm MT},D} (\bm{k}, \omega) 
              + \Pi^{Rij}_{{\rm DOS},D} (\bm{k}, \omega) \big] = 0 ,
\label{eq:MT-DOS_cancel_2SC}
\end{align}
i.e. the MT and DOS terms cancel out exactly in ${\rm Im}\Pi^{Rij}(\bm{k},\omega)$.

Equation~\eqref{eq:MT-DOS_cancel_2SC} is proved as follows. We manipulate 
$\tilde\Pi^{ij}_{{\rm MT},D} (k) 
+ \tilde\Pi^{ij}_{{\rm DOS},D} (k)$ as 
\begin{align}
\tilde\Pi^{ij}_{{\rm MT},D} (k) 
+ \tilde\Pi^{ij}_{{\rm DOS},D} (k) =&~
3 \int_q \tilde\Xi_D (q) \mathcal{R}^{ij}_D (q, k)
\nonumber \\
=&~3 \int \frac{d^3\bm{p}}{(2\pi)^3} R^{ij}_D(\bm{q}, \bm{k})
\oint_C\frac{dq_0}{2\pi i} \frac{{\rm coth} \frac{q_0}{2T}}{2} \tilde\Xi_D (\bm{q}, q_0)
\nonumber \\
=&~3 \int \frac{d^3\bm{p}}{(2\pi)^3} R^{ij}_D (\bm{q}, \bm{k})
\int \frac{d\omega'}{2\pi i} \frac{{\rm coth} \frac{\omega'}{2T}}{2} 
\bigg( \Xi^R_D(\bm{q}, \omega') - \Xi^A_D(\bm{q}, \omega') \bigg)
\nonumber \\
=&~3 \int \frac{d^3\bm{p}}{(2\pi)^3} R^{ij}_D (\bm{q}, \bm{k})
\int \frac{d\omega'}{2\pi} 
{\rm Im} \Xi^R_D (\bm{q}, \omega') \coth \frac{\omega'}{2T}.
\label{eq:MT-DOS_proof}
\end{align}
In the second equality, we have used the fact that ${\cal R}_D^{ij}(q,k)$ 
does not depend on Matsubara frequencies and transformed the Matsubara sum into the integral 
over the contour $C$ on the complex-energy plane that encircles the imaginary axis.
In the third equality, the contour is deformed so as to avoid the cut in $\tilde\Xi_D (q)$ on the real axis with $\Xi^A_D(\bm{q}, \omega)=\tilde\Xi_D(\bm{q}, i\nu_m)|_{i\nu_m\to\omega-i\eta}=\Xi^R_D(\bm{q}, \omega)^*$.
The last line of Eq.~\eqref{eq:MT-DOS_proof} shows that $\tilde\Pi^{ij}_{{\rm MT},D} (k) 
+ \tilde\Pi^{ij}_{{\rm DOS},D} (k)$ is real, since this expression only contains real functions. Moreover, Eq.~\eqref{eq:MT-DOS_proof} shows that $\tilde\Pi^{ij}_{{\rm MT},D} (k) 
+ \tilde\Pi^{ij}_{{\rm DOS},D} (k)$ does not depend on $k_0$. Therefore, its analytic continuation to the retarded function $\Pi^{Rij}_{{\rm MT},D} (\bm{k},\omega) 
+ \Pi^{Rij}_{{\rm DOS},D} (\bm{k},\omega)$ is also real, which proves Eq.~\eqref{eq:MT-DOS_cancel_2SC}.
A similar argument on the MT and DOS terms in the metallic superconductivity is found in Ref.~\cite{Larkin:book}.

To calculate the electric conductivity and the relaxation time, we need only the imaginary part of the spatial components, ${\rm Im} \Pi^{R ij} (\bm{k},\omega)$.
Equation~\eqref{eq:MT-DOS_cancel_2SC} thus shows that 
we only have to compute the AL term for this purpose.
The explicit form of ${\rm Im} \Pi^{Rij}_D (k)$
is obtained from Eq.~\eqref{eq:AL_2SC} by taking the analytic continuation $i\nu_l \rightarrow \omega + i\eta$ as
\begin{align}
{\rm Im} \Pi^{R ij}_D (\bm{k}, \omega) 
=&\ {\rm Im} \Pi^{R ij}_{{\rm AL}, D} (\bm{k}, \omega) \nonumber \\
=&~3 \int \frac{d^3q}{(2\pi)^3}
\Gamma_D^i (\bm{q}, \bm{q}+\bm{k}) \Gamma_D^j (\bm{q}+\bm{k}, \bm{q})
\int \frac{d\omega'}{2\pi} {\rm coth} \frac{\omega'}{2T}
 \nonumber \\
&\ {\rm Im} \Xi^R_D (\bm{q}, \omega')
\Big\{ {\rm Im} \Xi^R_D (\bm{q} + \bm{k}, \omega'+\omega) 
- {\rm Im} \Xi^R_D (\bm{q} - \bm{k}, \omega'-\omega) \Big\}.
\label{eq:AL-explicit_2SC}
\end{align}

To prove Eq.~\eqref{eq:MT-DOS_cancel_2SC}, we have used the fact that the vertex $R_D^{ij}(\bm{q},\bm{k})$ is a real 
function of momenta $\bm{q}$ and $\bm{k}$ only. 
In the above argument, 
this property of $R_D^{ij}(\bm{q},\bm{k})$ is obtained through 
the LE or TDGL approximation and the WT identities, which are well justified at small $\omega$ and $|\bm{k}|$ near $T_c$.
On the other hand, its validity may be violated at large energy or momentum. 
Nevertheless, it is expected that this possible violation 
does not significantly affect the critical enhancement of the transport coefficients 
near $T_c$ where the soft modes in the low energy-momentum region play crucial roles.

\subsection{Contribution of the soft modes of QCD-CP}
\label{sec:Self-energy_QCDCP}

Next, we calculate $\tilde\Pi_S^{\mu\nu}(k)$, i.e., the modification of the photon self-energy from the soft modes of the QCD-CP~\cite{Nishimura:2023oqn}.
As the calculational procedure goes in a similar way to the last subsection, duplicated descriptions will be omitted in what follows.

\begin{figure}[t]
\centering
\includegraphics[keepaspectratio, scale=0.45]{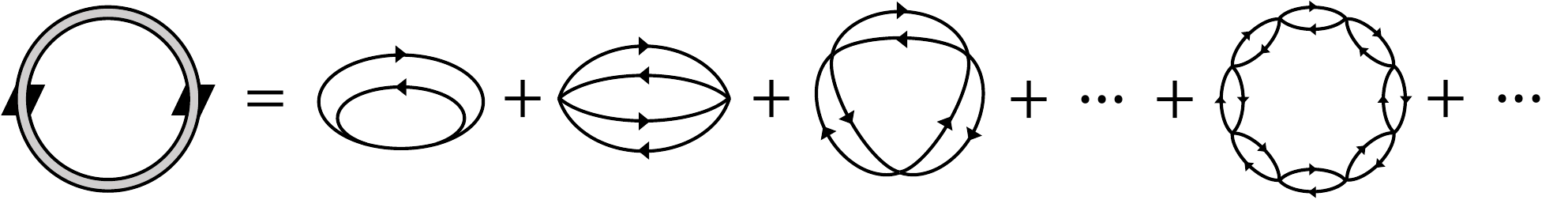}
\caption
{
Contribution of the soft mode of the QCD-CP to the thermodynamic potential.
}
\label{fig:Omega_QCDCP}
\end{figure}

\begin{figure*}[t]
\centering

\begin{tabular}{c@{\hspace{2pt}}c@{\hspace{2pt}}c@{\hspace{2pt}}c@{\hspace{2pt}}c}
        \includegraphics[keepaspectratio, scale=0.09]{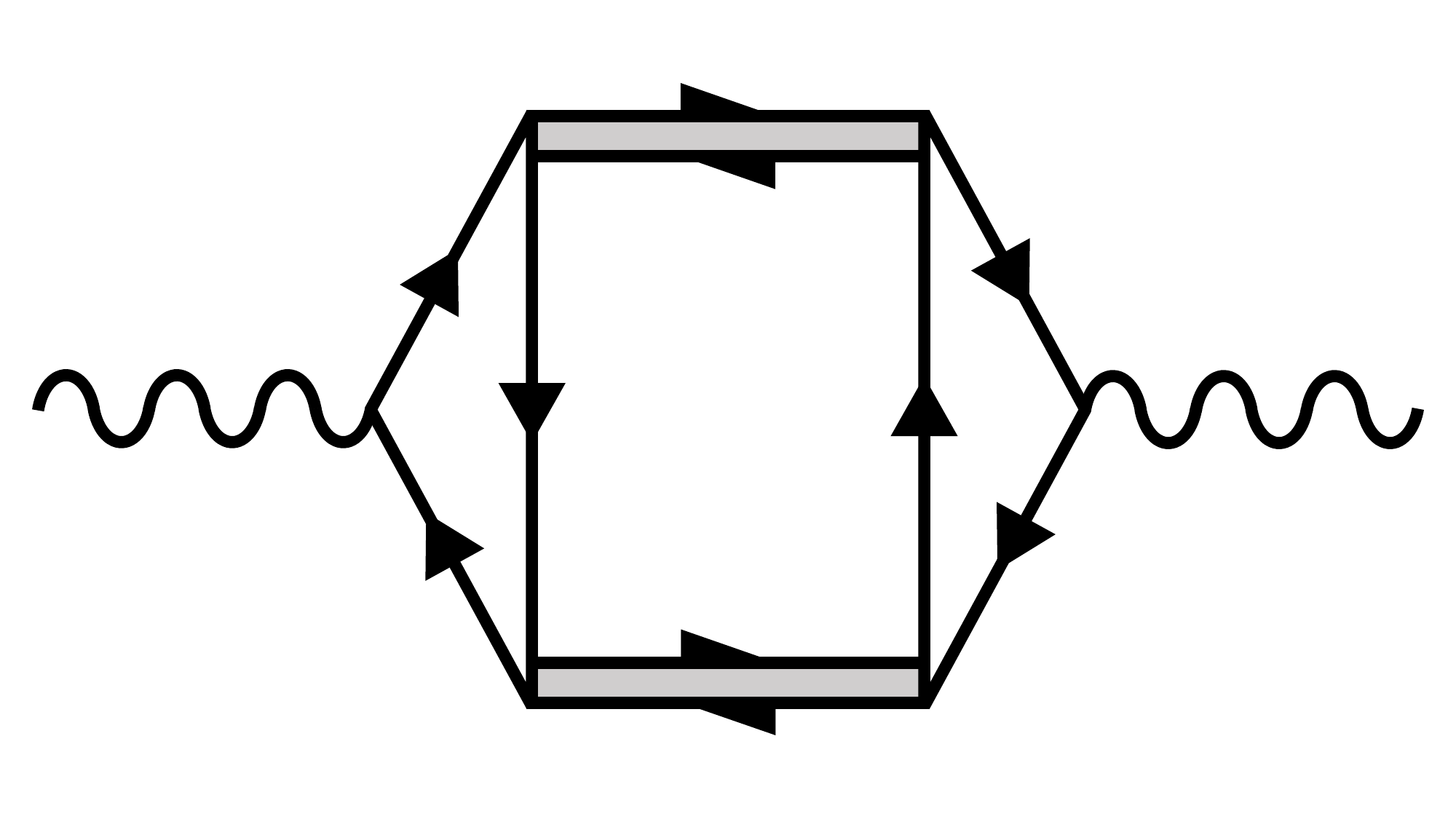} &
        \includegraphics[keepaspectratio, scale=0.09]{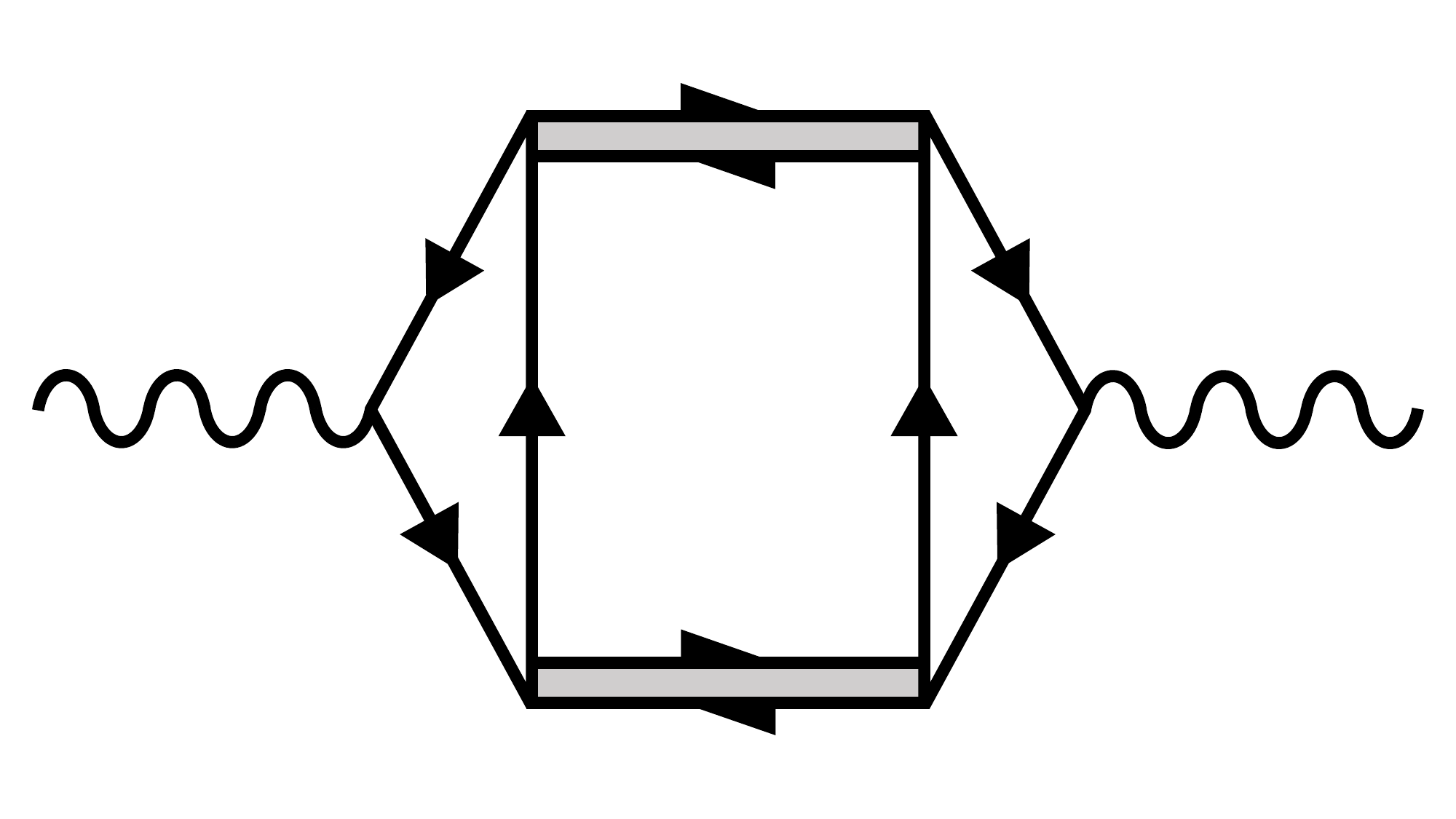} &
        \includegraphics[keepaspectratio, scale=0.09]{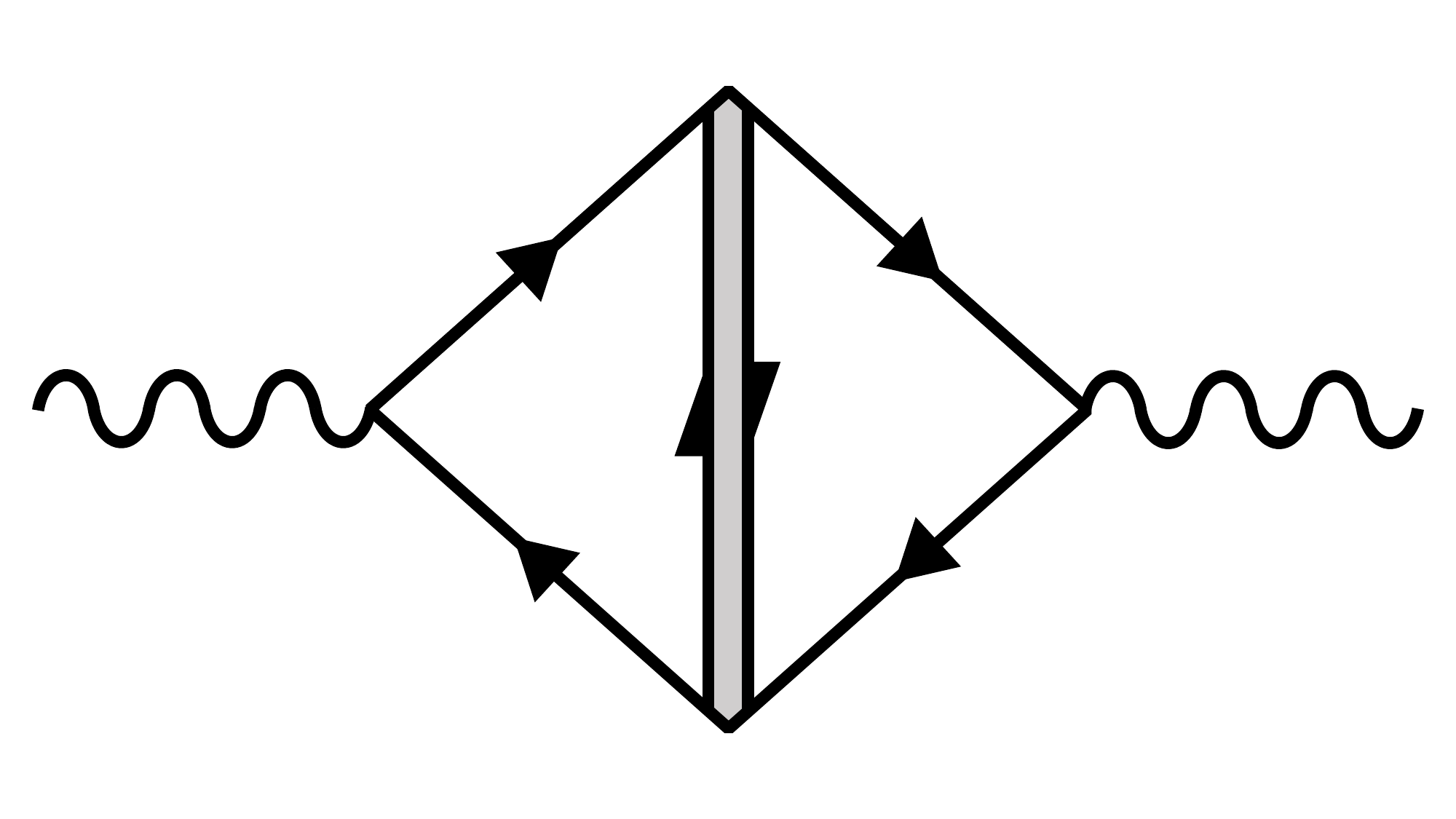} &
        \includegraphics[keepaspectratio, scale=0.09]{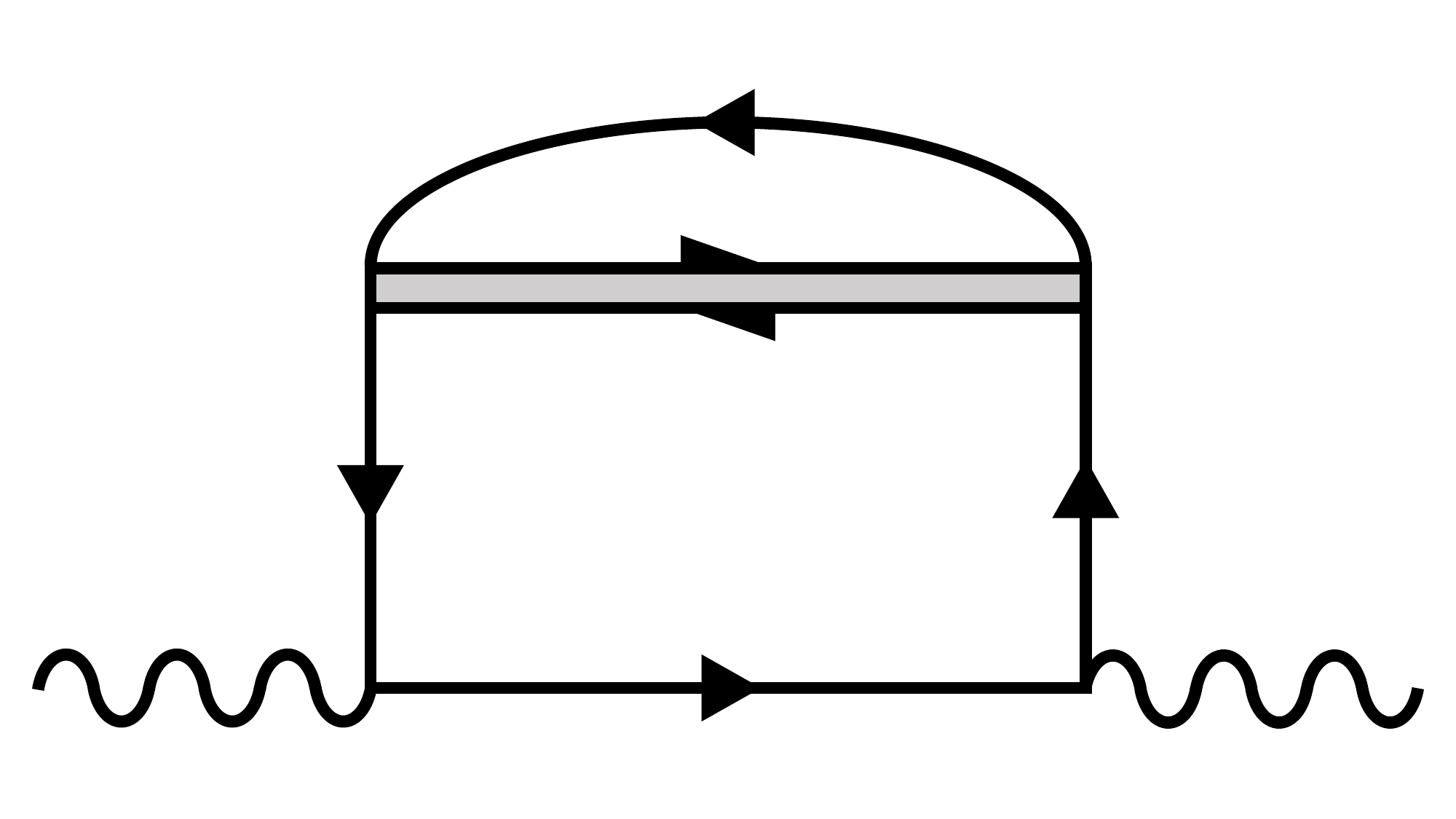} &
        \includegraphics[keepaspectratio, scale=0.09]{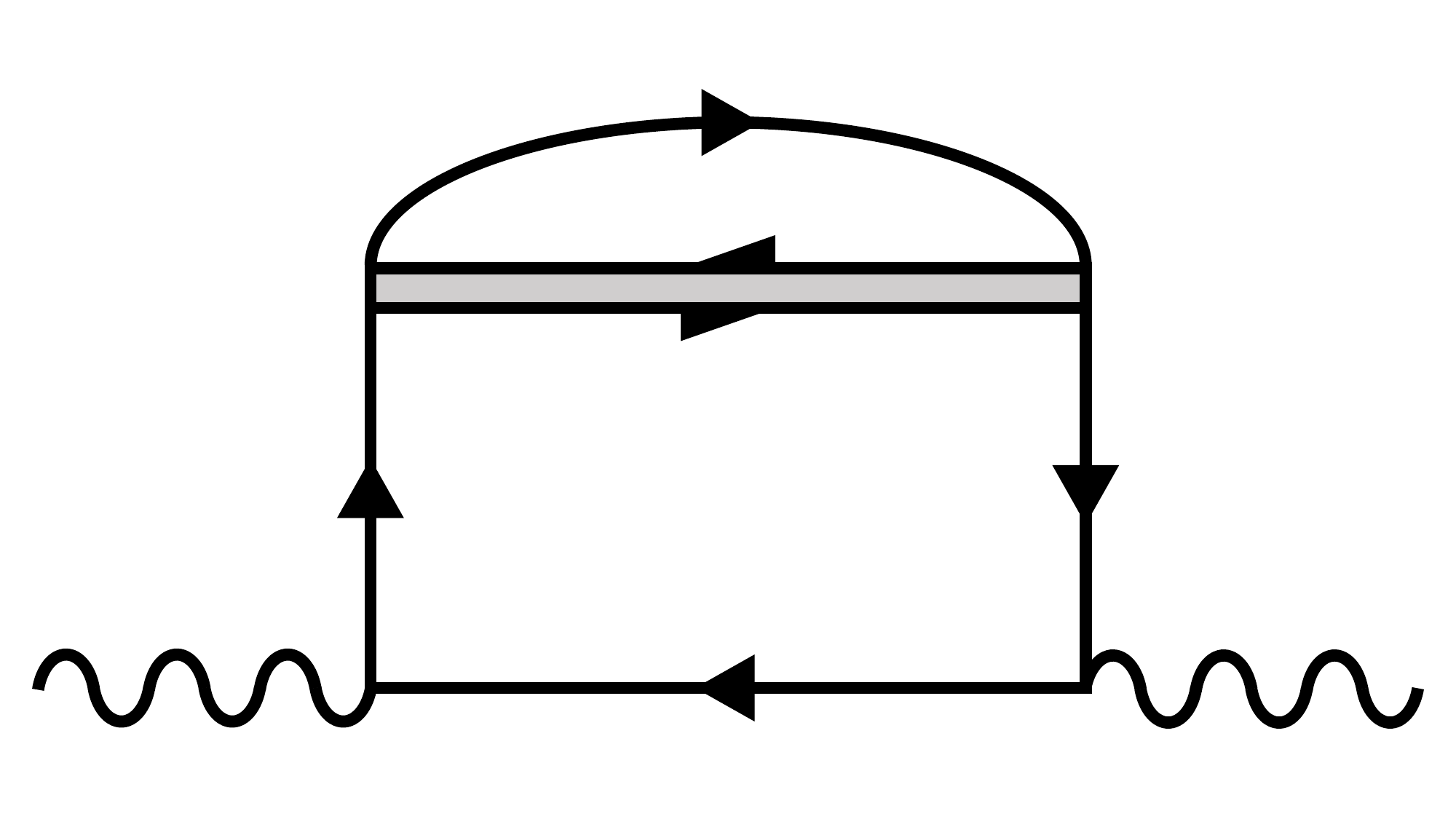}
        \\
        (a) & (c) & (e) & (g) & (i)
        \\
        \\
        \includegraphics[keepaspectratio, scale=0.09]{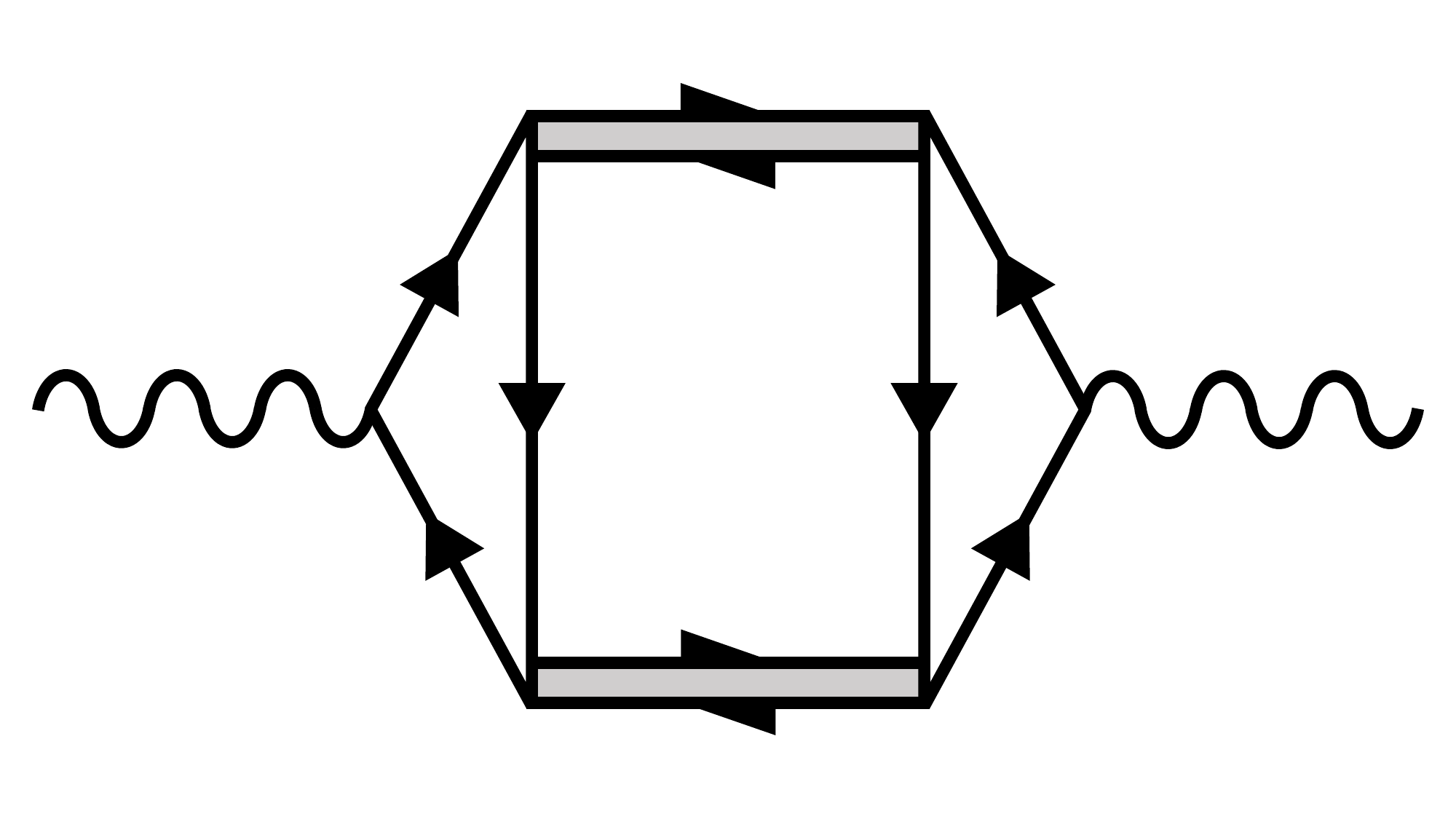} &
        \includegraphics[keepaspectratio, scale=0.09]{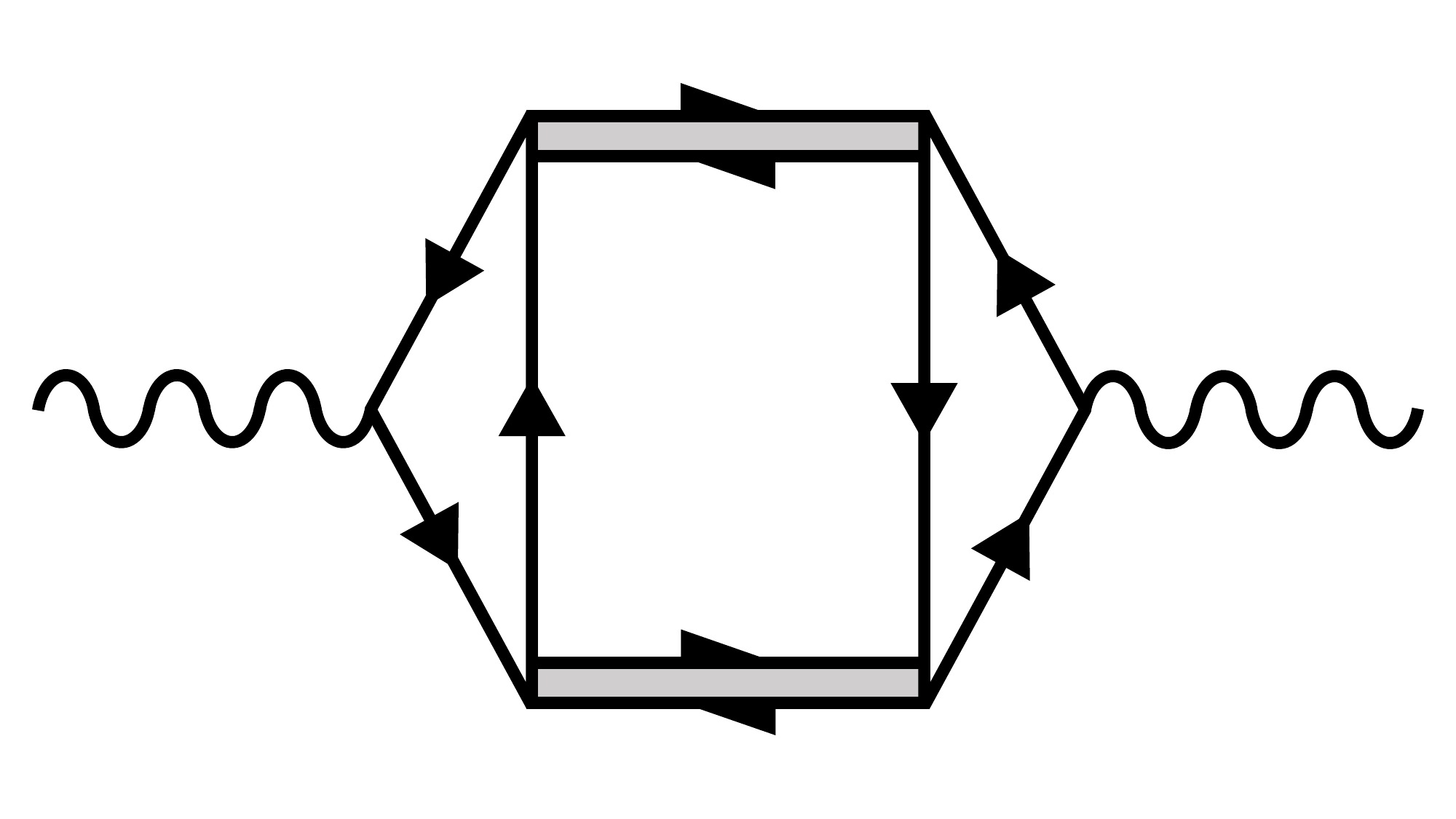} &
        \includegraphics[keepaspectratio, scale=0.09]{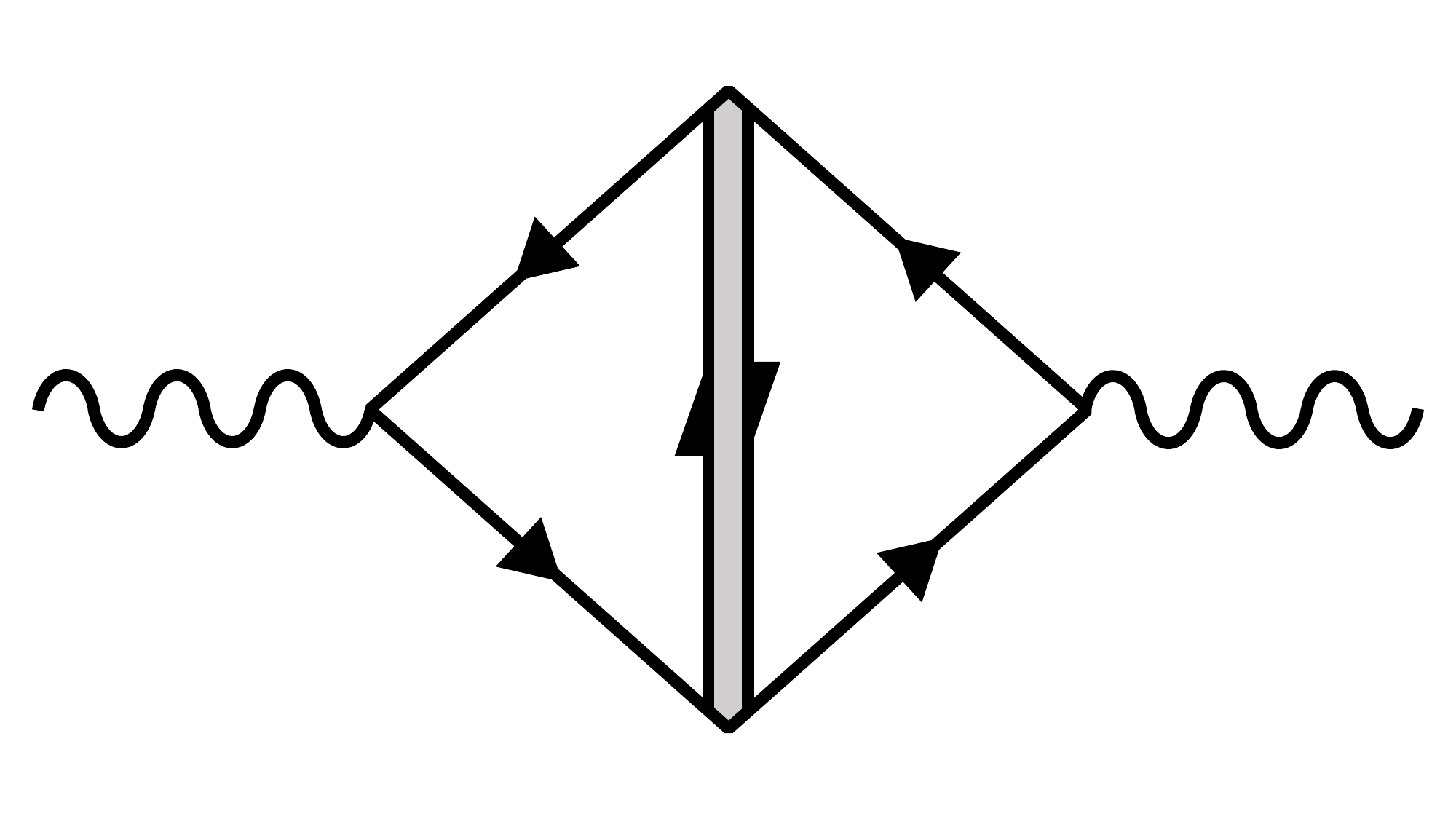} &
        \includegraphics[keepaspectratio, scale=0.09]{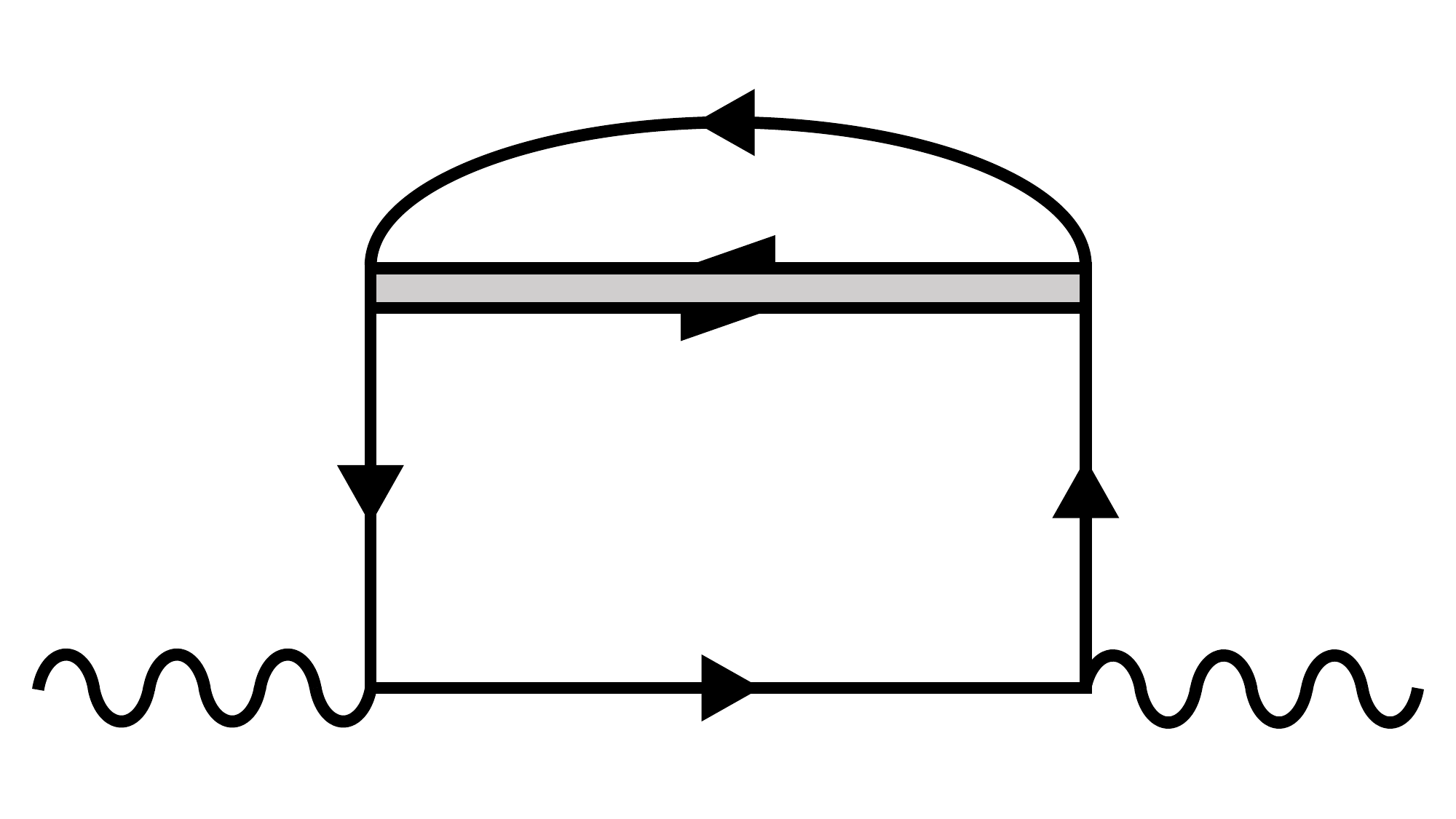} &
        \includegraphics[keepaspectratio, scale=0.09]{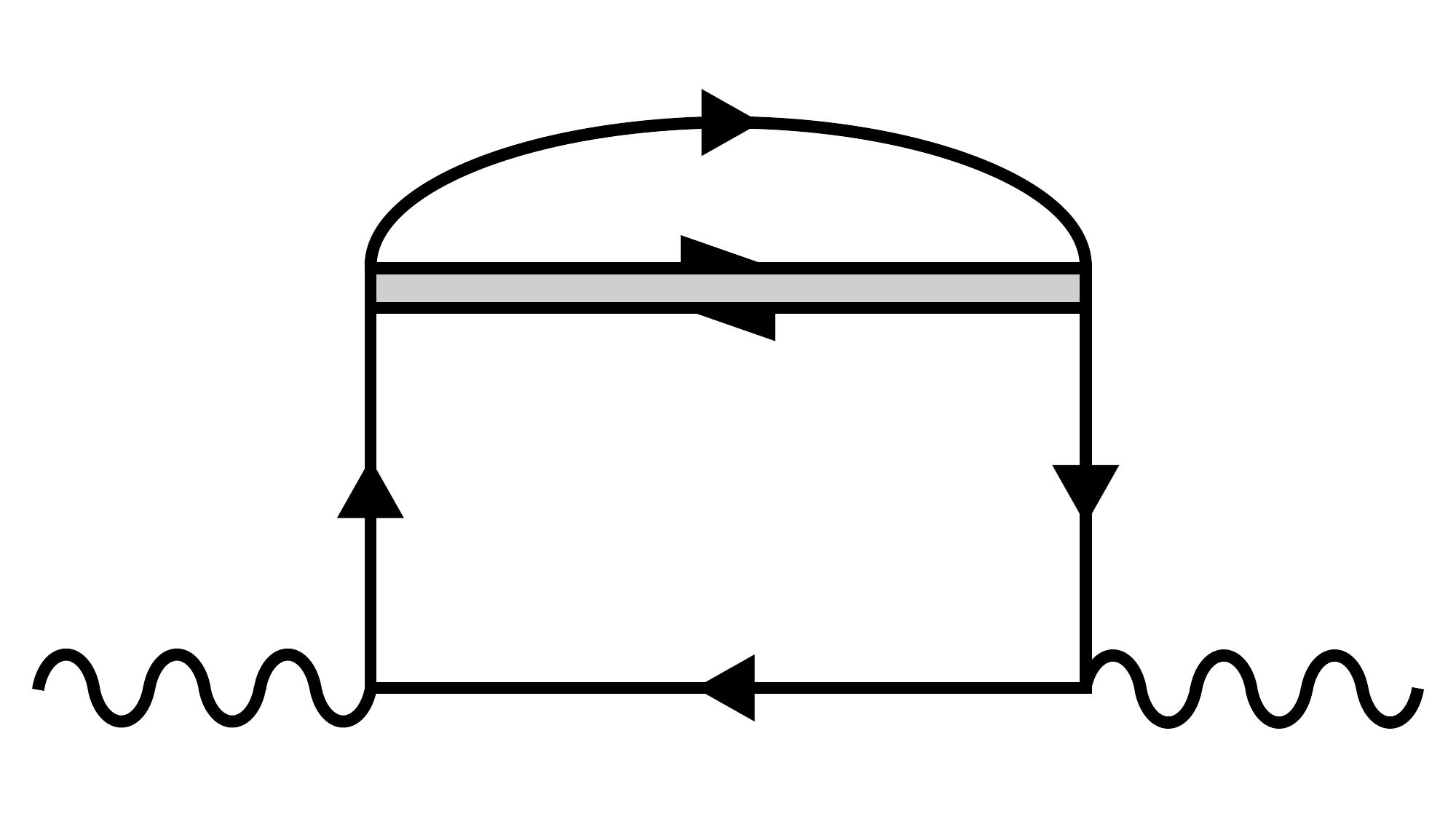}
        \\
        (b) & (d) & (f) & (h) & (j)
\end{tabular}
    \caption{
The diagrammatic representations of the Aslamazov-Larkin (a)--(d),
Maki-Thompson (e, f) and density of states (g)--(j) terms 
with the soft modes of the QCD-CP.
The single, double, and wavy lines are quarks, 
soft modes, and photon, respectively.
}
\label{fig:Self-energy_QCDCP}
\end{figure*}

As we did in Sec.~\ref{sec:Self-energy_2SC},
we start with the analysis 
of the one-loop diagram of the soft mode of the QCD-CP shown in Fig.~\ref{fig:Omega_QCDCP},
which is the lowest-order contribution to the thermodynamic potential
\begin{align}
\Omega_S = \int_p \ln [G_S \tilde\Xi_S^{-1}(p)].
\label{eq:Omega_S}
\end{align}
Attaching electromagnetic vertices at any different two points of quark lines in $\Omega_S$,
we obtain ten types of diagrams shown in Fig.~\ref{fig:Self-energy_QCDCP}, 
where the number of diagrams is 
larger than that in Fig.~\ref{fig:Self-energy_2SC}, 
because diagrams with different directions of the quark lines should be distinguished.
Because of similarities of the diagrams with those in the last subsection, we refer to 
the diagrams (a)--(d) as the AL, 
(e), (f) as the MT, and (g)--(j) as the DOS terms, respectively.
The respective contributions to the photon self-energy 
in the imaginary-time formalism
are denoted by 
\begin{align}
\tilde\Pi_{{\rm AL},S}^{\mu\nu} (k) &= \sum_{f=u,d} \int_q
\tilde\Gamma^\mu_f (q, q+k) \tilde\Xi_S (q+k) 
\tilde\Gamma^\nu_f (q+k, q) \tilde\Xi_S (q),
\label{eq:AL_QCDCP} \\
\tilde\Pi_{{\rm MT},S}^{\mu\nu} (k) &= \sum_{f=u,d} \int_q
\tilde\Xi_S (q) \ \mathcal{R}_{{\rm MT},f}^{\mu\nu} (q, k),
\label{eq:MT_QCDCP} \\
\tilde\Pi_{{\rm DOS},S}^{\mu\nu} (k) &= \sum_{f=u,d} \int_q
\tilde\Xi_S (q) \ \mathcal{R}_{{\rm DOS},f}^{\mu\nu} (q, k),
\label{eq:DOS_QCDCP} 
\end{align}
which in total gives $\tilde\Pi^{\mu\nu}_S (k)$
as 
\begin{align}
\tilde\Pi^{\mu\nu}_S (k) 
= \tilde\Pi^{\mu\nu}_{{\rm AL},S} (k)
+ \tilde\Pi^{\mu\nu}_{{\rm MT},S} (k) 
+ \tilde\Pi^{\mu\nu}_{{\rm DOS},S} (k).
\label{eq:Pi-fluc_QCDCP}
\end{align}
The vertex functions $\tilde\Gamma^\mu_{f} (q, k)$, 
$\mathcal{R}_{{\rm MT},f}^{\mu\nu} (q, k)$ and
$\mathcal{R}_{{\rm DOS},f}^{\mu\nu} (q, k)$ 
in Eqs.~(\ref{eq:AL_QCDCP})--(\ref{eq:DOS_QCDCP}) are given by
\begin{align}
\tilde\Gamma^\mu_f (q, q+k) &= 2 N_c e_f \sum_{s=\pm} s \int_p 
{\rm Tr} [\mathcal{G}_0 (p)\gamma^\mu\mathcal{G}_0 (p+sk)\mathcal{G}_0 (p-sq)],
\label{eq:Gamma_QCDCP} \\
\mathcal{R}_{{\rm MT},f}^{\mu\nu} (q, k) &= 4 N_c e^2_f \sum_{s=\pm} \int_p
{\rm Tr} [\mathcal{G}_0 (p) \gamma^\mu\mathcal{G}_0 (p+sk)
\mathcal{G}_0 (p+sk+sq) \gamma^\nu\mathcal{G}_0 (p+sq)],
\label{eq:RMT_QCDCP} \\
\mathcal{R}_{{\rm DOS},f}^{\mu\nu} (q, k) &= 2 N_c e^2_f \sum_{s, t=\pm} \int_p
{\rm Tr} [\mathcal{G}_0 (p)\gamma^\mu \mathcal{G}_0 (p+sk) \gamma^\nu    
\mathcal{G}_0 (p) \mathcal{G}_0 (q-tp)].
\label{eq:RDOS_QCDCP}
\end{align}
As before, one can explicitly check that these vertices satisfy the WT identities
\begin{align}
k_\mu \tilde{\Gamma}^\mu_f (q, q+k) &
= -e_f [\tilde{\Xi}_S^{-1} (q+k) - \tilde{\Xi}_S^{-1} (q)], 
\label{eq:Gamma-WT_QCDCP} \\
k_\mu \mathcal{R}^{\mu\nu}_f (q, k) &
= -e_f [\tilde{\Gamma}^\nu_f (q-k, q)-\tilde{\Gamma}^\nu_f (q, q+k)], 
\label{eq:R-WT_QCDCP}
\end{align}
where $\mathcal{R}^{\mu\nu}_f (q, k)=
\mathcal{R}_{{\rm MT},f}^{\mu\nu}(q, k) + \mathcal{R}_{{\rm DOS},f}^{\mu\nu}(q, k)$.
The WT identity of Eq.~\eqref{eq:Pi-fluc_QCDCP}
\begin{align}
k_\mu \tilde\Pi^{\mu\nu}_S (k) = 0,
\end{align}
is easily verified using Eqs.~\eqref{eq:Gamma-WT_QCDCP} and~\eqref{eq:R-WT_QCDCP}.

\begin{figure*}[t]
    \centering
    \includegraphics[keepaspectratio, scale=0.40]{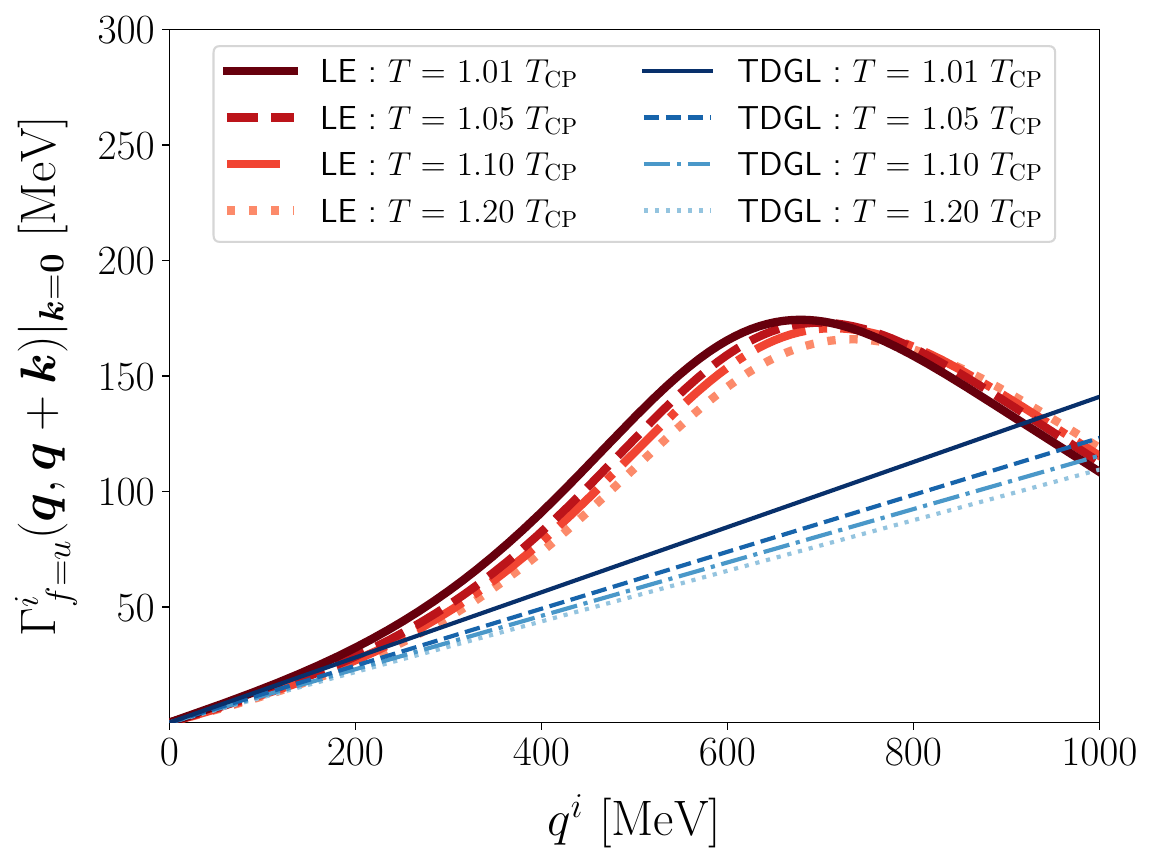}        \includegraphics[keepaspectratio, scale=0.40]{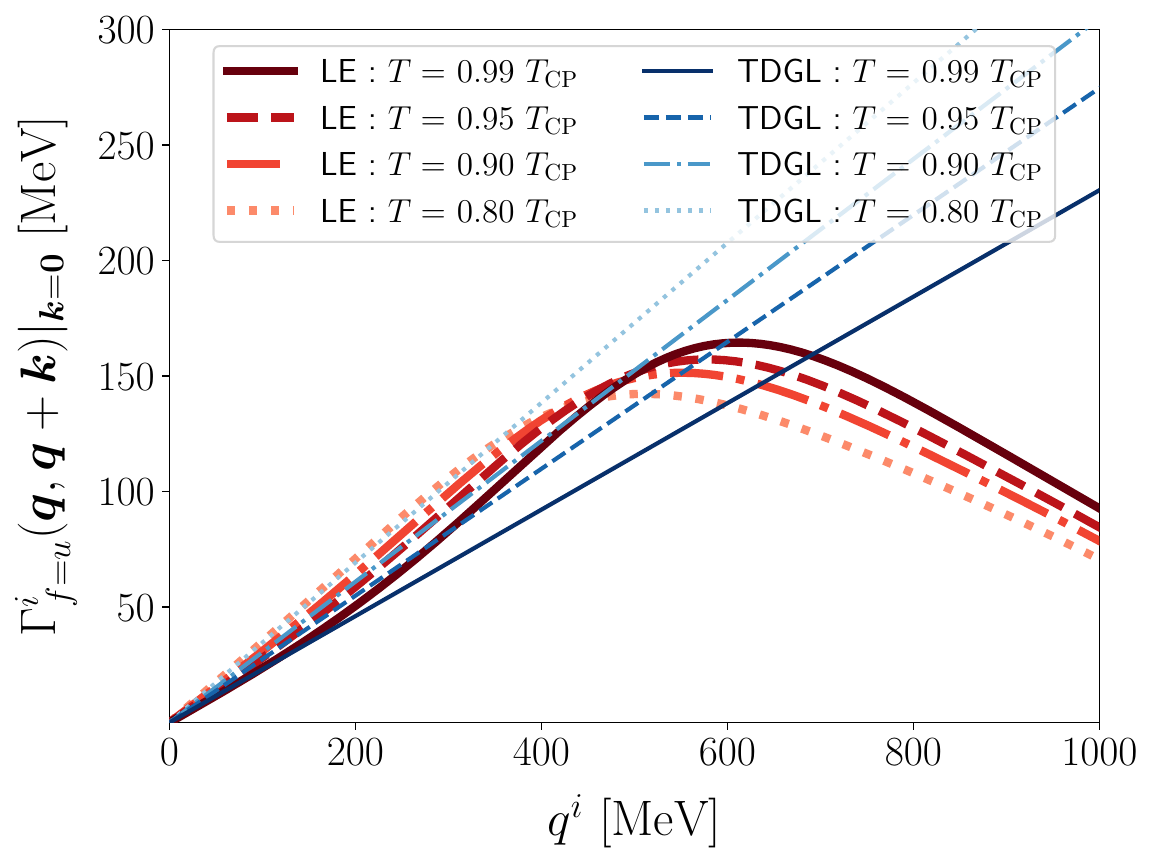}
\caption{
Momentum dependence of the vertex in the AL term 
$\tilde\Gamma^i_{f=u} (q, q+k)$ at $\mu = \mu_{\rm CP}$ 
with $q^i=|\bm{q}|$ and $|\bm{k}|=0$ for several temperatures.
The left and right panels show the results for $T>T_{\rm CP}$ and $T<T_{\rm CP}$, respectively.
The thick (red) and thin (blue) lines correspond to the LE and TDGL approximations, respectively.
} 
\label{fig:Gamma_QCDCP}
\end{figure*}

As we did in Sec.~\ref{sec:Self-energy_2SC}, 
we calculate Eqs.~(\ref{eq:AL_QCDCP})--(\ref{eq:DOS_QCDCP}) 
using the LE or TDGL approximation for $\Xi_S^R(\bm{q},\omega)$.
Then, the vertex functions $\tilde{\Gamma}^\mu_f (q, q+k)$ and $\mathcal{R}_{f}^{\mu\nu} (q, k)$ 
are determined so as to satisfy the WT identities
Eqs.~\eqref{eq:Gamma-WT_QCDCP} and~\eqref{eq:R-WT_QCDCP}.
In the LE approximation, this procedure leads to
\begin{align}
\tilde\Gamma^0_f (q, q+k) =& -e_f
\frac{{\cal C}_S (\bm{q}+\bm{k},q_0+k_0) - {\cal C}_S (\bm{q},q_0)}{k_0},
\label{eq:Gamma-0-omega_QCDCP} \\
\tilde\Gamma^i_f (q, q+k) =&~
\Gamma_f^i(\bm{q},\bm{q}+\bm{k})= e_f
A_S^{(1)}(\bm{q}+\bm{k}, \bm{q})  (2q+k)^i,
\label{eq:Gamma-i-omega_QCDCP} 
\end{align}
where
\begin{align}
A_S^{(1)} (\bm{q}_1, \bm{q}_2) 
= \frac{A_S (\bm{q}_1) - A_S (\bm{q}_2) }{ |\bm{q}_1|^2 - |\bm{q}_2|^2 },
\end{align}
and ${\cal C}_S(\bm{q},i\nu_l)$ is defined similarly to Eq.~\eqref{eq:calC}.
When the TDGL approximation~(\ref{eq:Xi-TDGL_QCDCP}) to $\Xi^R(\bm{q},\omega)$ is adopted, we obtain
\begin{align}
\Gamma^i_f (\bm{q}, \bm{q}+\bm{k}) &= e_f b_S (2q+k)^i.
\label{eq:Gamma-i-TDGL_QCDCP} 
\end{align}
We note that $\Gamma^i_f(\bm{q}, \bm{q}+\bm{k})$ is again a real function only of momenta 
both in the LE and TDGL approximations.
We also remark that 
the direct calculation of Eq.~\eqref{eq:Gamma_QCDCP} leads to the same expressions
in the small energy-momentum region.
In fact, from Eq.~\eqref{eq:Gamma_QCDCP}, one finds 
\begin{align}
\tilde\Gamma^\mu_f (q, q+k)|_{k=0} =&~2 N_c e_f \sum_{s=\pm} \int_p 
{\rm Tr} [\mathcal{G}_0 (p)\gamma^\mu\mathcal{G}_0 (p)\mathcal{G}_0 (p-sq)] 
= - e_f \frac{\partial \tilde\Xi^{-1}_S (q)}{\partial q_\mu}.
\label{eq:Gamma-exact_QCDCP}
\end{align}
Substituting Eqs.~\eqref{eq:ImQ_QCDCP} and \eqref{eq:ReQ_QCDCP} into this equation one finds that 
$\tilde\Gamma^0_f (q, q+k)|_{k=0}=\partial \tilde\Xi^{-1}_S (q) / \partial q_0 =$
$\partial \mathcal{Q}_S (q) / \partial q_0$ can contain a term proportional to $1/|\bm{q}|$ in the $\bm{q}\to0$ limit,
while $\tilde\Gamma^i_f (q, q+k)|_{k=0}=\partial \tilde\Xi^{-1}_S (q) / \partial q_i =$
$\partial \mathcal{Q}_S (q) / \partial q_i$ cannot.
Equations~\eqref{eq:Gamma-0-omega_QCDCP} and~\eqref{eq:Gamma-i-omega_QCDCP}, and hence Eq.~\eqref{eq:Gamma-i-TDGL_QCDCP}, are unique choices satisfying this constraint.
Moreover, the WT identity at $\bm{k}=\bm{0}$
\begin{align}
k_0 \tilde\Gamma^0_f (q, q+k)|_{\bm{k}=\bm{0}} 
= -e_f [\tilde\Xi^{-1}_S (\bm{q}, q_0+k_0) - \tilde\Xi^{-1}_S (\bm{q}, q_0)],
\label{eq:Gamma-0-WT_QCDCP}
\end{align}
with the LE approximation agrees with the $\bm{k}\to\bm{0}$ limit of Eq.~\eqref{eq:Gamma-0-omega_QCDCP}.
These analyses jusitfy Eqs.~\eqref{eq:Gamma-0-omega_QCDCP}, \eqref{eq:Gamma-i-omega_QCDCP}, \eqref{eq:Gamma-i-TDGL_QCDCP} in the small $k^\mu$ limit.

Figure~\ref{fig:Gamma_QCDCP} shows  the $q^i = |\bm{q}|$ dependence of 
Eqs.~\eqref{eq:Gamma-i-omega_QCDCP} and \eqref{eq:Gamma-i-TDGL_QCDCP} 
at $\bm{k}=\bm{0}$.
The figure shows that the two results agree with each other in the small $|\bm{q}|$ limit. 
For larger $|\bm{q}|$, the TDGL gives smaller values than those in the LE 
at $|\bm{q}|\lesssim800$~MeV for $T\gtrsim T_c$,
which indicates that the use of the TDGL 
tends to underestimate the non-critical 
contribution to $\tilde\Pi_D^{ij}(k)$ in the high temperature region.
This behavior is in contrast to the case for 2SC-PT shown in Fig.~\ref{fig:Gamma_2SC}, 
where the TDGL result overestimates the non-critical contributions.

Next, we construct the vertices of the MT and DOS terms
$\mathcal{R}^{\mu\nu}_f (q, k)$ with the same procedure as that given in Sec~\ref{sec:vertex_2SC}.
Through the WT-identity~(\ref{eq:R-WT_QCDCP}),
$\mathcal{R}^{\mu\nu}_f (q, k)$ is evaluated as
\begin{align}
\mathcal{R}^{00}_f (q, k) =& -e_f
\frac{\tilde\Gamma^0_f (q, q+k)-\tilde\Gamma^0_f (q-k, q)}{k_0}, 
\label{eq:R-00_QCDCP} \\
\mathcal{R}^{ij}_f (q, k) =&~
R^{ij}_f (\bm{q}, \bm{k}) =
4 e_f
\frac{\Gamma^i_f (\bm{q}, \bm{q}+\bm{k})-\Gamma^i_f (\bm{q}-\bm{k}, \bm{q})}
{(\bm{q}+\bm{k})^2 - (\bm{q}-\bm{k})^2} q^j.
\end{align}
Here, it is important that $\mathcal{R}^{ij}_f (q, k)$ 
is a real function only of momenta $\bm{q}$ and $\bm{k}$.
Using this property and the same procedure as we did in Eq.~\eqref{eq:MT-DOS_cancel_2SC}, one can show 
\begin{align}
{\rm Im} [\Pi^{Rij}_{{\rm MT}, S}(\bm{k}, \omega) 
            + \Pi^{Rij}_{{\rm DOS}, S}(\bm{k}, \omega)] = 0,
\label{eq:MT-DOS_cancel_QCDCP}
\end{align}
i.e. the cancellation of the MT and DOS terms in the spatial components of ${\rm Im}\Pi_S^{R\mu\nu}(\bm{k},\omega)$. 
Thus ${\rm Im} \Pi^{R ij}_S (\bm{k}, \omega)$ is again given solely by the AL term.

The explicit form of ${\rm Im} \Pi^{R ij}_S (\bm{k}, \omega)$ that 
will be used in the next section for the analysis of transport coefficients is given by taking 
the analytic continuations $k_0 = i\nu_l \rightarrow \omega + i\eta$ of Eqs.~\eqref{eq:AL_QCDCP} 
\begin{align}
{\rm Im} \Pi^{R ij}_S (\bm{k}, \omega)
=&~ {\rm Im} \Pi^{R ij}_{{\rm AL}, S} (\bm{k}, \omega) \nonumber \\
=&~ \sum_f \int \frac{d^3q}{(2\pi)^3}
\Gamma_f^i (\bm{q}, \bm{q}+\bm{k}) \Gamma_f^j (\bm{q}+\bm{k}, \bm{q}) \int \frac{d\omega'}{2\pi} \coth \frac{\omega'}{2T}
\nonumber \\
\times&~ {\rm Im} \Xi^R_S (\bm{q}, \omega')
\Big\{ {\rm Im} \Xi^R_S (\bm{q} + \bm{k}, \omega'+\omega) 
- {\rm Im} \Xi^R_S (\bm{q} - \bm{k}, \omega'-\omega) \Big\}.
\label{eq:AL-explicit_QCDCP}
\end{align}

\section{Transport coefficients}
\label{sec:Transport-coeff}

In this section, we calculate the electric conductivity $\sigma$ and associated relaxation time $\tau_\sigma$
near the 2SC-PT and QCD-CP using the photon self-energy obtained in the previous section.

We start with the spectral density at zero momentum 
\begin{align}
\rho (\omega) \equiv -\sum_{i=1}^3 {\rm Im} \Pi^{R ii} (\bm{0}, \omega).
\label{eq:spectral-density_definition}
\end{align}
It is reasonable to require that the low energy-momentum behavior of the electric current $j^0(\bm{x},t)$ in the rest frame of the medium obeys 
the Maxwell-Cattaneo equation 
\begin{align}
    \Big(\tau_\sigma\frac{\partial^2}{\partial t^2}+\frac\partial{\partial t}  - D\nabla^2 \Big)j^0(\bm{x},t)=0 ,
\end{align}
where the diffusion coefficient $D$ is
expressed in terms of the electric conductivity $\sigma$ and the electric susceptibility $\chi$ as $D=\sigma/\chi$.
Then, it can be shown~\cite{KADANOFF1963419} that 
the spectral density \eqref{eq:spectral-density_definition} at small $\omega$ is written as
\begin{align}
\rho (\omega) = 3 \frac{\sigma \omega}{\tau_\sigma^2 \omega^2 + 1}.
\label{eq:spectral-density_fluid}
\end{align}
Equation~\eqref{eq:spectral-density_fluid} tells us that  $\sigma$ and $\tau_\sigma$ can be extracted from $\rho(\omega)$ as 
\begin{align}
\sigma = \frac{1}{3} \frac{\partial \rho(\omega)}{\partial \omega} \bigg|_{\omega=0}
\hspace{0.5cm} {\rm and} \hspace{0.5cm}
\tau_\sigma = \sqrt{-\frac{1}{3!} \frac{\partial^3 \rho(\omega)}{\partial^3 \omega}
\bigg/~\frac{\partial \rho(\omega)}{\partial \omega} \bigg|_{\omega=0}}\; .
\label{eq:sigma-tau}
\end{align}

Corresponding to the fact that 
the photon self-energy consists of three contributions as given in Eq.~\eqref{eq:Pi-tot},
the spectral function~(\ref{eq:spectral-density_definition}) is also decomposed as 
\begin{align}
    \rho(\omega) = \rho_{\rm free}(\omega) + \rho_D(\omega) + \rho_S(\omega),
\end{align}
with 
\begin{align}
\rho_{\rm free}(\omega)=-\sum_i {\rm Im} \Pi_{\rm free}^{R ii} (\bm{0},\omega) ,
\qquad
\rho_D (\omega) = -\sum_i {\rm Im} \Pi_D^{R ii} (\bm{0}, \omega),
\qquad
\rho_S (\omega) = -\sum_i {\rm Im} \Pi_S^{R ii} (\bm{0}, \omega).
\label{eq:spectral-density_soft-mode}
\end{align}
Among them, $\rho_{\rm free}(\omega)$ does not contribute to $\sigma$ and $\tau_\sigma$ 
since $\rho_{\rm free}(\omega)=0$ for $|\omega|<2M$ and all the $\omega$ derivatives vanish at $\omega=0$.
Near the 2SC-PT, $\rho_D(\omega)$ dominates over $\rho_S(\omega)$ and 
the behavior of $\sigma$ and $\tau_\sigma$ is described only by $\rho_D(\omega)$, and vice versa.
In the following, therefore, we calculate the transport coefficients from $\rho_D(\omega)$ and $\rho_S(\omega)$ separately near the 2SC-PT and QCD-CP, respectively.

In Sec.~\ref{sec:Analytical-results}, we perform analytic calculations with the use of the TDGL approximation.
Then, the numerical results with the LE and TDGL approximations are shown in Sec.~\ref{sec:Numerical-results}.

\subsection{Critical exponents} 
\label{sec:Analytical-results}

Let us first investigate the limiting behaviors of $\sigma$ and $\tau_\sigma$ when the system approaches the 2SC-PT or QCD-CP. 
From Eqs.~\eqref{eq:AL-explicit_2SC} and~\eqref{eq:AL-explicit_QCDCP}, derivatives of $\rho_\gamma(\omega)$
($\gamma = D,\, S$) at $\omega=0$ are obtained as 
\begin{align}
\frac{\partial^n \rho_\gamma(\omega)}{\partial^n \omega} \bigg|_{\omega=0}
&= ~2\bar{N}_\gamma
\int \frac{d^3q}{(2\pi)^3} ~|\bm\Gamma_\gamma(\bm{q},\bm{q})|^2
\int d\omega' \coth\frac{\omega'}{2T} 
~{\rm Im}\Xi^R_\gamma(\bm{q}, \omega')
~\frac{\partial^n}{\partial^n \omega'} {\rm Im}\Xi_\gamma^R(\bm{q}, \omega') ,
\label{eq:rho_gamma_explicit}
\end{align}
with
\begin{align}
&\bar{N}_D = 3e_\Delta^2, 
&&
|\bm\Gamma_D(\bm{q},\bm{q})|^2 = \sum_i \Gamma_D^i(\bm{q},\bm{q})^2,
\nonumber \\
&\bar{N}_S = e_u^2+e_d^2,
&&
|\bm\Gamma_D(\bm{q},\bm{q})|^2 = \sum_{i,f} \Gamma_f^i(\bm{q},\bm{q})^2 .
\label{eq:overall-coeff}
\end{align}

\subsubsection{2SC-PT}

In the case of the 2SC-PT,
as $T$ approaches $T_c$ the integrand in Eq.~\eqref{eq:rho_gamma_explicit} diverges at $(|\bm{q}|,\omega')=(0,0)$ owing to $a_D\to0$ in this limit. Hence, the dominant contribution to Eq.~\eqref{eq:rho_gamma_explicit} gets to come from the origin. 
This justifies the use of the TDGL approximations~\eqref{eq:Xi-TDGL_2SC} and~\eqref{eq:Gamma-i-TDGL_2SC}. 
Moreover, one can safely replace $\coth (\omega'/2T)$ with $2T/\omega'$ and subsequently make the cutoff $\Lambda$ infinity in Eq.~\eqref{eq:rho_gamma_explicit}.
This manipulation leads to 
\begin{align}
\frac{\partial \rho_D (\omega)}{\partial \omega} \bigg|_{\omega=0}
=&~-\frac{9 e_\Delta^2 T}{16 \pi} 
\frac{1}{a_D^{1/2} b_D^{1/2}}
\frac{|c_D|^2}{{\rm Im} c_D},
\label{eq:deff-rho-1_2SC} \\
\frac{\partial^3 \rho_D (\omega)}{\partial^3 \omega} \bigg|_{\omega=0}
=&~\frac{27 e_\Delta^2 T}{512 \pi}
\frac{1}{a_D^{5/2} b_D^{1/2}}
\frac{|c_D|^6}{[{\rm Im} c_D]^3}.
\label{eq:deff-rho-3_2SC}
\end{align}
Inserting these results into Eq.~\eqref{eq:sigma-tau}, we obtain
\begin{align}
\sigma = -\frac{3 e_\Delta^2 T}{16 \pi} 
\frac{1}{a_D^{1/2} b_D^{1/2}}
\frac{|c_D|^2}{{\rm Im} c_D},
\qquad
\tau_\pi = \frac{|c_D|^2}{-8 a_D {\rm Im}c_D}.
\end{align}
Using Eq.~\eqref{eq:abc_simplified-TDGL}, their limiting behaviors are further approximated as
\begin{align}
\sigma \sim T \epsilon^{-1/2} , 
\qquad
\tau_\sigma \sim \frac1T \epsilon^{-1}.
\label{eq:coefficients-behavior_2SC}
\end{align}
This result shows that both the coefficients diverge at $T = T_c$ with critical exponents $-1/2$ and $-1$, respectively.
Equation~\eqref{eq:coefficients-behavior_2SC} also 
tells us 
that the magnitude of $\sigma$ and $\tau_\sigma$ do not have any explicit dependence on $\mu$ nor $G_D$ within the approximaiton~\eqref{eq:abc_simplified-TDGL}.
This implies that $\sigma$ and $\tau_\sigma$ are insensitive to $\mu$ around the 2SC-PT.
These analytic results will be confirmed numerically in Sec.~\ref{sec:Numerical-results}.
Using $\xi_D$ and $\tau_{\rm GL}$ introduced in Eq.~\eqref{eq:tau-xi_2SC}, Eq.~\eqref{eq:coefficients-behavior_2SC} 
is rewritten as
\begin{align}
\sigma \sim
T \frac{\tau_{\rm GL}}{\xi_D} 
\hspace{0.5cm} {\rm and} \hspace{0.5cm}
\tau_\sigma \sim
\tau_{\rm GL}.
\end{align}

We notice that the critical exponents obtained in Eq.~\eqref{eq:coefficients-behavior_2SC} 
correspond to the mean-field ones,
in accordance with the fact that they are obtained in the linear approximation 
for the photon self-energy as mentioned in Sec.~\ref{sec:Self-energy}.
It is expected that the non-linear effects will become so significant in the vicinity of the 2SC-PT that these mean-field results are altered
to those with critical exponents in the dynamical universality class. 
Although such higher-order effects are not incorporated in the present study, it is worth mentioning that the universal arguments are not powerful enough to constrain the overall coefficient of $\sigma$ and $\tau_\sigma$,
whereas the present analysis gives their magnitude quantitatively.

\subsubsection{QCD-CP}

Next, we consider the case of the QCD-CP. As was done in the case of the 2SC-PT, we first replace $\coth (\omega'/2T)$ with $2T/\omega'$ 
and make the cutoff $\Lambda$ infinity in Eq.~\eqref{eq:rho_gamma_explicit} 
with $\gamma=S$. 
The use of the TDGL approximation~\eqref{eq:Xi-TDGL_QCDCP} 
and~\eqref{eq:Gamma-i-TDGL_QCDCP} then allows one to perform the integration explicitly, which leads to 
\begin{align}
\frac{\partial \rho_S (\omega)}{\partial \omega} \bigg|_{\omega=0}
=&~\frac{(e_u^2+e_d^2) T}{2 \pi^3} 
\frac{{\rm Im} c_S}{a_S} {\rm tan}^{-1} \frac{{\rm Im} c_S}{a_S},
\label{eq:deff-rho-1_QCDCP} \\
\frac{\partial^3 \rho_S (\omega)}{\partial^3 \omega} \bigg|_{\omega=0}
=&~\frac{(e_u^2+e_d^2) T}{2 \pi^3} 
\frac{3 b_S [{\rm Im} c_S]^3}{4 a_S^4} {\rm tan}^{-1} \frac{{\rm Im} c_S}{a_S}.
\label{eq:deff-rho-3_QCDCP}
\end{align}

As we have seen in Eq.~\eqref{eq:a_S},
$a_S$ vanishes at the QCD-CP with the exponent dependent on the direction to approach the CP,
while $b_S$ and $c_S$ stay finite.
One thus can find the exponents of the divergences 
\begin{align}
\sigma \sim \frac{1}{a_S} &\sim
\begin{cases}
~\epsilon_{\rm CP}^{-1} &\text{along the first-order PT or crossover transition lines,} \\
~\epsilon_{\rm CP}^{-2/3} & \text{otherwise,} 
\end{cases}
\label{eq:sigma-behavior_QCDCP} \\
\tau_\sigma \sim \frac{1}{a_S^{3/2}} &\sim
\begin{cases}
~\epsilon_{\rm CP}^{-3/2} & \text{along the first-order PT or crossover transition lines,} \\
~\epsilon_{\rm CP}^{-1} & \text{otherwise.} 
\end{cases}
\label{eq:tau-behavior_QCDCP}
\end{align}

\subsection{Numerical results}
\label{sec:Numerical-results}

\begin{figure*}[t]
\centering
\includegraphics[width=0.87\textwidth]{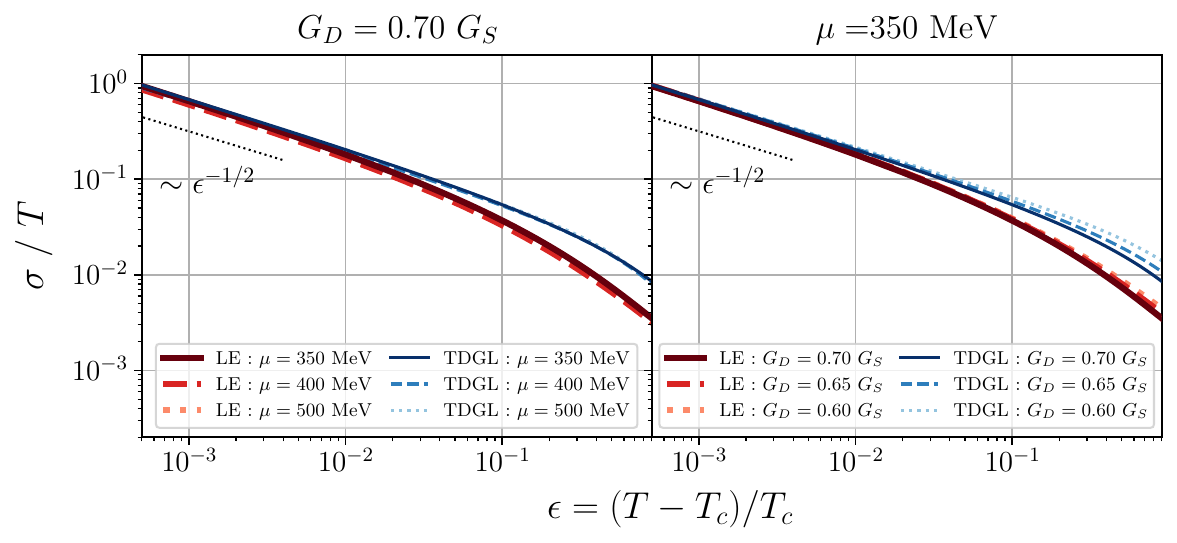}
\\
\includegraphics[width=0.87\textwidth]{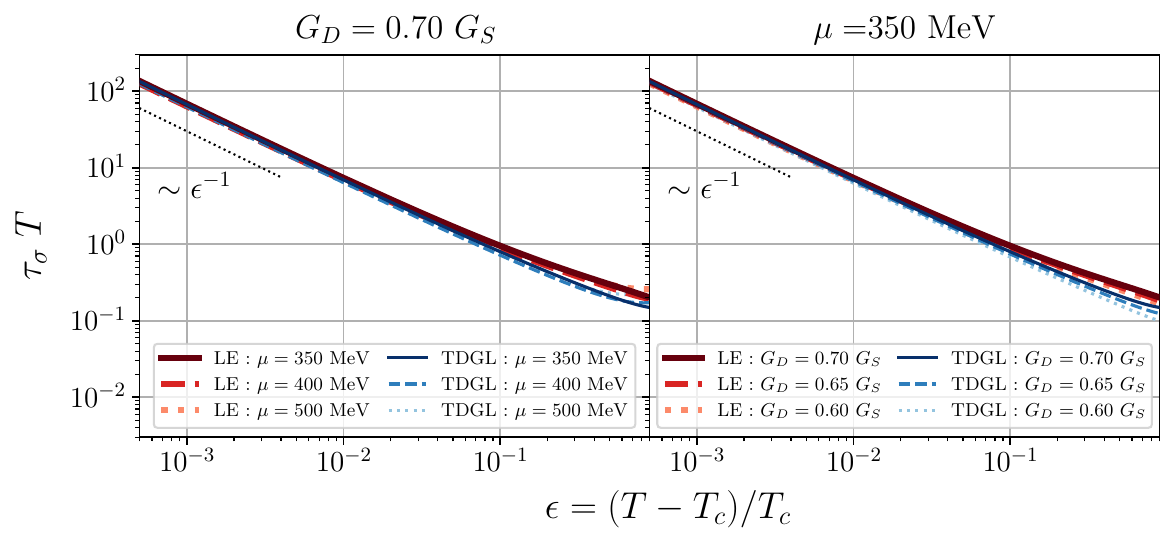}
\caption{
Electric conductivity $\sigma$ and associated relaxation time $\tau_\sigma$ near the 2SC-PT for several values of $\mu$ and $G_D$.
The thick-red and thin-blue lines are the results of the LE and TDGL approximations, respectively.
In the left panels, the lines are plotted 
at $\mu = 350$, $400$, and $500$~MeV with fixed $G_D / G_S= 0.7$,
while the right panels show the results at $G_D / G_S = 0.70$, $0.65$, and $0.60$ for $\mu = 350$~MeV.
The dotted lines indicate the critical exponents $\epsilon^{-1/2}$ (upper panels) and $\epsilon^{-1}$ (lower panels).
}
\label{fig:sigma-tau_2SC}
\end{figure*}

In this subsection, we study the behavior of $\sigma$ and $\tau_\sigma$ numerically with the LE and TDGL approximations. 

First, let us see the behavior of $\sigma$ and $\tau_\sigma$ near the 2SC-PT.
Figure~\ref{fig:sigma-tau_2SC} shows $\sigma/T$ (upper panels) and $\tau_\sigma T$ (lower panels) obtained from $\rho_D(\omega)$ as functions of $\epsilon=(T-T_c)/T_c$.
The left panels show the results for $\mu=350,400,500$~MeV with fixed $G_D/G_S=0.7$, while in the right panels the value of $G_D$ is varied at $\mu=350$~MeV.
The thick-red (thin-blue) lines represent the results 
obtained with the LE (TDGL) approximation.
The figure shows that $\sigma / T$ and $\tau_\sigma T$ grows as $T$ approaches $T_c$ with the exponents in Eq.~\eqref{eq:coefficients-behavior_2SC} indicated by the dotted lines in the figure.
One also sees that the LE and TDGL results 
approach each other in this limit, 
while their difference grows as $\epsilon$ becomes larger.
For large values of $\epsilon$, the TDGL result tends 
to overestimate the LE one,
which is mainly attributed to the behavior of the vertex function as discussed in Sec.~\ref{sec:vertex_2SC}.
Another interesting feature found in Fig.~\ref{fig:sigma-tau_2SC} is that the numerical results of $\sigma/T$ and $\tau_\sigma T$ are insensitive to $\mu$ and $G_D$, 
as anticipated from the analytical results in 
Eq.~\eqref{eq:coefficients-behavior_2SC}.

\begin{figure*}[t]
  \centering
  \includegraphics[width=0.95\textwidth]{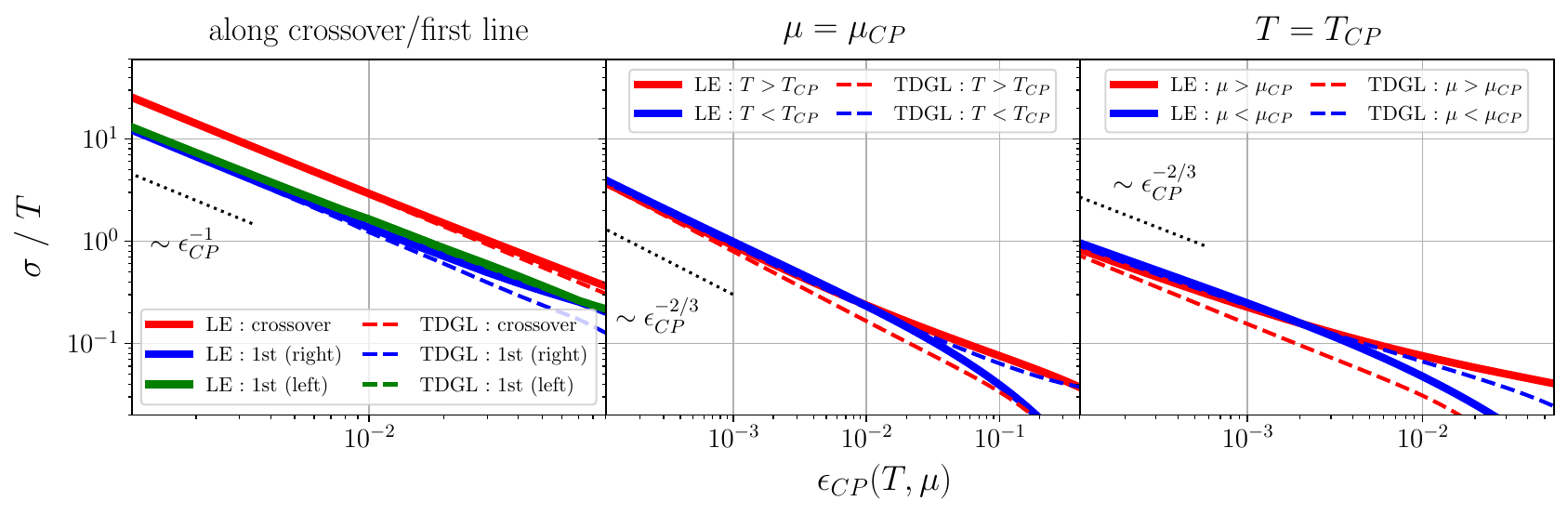}
  \\
  \includegraphics[width=0.95\textwidth]{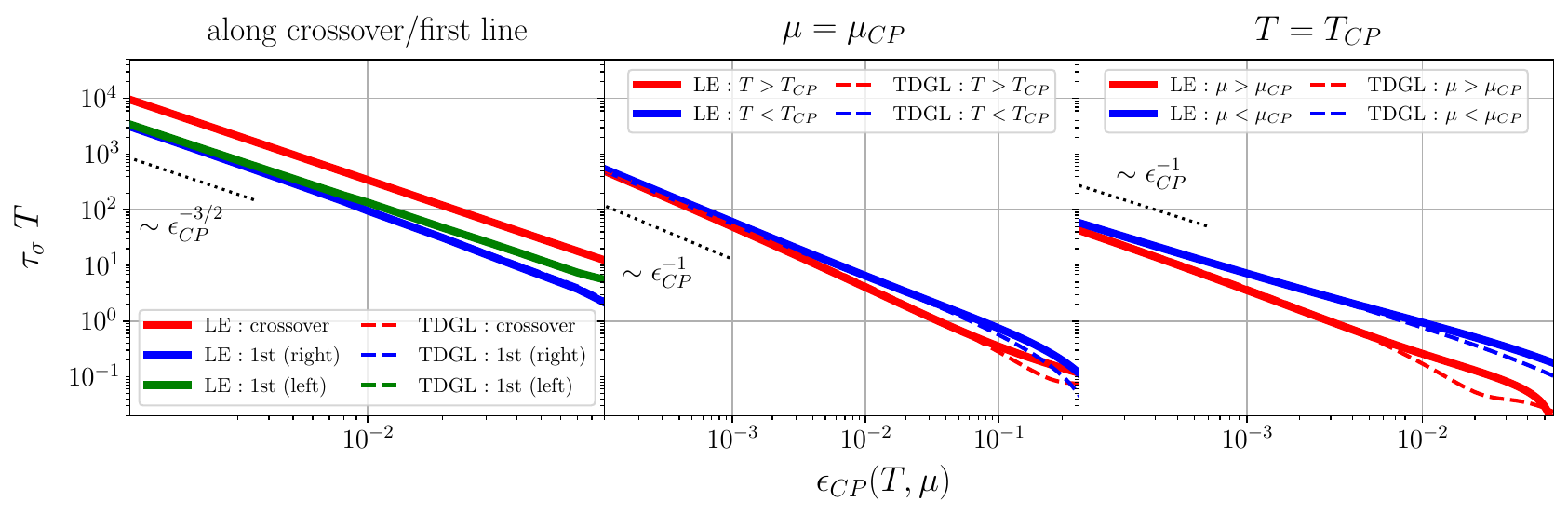}
\caption
{
Electric conductivity $\sigma$ and associated relaxation time $\tau_\sigma$ near the QCD-CP.
In the left panels, $T$ and $\mu$ are varied along the transition line. 
The middle and right panels show the results with fixed $\mu = \mu_{\rm CP}$ 
and $T = T_{\rm CP}$, respectively.
The dotted line in each panel represents the critical exponents in Eqs.~\eqref{eq:sigma-behavior_QCDCP} and~\eqref{eq:tau-behavior_QCDCP}.
}
\label{fig:sigma-tau_QCDCP}
\end{figure*}

Next, we show the numerical results for the QCD-CP in 
Fig.~\ref{fig:sigma-tau_QCDCP}.
The upper and lower panels show the results of $\sigma / T$ and $\tau_\sigma T$, respectively, 
obtained from $\rho_S(\omega)$ as functions of $\epsilon_{\rm CP}$.
In the left panels, $T$ and $\mu$ are varied along the phase transition line.
For the first-order transition side, the results on the transition line are shown for the two coexisting states. 
For the crossover side with $T>T_{\rm CP}$, the transition line is defined by the point at which 
the chiral susceptibility,
$\chi_M=\partial^2\Omega/\partial M^2$,
takes the maximum for a given temperature $T$.
In the middle panels, we set $\mu=\mu_{\rm CP}$ and vary $T$, while in the right panels $\mu$ is varied with fixed $T=T_{\rm CP}$.
The thick and thin lines are the results obtained by
the LE and TDGL approximations, respectively.
The thin-dotted lines indicate the critical exponents in Eqs.~\eqref{eq:sigma-behavior_QCDCP} and~\eqref{eq:tau-behavior_QCDCP}.
The figure shows that the numerical results well reproduce these critical behaviors near the QCD-CP.

\begin{figure}[t]
\centering
\includegraphics[keepaspectratio, scale=0.75]{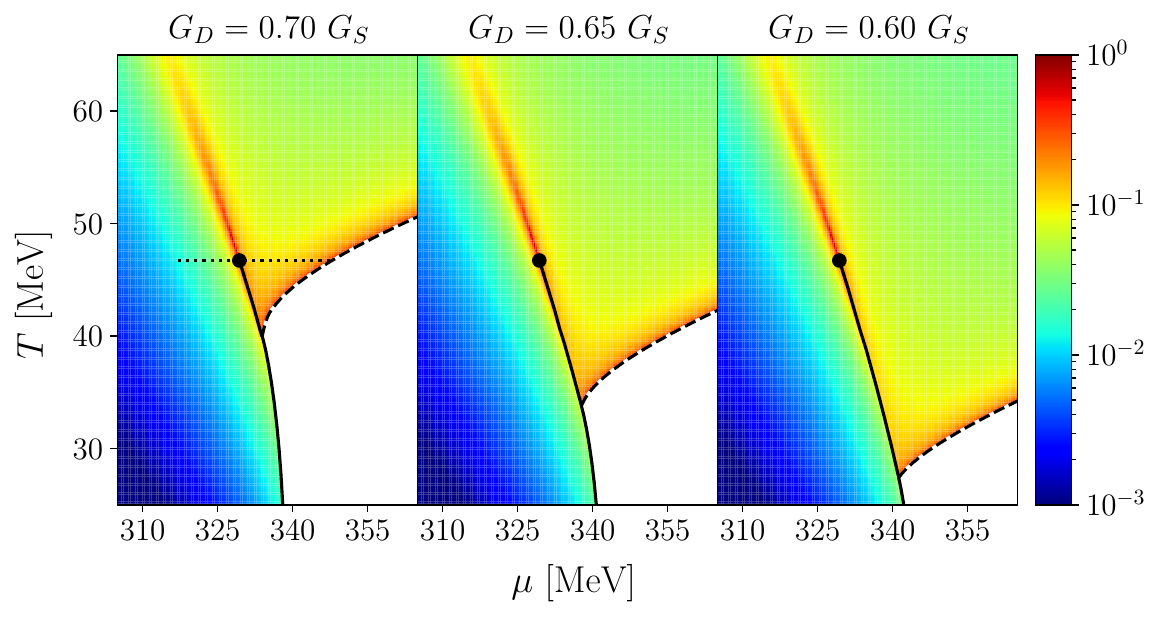}
\caption{
Contour maps of $\sigma/T$ on the $T$--$\mu$ plane 
around the CP with $G_D / G_S= 0.70$, $0.65$ and $0.60$.
The solid and dashed lines are the first-order and second-order phase transitions, respectively.
}
\label{fig:sigma_3d}
\end{figure}

Next, to see the behavior of $\sigma / T$ on the phase diagram on the $T$--$\mu$ plane including the effects of the QCD-CP and 2SC-PT simultaneously,
we show in Fig.~\ref{fig:sigma_3d} the contour color maps of $\sigma / T$ 
on the $T$--$\mu$ plane around the first-order transition line for three values of the diquark couplings $G_D / G_S= 0.70$, $0.65$ and $0.60$ obtained by the LE approximation.
The solid and dashed lines show the first-order transition and the second-order 2SC-PT, respectively.
Since our formalism does not apply to the 2SC phase where the diquark condensate has a nonzero expectation value $\Delta\ne0$, 
this phase is left blank in the figure.
One finds that $\sigma/T$ is enhanced around the QCD-CP and 2SC-PT.
A significant enhancement due to the presence of the QCD-CP occurs along the critical line parallel to the first-order transition line.
For $G_D/G_S=0.7$, this enhancement is almost connected to the one from the 2SC-PT. 
However, as $G_D$ decreases these two enhancements move away from each other as $T_c$ of the 2SC-PT is lowered.

Since the electric conductivity is related to the low energy-momentum behavior of the DPR,
Fig.~\ref{fig:sigma_3d} 
suggests that the dilepton yield in the HIC is enhanced when the medium created by the collisions passes through the red color region in the figure~\cite{Nishimura:2022mku,Nishimura:2023oqn}.
In Ref.~\cite{Nishimura:2023not}, it is shown that the dilepton yield is 
sensitive to the trajectory
and the enhancements due to the 2SC-PT and QCD-CP would be distinguishable in the beam-energy scan.

\begin{figure}[t]
\centering
\includegraphics[keepaspectratio, scale=0.44]{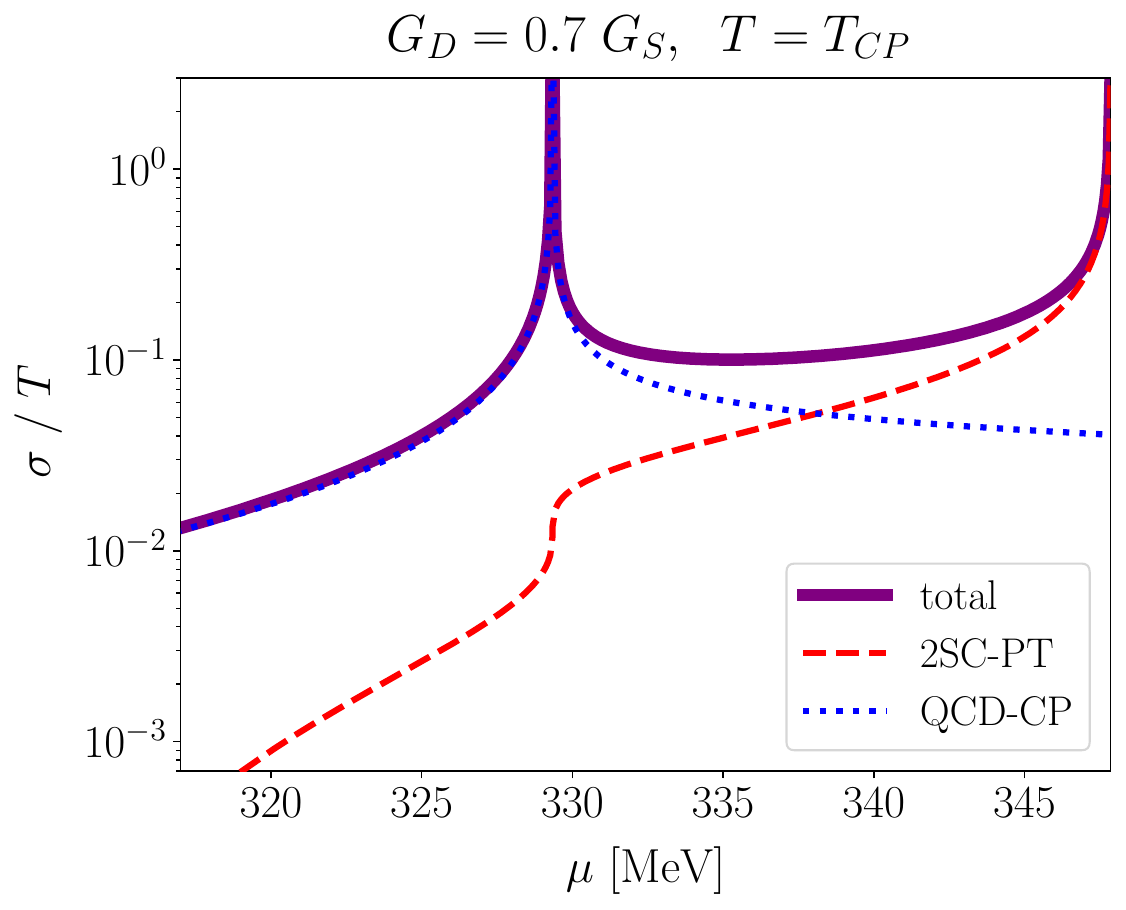}
\caption{
Behavior of $\sigma/T$ at $T = T_{\rm CP}$ and $G_D/G_S=0.7$ (dotted line in the left panel of Fig.~\ref{fig:sigma_3d}.
The dashed and dotted lines show the contributions from the 2SC-PT and QCD-CP soft modes, respectively. 
}
\label{fig:sigma_compare}
\end{figure}

Finally, in Fig.~\ref{fig:sigma_compare} we show the behavior of $\sigma/T$ as a function of $T$ at $\mu=\mu_{\rm CP}$, i.e. along the dotted line in the left panel of Fig.~\ref{fig:sigma_3d} at $G_D/G_S=0.7$.
In the figure, the solid line represents the total $\sigma$, 
while the dashed and dotted lines are the contributions from 
the soft modes of the 2SC-PT and the QCD-CP, respectively.
The contribution of the QCD-CP is divergent at $\mu_{\rm CP} \simeq 329~{\rm MeV}$.
On the other hand, the contribution of the 2SC-PT 
is divergent at $\mu \simeq 347~{\rm MeV}$ where
the transition point of the 2SC at $T = T_{\rm CP}$ exists.
One also finds that the contribution of the 2SC-PT is drastically decreasing around $\mu = \mu_{\rm CP}$.
This behavior is explained 
as follows. In our formalism, 
the soft mode of the 2SC-PT has a spectral strength in the region~(1) in Fig.~\ref{fig:Support-region}, i.e.
$|\omega + 2\mu| > \sqrt{\bm{k}^2 + 4M^2}$ as in Eq.~\eqref{eq:ImQ_2SC}.
Therefore, the soft mode can exist only for $\mu>M$, where the origin is inside the support~(1). Even for $\mu>M$, the kinematic range allowed for the soft mode is restricted as $\mu-M$ becomes smaller, which acts to suppress its contribution to $\sigma/T$.

\section{Summary}
\label{sec:Summary}

In the present article,
we have explored how the electric conductivity $\sigma$ and the associated relaxation time $\tau_\sigma$ are affected by the soft modes of the 2SC-PT and QCD-CP near these phase transitions. 
The basic motivation lies in the fact that if the phase transition is of second order there should exist the soft mode, which is the collective excitation associated with the 
fluctuations of the order parameter that becomes massless at the critical point. 
After clarifying the properties of the soft modes of these phase transitions with an emphasis on their similarities and differences, their effect on the photon self-energy for each transition is calculated with a similar formalism as in our previous study~\cite{Nishimura:2022mku,Nishimura:2023oqn}. 
The effects of the soft modes on the photon self-energy are taken into account using the idea developed in condensed matter physics to describe the para-conductivity in metals. We calculated the Aslamazov-Larkin (AL), Maki-Thompson (MT), and Density of states (DOS) terms for the photon self-energy in a gauge-invariant manner.
We have shown that the contributions of the MT and DOS terms to the self-energy cancel out with each other in the imaginary part of the spatial component, hence their calculation is not necessary for the analysis of the transport coefficients.
We have shown that $\sigma$ and $\tau_{\sigma}$ are both divergent at the 2SC-PT and QCD-CP with a power-like behavior with different critical exponents.

The present work can be extended in various directions. 
Although the present analysis is restricted to the precursory phenomena in the normal phase with vanishing diquark condensates, the same analysis could be performed
within the CSC 
phase~\cite{Jaikumar:2001jq}, 
where it is necessary to take into account the effects of the nonzero 
diquark condensates.
Another interesting subject is to 
incorporate the coupling between the soft mode of the QCD-CP 
and the density fluctuations and/or other fluid-dynamical fluctuations. 
It is known that the soft mode of the QCD-CP is a diffusive mode due to the coupling to the density fluctuations~\cite{Fujii:2003bz,Fujii:2004jt,Son:2004iv}, 
while this effect is not included in our formalism. 
To investigate this coupling in a systematic way in the NJL model,
the introduction of the vector interaction~\cite{Kitazawa:2001ft} would be indispensable. 
It would be intriguing to investigate the effects of such couplings on the various transport coefficients.

We are aware that our analysis is not self-consistent in the sense that the effects of the soft modes are not incorporated into the thermodynamic potential nor the quark propagators that compose the soft modes.
Because of this incompleteness, we have refrained from discussing the critical behavior of static observables, such as the quark-number susceptibility $\chi$, whereas the consistent analysis of $\chi$ 
enables us to calculate interesting observables such as the diffusion coefficient $D=\sigma/\chi$ 
and the shock velocity $c_s=(D/\tau_\sigma)^{1/2}$ near the critical points.
We leave these analyses for a future study.

Finally, let us briefly discuss the possibility of
an experimental confirmation of the 
present theoretical work in the HIC.
The electric conductivity $\sigma$
is related to the low energy-momentum limit of the DPR, and accordingly the DPR in the low invariant-mass region should virtually provide us with the value of $\sigma$. 
Our result in Fig.~\ref{fig:sigma_3d} suggests that there are two ``hot spots'' of the DPR on the QCD phase diagram 
reflecting the QCD-CP and 2SC-PT~\cite{Nishimura:2022mku,Nishimura:2023oqn}.
This structure indicates that the enhancements of the DPR due to the QCD-CP and 2SC-PT
can be observed in separate regions in the beam-energy scan~\cite{Nishimura:2023not}.
With the aid of the recent and forthcoming developments of experimental technologies, it would be possible to extract the electric conductivity directly in the HIC~\cite{ALICE:2022wwr}.
It is to be noted, however, that for obtaining clear-cut data it is indispensable to quantify other medium effects affecting the DPR~\cite{Rapp:2009yu,Laine:2013vma,Ghiglieri:2014kma,Kaczmarek:2022ffn,Sakai:2023fbu}.
We also note that important low-energy phenomena around the critical points include the critical slowing down, 
which is a nonequilibrium effect of the second-order phase transitions.
Although this effect will be significant near the critical point, it has not been considered throughout this study since our formalism assumes a thermal equilibrium.
Investigating the non-equilibrium effects of the critical fluctuations is another challenging extension of the present study.

\section*{Acknowledgements}
The authors thank Hirotsugu Fujii, Akira Ohnishi, and Sanjay Reddy for their valuable comments.
T.~N. thanks JST SPRING (Grant No.~JPMJSP2138) and
Multidisciplinary PhD Program for Pioneering Quantum Beam Application.
This work was supported in part by JSPS KAKENHI (Nos.~19H05598, 22K03619, 23H04507) and by the Center for Gravitational Physics and Quantum Information (CGPQI) at Yukawa Institute for Theoretical Physics.

\appendix

\section{Coefficients of the TDGL approximation}
\label{sec:TDGL}

In this appendix, we derive Eqs.~\eqref{eq:a-tilde_D}--\eqref{eq:c-i_D}.

First, from Eq.~\eqref{eq:a_D} $\tilde a_D$ is given by
\begin{align}
\tilde a_D = \frac{N_f (N_c-1)}{2 \pi^2 T} 
\int_{-\Lambda-\mu}^{\Lambda-\mu} dp (p+\mu)^2 \cosh^{-2} \frac p{2T} \bigg|_{T=T_c}.
\label{eq:a_D1}
\end{align}
By taking the limit $\Lambda\to\infty$, one obtains
\begin{align}
\tilde a_D = \frac{N_f (N_c-1)\mu^2}{2 \pi^2}  
\Big( C_0 + 2 C_1 \frac T\mu + C_2 \frac{T^2}{\mu^2} \Big) \bigg|_{T=T_c},
\end{align}
with $C_n = \int_{-\infty}^{\infty} dx x^n \cosh^{-2} (x/2)$. Equation.~\eqref{eq:a-tilde_D} is then obtained by substituting $C_2 = 4 \pi^2 / 3$, $C_1=0$, and
$C_0 = 4$.

Next, from Eq.~\eqref{eq:abc_TDGL} $b_D$ is represented explicitly as
\begin{align}
b_D = \ \frac{N_f(N_c-1)}{4\pi^2} 
\bigg(
& \frac{4T}{3\mu} \sum_{s=\pm} s \log \cosh \frac{\Lambda - s\mu}{2T}
+ \int_{-\Lambda-\mu}^{\Lambda-\mu} \frac{dp}{p} \tanh \frac p{2T}
\nonumber \\
&+ \frac{1}{12T^2} \int_{-\Lambda-\mu}^{\Lambda-\mu} \frac{dp}{p} (p+\mu)^2 
\tanh \frac p{2T} \cosh^{-2} \frac p{2T} 
\bigg).
\label{eq:b_D1}
\end{align}
In the limit $\Lambda\to\infty$, the first term in the bracket in Eq.~\eqref{eq:b_D1} becomes 
\begin{align}
 \lim_{\Lambda\to\infty} \frac{4T}{3\mu} \sum_{s=\pm} s \log \cosh \frac{\Lambda - s\mu}{2T}
=  \lim_{\Lambda\to\infty} \frac{4T}{3\mu} \bigg( \frac{\Lambda-\mu}{2T}-\frac{\Lambda+\mu}{2T} \bigg)
= - \frac{4}{3} .
\label{eq:b1}
\end{align}
The second term includes a logarithmic divergence in this limit.
In order to isolate the divergence, we rewrite this term by the integral by parts as 
\begin{align}
\int_{-\Lambda-\mu}^{\Lambda-\mu} \frac{dp}{p} \tanh \frac p{2T}
=\ \sum_{s=\pm} \log \frac{\Lambda-s\mu}{2T} \tanh \frac{\Lambda-s\mu}{2T}
-\frac{1}{2T} \int_{-\Lambda-\mu}^{\Lambda-\mu} dp \log \frac {|p|}{2T} \cosh^{-2} \frac p{2T} .
\nonumber
\end{align}
Then, using the formula $\int_{-\infty}^{\infty} dx \log x \cosh^{-2} x
= -2\gamma_{\rm E} + 2\log\pi/4 \simeq - 1.63756$
and $\lim_{\Lambda\to\infty} \tanh((\Lambda\mp\mu)/2T) = 1$, we obtain 
\begin{align}
\lim_{\Lambda\to\infty} \ \int_{-\Lambda-\mu}^{\Lambda-\mu} \frac{dp}{p} \tanh \frac p{2T}
= \log \frac{\Lambda^2 - \mu^2}{4 T^2} + 2\gamma_{\rm E} - 2\log \frac{\pi}{4} .
\label{eq:b2}
\end{align}
The third term is convergent in the limit $\Lambda\to\infty$ and calculated to be
\begin{align}
\lim_{\Lambda\to\infty} \ \int_{-\Lambda-\mu}^{\Lambda-\mu} \frac{dp}{p} (p+\mu)^2 
\tanh \frac p{2T} \cosh^{-2} \frac p{2T} 
= \mu^2 \bigg( T_{-1} (T)  + 2 T_{0} (T) \frac T\mu + T_{1} (T) \frac {T^2}{\mu^2} \bigg),
\nonumber
\end{align}
with $T_n (T) = \int_{-\infty}^\infty dx x^n \tanh x \cosh^{-2} x$.
The values $T_{1} (T) = 4$, $T_0 (T) = 0$, and $T_{-1} (T) = 7\zeta(3)/\pi^2$ together with Eq.~\eqref{eq:b1} and~\eqref{eq:b2} lead to Eq.~\eqref{eq:b_D}.

Equation~\eqref{eq:c-r_D} is also obtained from 
\begin{align}
{\rm Re}\ c_D =& \ \frac{N_f(N_c-1)}{4\pi^2} 
\bigg( 
2\Lambda^2 \sum_{s=\pm} \frac{s}{\Lambda-s\mu} \tanh \frac{\Lambda-s\mu}{2T}
\nonumber \\
& \qquad
- 4 \int_{-\Lambda-\mu}^{\Lambda-\mu} \frac{dp}{p} (p+\mu) \tanh \frac p{2T}
- \frac{1}{T} \int_{-\Lambda-\mu}^{\Lambda-\mu} \frac{dp}{p} (p+\mu)^2 \cosh^{-2} \frac p{2T}
\bigg) .
\label{eq:c_DR1}
\end{align}
With the same manipulations as above, Eq.~\eqref{eq:c_DR1} is calculated in the limit $\Lambda\to\infty$ as
\begin{align}
{\rm Re}\ c_D = \ \frac{N_f(N_c-1)}{4\pi^2} 
\bigg(
4&\mu \bigg[ 1 - \log \frac{\Lambda^2 - \mu^2}{4 T^2} - 2\gamma_{\rm E} + 2\log \frac{\pi}{4} \bigg] 
\nonumber \\
-& \frac{1}{T} ( T^2 C_{1} (T) + 2T\mu C_0 (T) + \mu^2 C_{-1} (T) ) 
\bigg) .
\end{align}

We emphasize that only the approximation to obtain these results is the $\Lambda\to\infty$ limit. Therefore, these results are well justified for $\Lambda\mp\mu\gg T$.

\section{Analysis for the exponents of $a_S$}
\label{sec:pole}

In this appendix, we discuss the effective potential and susceptibility near the critical point (CP) in the MFA based on the Landau theory. This allows us to understand the behavior of $a_S$ in Eq.~\eqref{eq:a_S}. 

We start from the Landau free energy density
\begin{align}
f(t,h;\Psi) = f_0 + \frac{At}2 \Psi^2 + \frac{B}{4} \Psi^4 - h \Psi,
\label{eq:Landau-energy}
\end{align}
of the order parameter $\Psi$
for the reduced temperature $t$ and the external magnetic field $h$ with $A>0$, $B>0$. The CP is located at $(t,h)=(0,0)$.
In the MFA, the physical value of the order parameter $\Psi=\bar\Psi$ is determined as the minimum of $f(\Psi)$ satisfying
\begin{align}
\frac{\partial f}{\partial \Psi} \bigg|_{\Psi = \bar\Psi} 
= At \bar\Psi + B \bar\Psi^3 - h = 0.
\label{eq:gap-eq}
\end{align}

Now, let us consider the behavior of the second-order derivative of $f(\Psi)$ 
\begin{align}
a = \frac{\partial^2 f}{\partial^2 \Psi} \bigg|_{\Psi = \bar\Psi} 
= At + 3 B \bar\Psi^2 ,
\label{eq:pole}
\end{align}
when the system approaches the CP.

First, for $h=0$ and $t>0$ we have $\bar\Psi=0$ and
\begin{align}
    a = At.
\label{eq:pole-1_CO}
\end{align}
For $t<0$, one has $\bar\Psi = (-A/B)^{1/2}$ and 
\begin{align}
a = At + 3 B (-At/B) = - 2 At = 2A|t|.
\label{eq:pole-1_1st}
\end{align} 
Notice that Eq.~\eqref{eq:pole-1_1st} is twice larger than Eq.~\eqref{eq:pole-1_CO} at the same $|t|$.

Next, when we take the limit $h\to0$ with $t=0$, one finds $\bar\Psi=(h/B)^{1/3}$ 
from Eq.~\eqref{eq:gap-eq} and thus
\begin{align}
a = 3 B^{1/3} h^{2/3} \sim h^{2/3}.
\label{eq:pole-2/3}
\end{align}
Finally, when the CP is approached linearly, i.e. $h\to0$ with $(t,h)=(ch,h)$ for a constant $c$, from Eq.~\eqref{eq:gap-eq} we have
\begin{align}
    h = \frac{B\bar\Psi^3}{1-Ac\bar\Psi} \xrightarrow[h\to0]{} 
    B\bar\Psi^3.
    \label{eq:pole_c}
\end{align}
Equation~\eqref{eq:pole_c} means that $\bar\Psi\sim(h/B)^{1/3}$ in this limit, and hence the behavior of $a$ is the same as Eq.~\eqref{eq:pole-2/3} even in this case.

In the vicinity of the QCD-CP, the singular part of the thermodynamic potential is given by Eq.~\eqref{eq:Landau-energy} with the linear mapping
\begin{align}
    \begin{pmatrix}
        t \\ h
    \end{pmatrix}
    = M
    \begin{pmatrix}
        T-T_{\rm CP}  \\
        \mu-\mu_{\rm CP}     
    \end{pmatrix},
\end{align}
with a matrix $M$.
This means that the parameter $a_S$ in the MFA is given by Eq.~\eqref{eq:a_S}.

\bibliography{reference.bib}






\end{document}